\documentclass[11pt]{article}
\pdfoutput=1

\usepackage{jheppub} 

\usepackage{amsmath,amssymb,epsfig,amsfonts}
\usepackage{graphicx}
\usepackage{epsf}
\usepackage{makeidx}
\usepackage{multirow}
\usepackage[all]{xy}
\usepackage{hyperref}
\usepackage{verbatim} 
\usepackage{rotating}
\usepackage{xcolor}
\usepackage{multirow}
\usepackage{bm}
\usepackage{cleveref}
\usepackage{slashed}
\usepackage[normalem]{ulem}

\usepackage{graphicx}
\usepackage{latexsym,amsmath,amsfonts,amssymb}
\usepackage{mathrsfs}
\usepackage{bbm}
\usepackage{bm}
\usepackage{subfigure}
\usepackage{paralist}

\usepackage{tikz}
\usepackage{diagbox}
\usetikzlibrary{positioning}
\usetikzlibrary{calc}
\usetikzlibrary{decorations.pathreplacing,calligraphy}

\usepackage{xstring}
\usetikzlibrary{decorations.pathmorphing} 
\usetikzlibrary{decorations.markings} 
\usetikzlibrary{arrows} 
\usetikzlibrary{shapes} 
\usetikzlibrary{matrix} 
\usetikzlibrary{positioning} 
\usepackage[english]{babel} 
\usepackage[autostyle]{csquotes}

\tikzstyle{brane}=[draw]
\tikzset{D7/.style={circle, draw=black, inner sep=0pt, fill=white, minimum size=3mm}}
\tikzset{hasse/.style={circle, fill,inner sep=2pt}}
\tikzset{flavor/.style={regular polygon,fill=white,regular polygon sides=4,inner sep=2.5pt, draw}}
\tikzset{gauge/.style={circle, draw,inner sep=2.5pt}}
\tikzset{gaugeb/.style={circle, draw,fill=black,inner sep=2.5pt}}
\tikzset{gauger/.style={circle, draw,fill=cyan,inner sep=2.5pt}}
\tikzset{gaugeg/.style={circle, draw,fill=red,inner sep=2.5pt}}
\tikzset{bd/.style={circle, draw=black, inner sep=0pt, fill=black, minimum size=2mm}}
\tikzset{wd/.style={circle, draw=black, inner sep=0pt, fill=white, minimum size=2mm}}
\tikzset{Dynkin/.style={circle, draw=black, inner sep=0pt, fill=white, minimum size=2mm}}
\tikzstyle{ligne}=[draw, thick] 
\tikzset{doublearrow/.style={ draw=black!75, color=black!75, thick, double distance=3pt, }}

\numberwithin{equation}{section}  

\graphicspath{ {figs/} }

\newcommand{\be}{\begin{equation}}
\newcommand{\ee}{\end{equation}}
\newcommand{\ba}{\begin{aligned}}
\newcommand{\ea}{\end{aligned}}

\def\unit{{1\kern-.65ex {\rm l}}}
\def\1{{1\kern-.65ex {\rm l}}}

\newcount\hour \newcount\minute
\hour=\time \divide \hour by 60
\minute=\time
\def\now{%
\ifnum \hour<13
  \ifnum \hour=0 \advance \hour by 12 \number\hour:\else \number\hour:\fi%
     \ifnum \minute<10 0\fi%
     \number\minute%
\ A.M.%
\else \advance \hour by -12 \number\hour:%
  \ifnum \minute<10 0\fi%
  \number\minute%
  \ P.M.%
\fi%
}

\makeatother

\def\mb{\mathbb}
\def\mbf{\mathbf}
\def\mc{\mathcal}
\def\mk{\mathfrak}
\def\bp{\begin{pmatrix}}
\def\ep{\end{pmatrix}}

\def\la{\langle}
\def\ra{\rangle}
\def\ptl{\partial}

\newcommand{\bea}{\begin{equation} \begin{aligned}}
 \newcommand{\eea}{\end{aligned} \end{equation}}
\newcommand{\bit}{\begin{itemize}} 
\newcommand{\eit}{\end{itemize}}

\renewcommand{\t}{\widetilde }

\newcommand{\h}{\widehat}

\newcommand{\MG}{{\mathbf X}} 

\newcommand{\FT}{{\mathcal{T}_\MG^{\rm 5d}}} 

\newcommand{\FTfour}{{\mathscr{T}_\MG^{\rm 4d}}} 

\newcommand{\EQfour}{{\text{EQ}^{(4)}}} 
\newcommand{\MQfour}{{\text{MQ}^{(4)}}} 
\newcommand{\EQfive}{{\text{EQ}^{(5)}}} 
\newcommand{\MQfive}{{\text{MQ}^{(5)}}} 

\tikzstyle{brane}=[draw]
\tikzset{D7/.style={circle, draw=black, inner sep=0pt, fill=white, minimum size=3mm}}
\tikzset{hasse/.style={circle, fill,inner sep=2pt}}
\tikzset{flavor/.style={regular polygon,fill=white,regular polygon sides=4,inner sep=2.5pt, draw}}
\tikzset{gauge/.style={circle, draw,inner sep=2.5pt}}
\tikzset{gaugeb/.style={circle, draw,fill=black,inner sep=2.5pt}}
\tikzset{gauger/.style={circle, draw,fill=cyan,inner sep=2.5pt}}
\tikzset{gaugeg/.style={circle, draw,fill=red,inner sep=2.5pt}}
\tikzset{SUd/.style={circle, draw=black, inner sep=0pt, fill=yellow, minimum size=2mm}}
\tikzset{bd/.style={circle, draw=black, inner sep=0pt, fill=black, minimum size=2mm}}
\tikzset{wd/.style={circle, draw=black, inner sep=0pt, fill=white, minimum size=2mm}}
\tikzset{Dynkin/.style={circle, draw=black, inner sep=0pt, fill=white, minimum size=2mm}}
\tikzstyle{ligne}=[draw, thick] 
\tikzset{doublearrow/.style={ draw=black!75, color=black!75, thick, double distance=3pt, }}

\begin{document}

\baselineskip=18pt  
\numberwithin{equation}{section}  
\allowdisplaybreaks  


%
%


\thispagestyle{empty}

\vspace*{0.8cm} 
\begin{center}
{{\Huge  5d SCFTs from Isolated Complete Intersection Singularities
}}

\vspace*{1.5cm}

 {\large Jisheng Mu$^1$, Yi-Nan Wang$^{1,2}$, Hao N. Zhang$^1$}

\vspace*{1.0cm}

\smallskip
{\it $^1$ School of Physics,\\
Peking University, Beijing 100871, China\\ }

\smallskip

{\it  $^2$ Center for High Energy Physics, Peking University,\\
Beijing 100871, China}

\vspace*{0.8cm}
\end{center}

\vspace*{.5cm}

\noindent

In this paper, we explore the zoo of 5d superconformal field theories (SCFTs) constructed from M-theory on Isolated Complete Intersection Singularities (ICIS). We systematically investigate the crepant resolution of such singularities, and obtain a classification of rank $\leqslant 10$ models with a smooth crepant resolution and smooth exceptional divisors, as well as a number of infinite sequences with the same smoothness properties. For these models, we study their Coulomb branch properties and compute the flavor symmetry algebra from the resolved CY3 and/or the magnetic quiver. We check the validity of the conjectures relating the properties of the 5d SCFT and the 4d $\mathcal{N}=2$ SCFT  from IIB superstring on the same singularity. When the 4d $\mathcal{N}=2$ SCFT has a Lagrangian quiver gauge theory description, one can obtain the magnetic quiver of the 5d theory by gauging flavor symmetry, which encodes the 5d Higgs branch information. Regarding the smoothness of the crepant resolution and integrality of 4d Coulomb branch spectrum, we find examples with a smooth resolved CY3 and smooth exceptional divisors, but fractional 4d Coulomb branch spectrum. Moreover, we compute the discrete (higher)-symmetries of the 5d/4d SCFTs from the link topology for a few examples.
\newpage

\tableofcontents

\newpage


\section{Introduction}

String/M-/F-theory provides powerful tools of defining and studying many unconventional supersymmetric quantum field theories. In particular, M-theory on Calabi-Yau (canonical) threefold singularities $X$ realize a large class of 5d $\mc{N}=1$ superconformal field theories (SCFTs) $\FT$~\cite{Seiberg:1996bd,Morrison:1996xf,Intriligator:1997pq,Aharony:1997bh,Benini:2009gi,Kim:2012gu,Bergman:2013aca,Zafrir:2014ywa,Hayashi:2015zka,Xie:2017pfl,Ferlito:2017xdq,Jefferson:2017ahm,Hayashi:2018bkd,Hayashi:2018lyv,Jefferson:2018irk,Bhardwaj:2018vuu,Closset:2018bjz,Cabrera:2018jxt,Apruzzi:2018nre,Bhardwaj:2018yhy,Apruzzi:2019vpe,Apruzzi:2019opn,Apruzzi:2019enx,Apruzzi:2019kgb,Bhardwaj:2019ngx,Bhardwaj:2019jtr,Bhardwaj:2019fzv,Bhardwaj:2019xeg,Hayashi:2019jvx,Saxena:2020ltf,Bhardwaj:2020gyu,Eckhard:2020jyr,Morrison:2020ool,Collinucci:2020jqd,Closset:2020scj,vanBeest:2020kou,Bhardwaj:2020ruf,Bhardwaj:2020avz,VanBeest:2020kxw,Kim:2020hhh,Hayashi:2021pcj,Apruzzi:2021vcu,Collinucci:2021ofd,vanBeest:2021xyt,Acharya:2021jsp,Tian:2021cif,Closset:2021lwy,Kim:2021fxx,DelZotto:2022fnw,Collinucci:2022rii,Xie:2022lcm,DeMarco:2022dgh,Bourget:2023wlb,DeMarco:2023irn}. It was long known that one can read off the Coulomb branch information of the 5d SCFT from the crepant resolution of such singularities. On the other hand, one can engineer 4d $\mc{N}=2$ SCFTs $\FTfour$ from IIB superstring theory on isolated canonical threefold singularities $X$ as well~\cite{Shapere:1999xr,Cecotti:2011rv,Alim:2011ae,Xie:2012hs,DelZotto:2015rca,Xie:2015rpa,Wang:2016yha,Chen:2016bzh,Chen:2017wkw,Closset:2020scj,Closset:2020afy,Closset:2021lwy}. The Coulomb branch spectrum of the 4d SCFT can be computed from the deformation properties of the singularity.

In \cite{Closset:2020scj,Closset:2020afy,Closset:2021lwy}, a novel relation between the Coulomb branches and Higgs branches of $\FT$ and $\FTfour$ was proposed, and we denote it as the 5d-4d relation. Assuming that we have a quiver gauge theory description of the Coulomb branch of $\FTfour$, after the dimensional reduction to 3d and gauging $U(1)^f$ (where $f$ is the flavor rank of $\FTfour$), one obtains the magnetic quiver of $\FT$ which encodes the Higgs branch properties of $\FT$~\cite{Cremonesi:2015lsa,Ferlito:2017xdq,Cabrera:2018jxt,Bourget:2019aer,Akhond:2020vhc,Bourget:2020gzi,vanBeest:2020kou,VanBeest:2020kxw,Bourget:2021jwo}\footnote{Another similar relation was observed for the Coulomb branch of $\FT$ and Higgs branch of $\FTfour$, but we will not investigate these in this paper.}. We observed such relation to hold for many hypersurface examples of $X$. The cases of non-Lagrangian $\FTfour$ are much more complicated, which were partially explored in \cite{Closset:2020afy}. 

From the hypersurface examples, a curious conjecture relating the smoothness of the crepantly resolved $\widetilde{X}$ and the Coulomb branch spectrum of $\FTfour$ was proposed~\cite{Closset:2021lwy}: when there exists a smooth $\widetilde{X}$ with smooth compact divisors $S_i$, the Coulomb branch spectrum of $\FTfour$ is integral. On the other hand, if the Coulomb branch spectrum of $\FTfour$ is integral, there exists a smooth $\widetilde{X}$. We call these ``smoothness-integrality'' conjectures.

In this paper, we extend the philosophy of the 5d-4d relation in \cite{Closset:2020scj,Closset:2020afy,Closset:2021lwy} to the cases of isolated complete intersection canonical threefold singularities (ICIS) $X$, defined as the intersection of two quasi-homogeneous equations in the $\mb{C}^5$ ambient space. Such singularities were classified into 303 classes in \cite{Chen:2016bzh}. For a number of these classes, the $\FTfour$ interpretations were also discussed in \cite{Wang:2016yha}. From the Lagrangian quiver $\FTfour$, one can construct the magnetic quiver 
 $\MQfive$ of $\FT$ using the 5d-4d relation. 
 
 For the Coulomb branch data of $\FT$, we innovated a systematic crepant resolution algorithm for the ICIS examples, based on weighted blow-ups of the $\mb{C}^5$ ambient space. For many examples of a smooth resolved $\widetilde{X}$, we computed the triple intersection numbers and other physical information that can be read off from the Coulomb branch. In particular, for the cases of a smooth $\widetilde{X}$ (with smooth compact divisors $S_i$) and $b_3(\widetilde{X})=0$, we present the details of the infinite sequences and sporadic examples with $r\leqslant 10$ in section \ref{sec:smooth-zerob3}. In these cases, we systematically studied the UV flavor symmetry algebra $G_F$ with two methods:
\begin{itemize}
\item If there exists a good unitary quiver $\MQfive$, we can compute the Hasse diagram of the 5d Higgs branch from quiver subtraction~\cite{Cabrera:2018ann}, and the non-abelian factors $G_{F,nA}\subset G_F$ can be read off from the bottom layer of the Hasse diagram.
\item If the previous approach is unavailable, we try to find the possible additional flavor curves on the compact divisors $S_i$. M2-branes wrapping these flavor curves would lead to W-bosons for the non-abelian flavor symmetry factors $G_{F,nA}$. We describe the methodology in detail in section~\ref{sec:flavor}.
\end{itemize}
 
 We also summarize the physical data of many smooth examples with $b_3(\widetilde{X})>0$ in appendix \ref{app:tables}. In particular, we discuss the detail of the Class (41) in appendix \ref{sec:41}, which is a case where the 4d description $\FTfour$ has $a>c$.

 Interestingly, we found a counter-example to the smoothness-integrality conjecture in \cite{Closset:2021lwy}, which is the infinite sequence Class (209). The equation is
 \be
\begin{cases}
&z_1 z_2+z_3 z_4=0 \\
&z_1 z_3+z_2^n z_3+z_4^3+z_5^3=0\,.
\end{cases}
\ee
When $n=2k+1$, the resolution is smooth, while the Coulomb branch spectrum is
\be
\ba
&1^5,\underbrace{\frac{3}{2}^2,2,\frac{5}{2}^2,3^2,......,\frac{2n-3}{2}^2,n-1,\frac{2n-1}{2}^2,n^2},\frac{2n+1}{2}^2,n+1,\frac{2n+3}{2}^3,\cr
&\underbrace{{n+2}^2,n+3,\frac{2n+7}{2},......,{2n+1}^2,2n+2,\frac{4n+5}{2}},
\underbrace{\frac{4n+9}{2},\frac{4n+13}{2},......,\frac{6n+3}{2}}\,,
\ea
\ee
which is not integral. We discuss this example in detail in section~\ref{sec:209}. 

On the other hand, we also find examples with an integral 4d CB spectrum but a smooth $\widetilde{X}$ with singular divisors. We list such examples in appendix~\ref{app:sin-div}.

 Furthermore, we computed the topology of the link Sasaki-Einstein $L_5=\partial X$ for a few examples in section~\ref{sec:linktop}. The torsion subgroup of $H_2(L_5,\mb{Z})$ gives the 1-form symmetry of $\FTfour$ as well the 0-form/3-form symmetry of $\FT$~\cite{Closset:2020scj,Bhardwaj:2023kri,Luo:2023ive}. Nonetheless, a general formula of the link topology for an ICIS has not been developed in the mathematics literature and it is subject to future study.

 The structure of the paper is as follows. In section \ref{sec:ICIS} we review the mathematics of ICIS,  describe the systematic crepant resolution algorithm proposed in this paper, as well as the computation of link topology that apply to a limit number of cases. In section \ref{sec:4d-5d-phys} we discuss the physics of $\FT$ and $\FTfour$, including the basic physics dictionaries, the relation between the $\FT$ and $\FTfour$ and the procedure to read off non-abelian flavor symmetry algebra from the resolution. We present the detailed discussion of the Class (1) example. In section \ref{sec:smooth-zerob3} we give the details of the models with a smooth $\widetilde{X}$, smooth compact divisors and $b_3=0$. We show the triple intersection numbers of $\widetilde{X}$ with figures, and present the physical data and the calculation of flavor symmetry $G_F$ for $\FT$. In section \ref{sec:discussions} we discuss about some future directions.

 In appendix~\ref{app:tables} we list the tables of smooth models with $b_3\geqslant 0$, for infinite sequences and sporadic cases. In appendix~\ref{app:link-top} we review the algorithm of computing the link topology. In appendix~\ref{app:sin-div} we list a few examples with integral 4d CB spectrum and singular exceptional divisors in $\widetilde{X}$. In appendix~\ref{sec:41} we write down the details of the Class (41) example with $a>c$.
 
\section{Isolated Complete Intersection Singularities}
\label{sec:ICIS}

In this paper, we consider canonical threefold singularities $X$ in form of a complete intersection of two equations in $\mb{C}^5$:
\be
\begin{cases}
&f_1(z_1,z_2,z_3,z_4,z_5)=0\\
&f_2(z_1,z_2,z_3,z_4,z_5)=0
\end{cases}
\ee
We restrict ourselves to isolated singularities, which means that $X$ is only singular at the origin $z_1=z_2=z_3=z_4=z_5=0$. Note that in the cases of complete intersections, the singular condition means that the Jacobian matrix $\{\ptl f_i/\ptl z_j\}$ is degenerate (has rank $<2$).

We further require that the two equations are quasi-homogeneous, such that we can assign weights $q_i$ to each variable $z_i$, and the equations $(f_1,f_2)$ has weights $(1,d)$ respectively ($q_i$, $d\in\mb{Q}$). Under these conditions, the 303 classes of canonical threefold ICIS singularities are classified in \cite{Chen:2016bzh}. We will refer to the ``Class'' of these singularities using the notations of \cite{Chen:2016bzh}. The deformations of these singularities are well known, which will be discussed in section~\ref{sec:4d-def}. On the other hand, the crepant resolutions of such singularities are much less studied, and we will develop a systematic resolution algorithm for the physics purposes.

\subsection{Crepant resolution}
\label{sec:CR}

Mathematically, a crepant resolution of a canonical singularity $X$ is a birational map $\phi:\widetilde{X}\rightarrow X$ which leaves the canonical divisor $K_X$ invariant: $\phi^*(K_X)=K_{\widetilde{X}}$, such that the supersymmetry of the geometrically engineered QFT is fully preserved. Note that for many singularities $X$, $\widetilde{X}$ would still have Q-factorial terminal singularities after a crepant resolution, and such cases are generally allowed. We would not consider the further crepant small resolutions of terminal singularities, since we are mostly interested in the properties of exceptional divisors in $\widetilde{X}$.

More specifically, we formulate a crepant resolution of an ICIS as a sequence of toric blow-ups of the $\mb{C}^5=\mb{C}[z_1,z_2,z_3,z_4,z_5]$ ambient space\footnote{We will not consider more general types of crepant resolution in this paper.}. Note that before the blow-ups, the ambient space $\mb{C}^5=\mb{C}[z_1,z_2,z_3,z_4,z_5]$ is described by a toric fan with rays
\be
V_1=(1,0,0,0,0)\ ,\ V_2=(0,1,0,0,0)\ ,\ V_3=(0,0,1,0,0)\ ,\ V_4=(0,0,0,1,0)\ ,\ V_5=(0,0,0,0,1)\,
\ee
and the 5d cone $V_1 V_2 V_3 V_4 V_5$. 
Each ray $V_i$ corresponds to the non-compact divisors $D_i$: $z_i=0$.

Using the notations of \cite{Lawrie:2012gg,Apruzzi:2019opn,Tian:2021cif,Closset:2021lwy},  each weighted blow-up of the locus $x_1=x_2=\dots=x_k=0$ ($x_i$ can be any global toric coordinate) of the toric fivefold ambient space is denoted as
\be
(x_1^{(w_1)},x_2^{(w_2)},\dots,x_k^{(w_k)};\delta_j)\,.
\ee
The blow-up weights are $(w_1,w_2,\dots,w_k)$, which are omitted if they all equal to one. $\delta_j=0$ is the exceptional divisor in the toric ambient space.

If the toric ray of each divisor $x_i=0$ is denoted as $v_i$, then the new ray $\tilde{v}$ of the exceptional divisor $\delta_j=0$ after the blow-up is given by
\be
\tilde{v}=\sum_{i=1}^k w_i v_i\,.
\ee

After one substitute into the equations $f_1$ and $f_2$
\be
x_i\rightarrow x_i'\delta_j^{w_i}\quad (i=1,\dots,k)\,,
\ee
$f_1$ and $f_2$ vanishes to order $d_1$ and $d_2$ at $\delta_j=0$. We define
\be
f_1=f_1'\delta_j^{d_1}\ ,\ f_2=f_2'\delta_j^{d_2}\,.
\ee

Now we derive the condition for such a  step of crepant resolution $\phi:X'\rightarrow X$. Denote each divisor  $x_i=0$ of $X$ by $D_i$, the new divisor $x_i'=0$ of $X'$ corresponds to $\phi^*(D_i)=D_i-w_i E$ after the blow-up. We denote the divisor class of $f_i=0$ in the toric ambient space  $T_\Sigma$ by $F_i$. Before the blow-up, from the adjunction formula the canonical class of $X$ is given by
\be
K_X=(K_\Sigma+F_1+F_2)\cdot F_1\cdot F_2.
\ee
The Calabi-Yau condition of $X$ applies that 
\be
K_\Sigma=-F_1-F_2\,.
\ee
After the blow up, $K_\Sigma$ is transformed as
\be
K_{\Sigma'}=\phi^*(K_\Sigma)+\sum_{i=1}^k w_i E-E\,.
\ee
Hence the canonical class of $X'$ is
\be
K_{X'}=(K_{\Sigma'}+F_1'+F_2')\cdot F_1'\cdot F_2'\,,
\ee
which is trivial if and only if 
\be
\label{CY3-cond2}
K_{\Sigma'}+F_1'+F_2'=0\,.
\ee
Since the proper transformation of $F_i$ is given by
\be
\label{CI-proper-transform}
F_i'=\phi^*(F_i)-d_i E\,,
\ee
the condition for a crepant resolution is
\be
\sum_{i=1}^k w_i-1=d_1+d_2\,.\label{CI-crepant}
\ee
After the blow-up, a new SR ideal generator $x_1 x_2\dots x_k$ in the ambient space will be added. In the new toric fan, the $k$-dimensional cone $v_1 v_2\dots v_k$ in $\Sigma$ is subdivided into $\tilde{v} v_2\dots v_k$, $v_1 \tilde{v}\dots v_k$, $v_1 v_2\dots v_{k-1}\tilde{v}$. The same subdivision applies to any higher-dimensional cone in $\Sigma$ that contains $v_1 v_2\dots v_k$ as a sub-cone.

Finally, the triple intersection numbers of divisors $S_i$ in $\widetilde{X}$ can be computed from the ambient space divisors $S_{\Sigma',i}$ as
\be
S_i\cdot S_j\cdot S_k=F_1'\cdot F_2'\cdot S_{\Sigma',i}\cdot S_{\Sigma',j}\cdot S_{\Sigma',k}\,.
\ee
After the blow-up we will typically replace the notations of $F'$, $x'$ and $f'$ by $F$, $x$, $f$ for simplifications.

In the cases of isolated hypersurface singularities, we encounter terminal singularities $X$ that give rise to rank-0 5d SCFTs~\cite{Closset:2020scj,Closset:2020afy,Closset:2021lwy}. For the cases of ICIS, there is at least one compact divisor and $r\geqslant 1$. The reason is that $f_1$ and $f_2$ always vanish to at least order two in $z_1$, $z_2$, $z_3$, $z_4$ and $z_5$, and hence it is possible to perform the blow-up $(z_1,z_2,z_3,z_4,z_5;\delta_1)$ in the resolution sequence. Without loss of generality, if there is a linear term $z_1$ in $f_1$, such as 
\be
f_1=z_1+\tilde{f}_1\,,
\ee
then one can plug $z_1=-\tilde{f}_1$ into the second equation $f_2$, and the singularity becomes a hypersurface, which contradicts the definition of an ICIS.

Similarly, for the cases of ICIS, if there is a residual terminal singularity after the crepant resolution, its local form can always be rewritten into a hypersurface singularity, see the discussions of Class (1) in section~\ref{sec:class1} when $n=2k+1$.

\subsection{Computation of resolution sequences}
\label{sec:resol}

To systematically compute the resolution sequence, we propose the following two steps:
\begin{enumerate}
\item{From the defining equations $f_1=f_2=0$, write down the rays in the toric ambient space $T_{\Sigma'}$ after the full crepant resolution (not including small resolutions).}
\item{Generate the blow up sequence on the toric ambient space $\mb{C}^5$ from the list of rays obtained in step 1.}
\end{enumerate}

\paragraph{Step 1}

The rays of the toric ambient space $T_{\Sigma'}$ after the blow-ups include

\begin{enumerate}
\item $V_1=(1,0,0,0,0)$, $V_2=(0,1,0,0,0)$, $V_3=(0,0,1,0,0)$, $V_4=(0,0,0,1,0)$ , $V_5=(0,0,0,0,1)$, which are associated to non-compact divisors $D_1$, $D_2$, $D_3$, $D_4$ and $D_5$;

\item $\{v_i\}$ associated to the exceptional divisors $\{E_i\}$. For an isolated singularity, the resolution sequence should not produce non-compact exceptional divisors. Hence for any $v_i=(v_{i,1},v_{i,2},v_{i,3},v_{i,4},v_{i,5})$, we require that $v_{i,k}>0$ $(k=1,2,3,4,5)$.
\end{enumerate}

There are linear equivalence relations
\be
\label{linear-equiv}
D_k+\sum_i v_{i,k}E_i=0\quad (k=1,2,3,4,5)\,.
\ee

The anticanonical divisor of $T_{\Sigma'}$ is 
\be
-K_{\Sigma'}=D_1+D_2+D_3+D_4+D_5+\sum_i E_i\,.
\ee

Due to (\ref{CY3-cond2}), one should decompose it into
\be
-K_{\Sigma'}=F_1'+F_2'\,,
\ee
where $F_j'$ $(j=1,2)$ corresponds to the divisor class of the equation $f_j'=0$ after the resolution.

We decompose $F_j'$ into
\be
\label{Fj-decomp}
F_j'=-\sum_i a_{i,j}E_i\quad (j=1,2)\,,
\ee
where $a_{i,j}$ are undetermined integer coefficients at this stage. 
Note that we can use the five linear equivalence relations (\ref{linear-equiv}) to remove $D_i$ $(i=1,2,3,4,5)$ in the above expression.

With $a_{i,j}$, the generators for the line bundle $\mc{O}(F_j')$ of $T_{\Sigma'}$ correspond to the lattice points $u\in P_j$ in the $M$-lattice of the toric ambient fivefold. The polytope $P_j$ is defined as
\be
P_j=\{u=(u_1,u_2,u_3,u_4,u_5)\in \mb{Z}^5|\la u,v_i\ra\geqslant a_{i,j}\  (\forall v_i)\ ,\ u_k\geqslant 0\ (k=1,2,3,4,5)\}\ (j=1,2)\,.
\ee
Each lattice point $u$ can be naturally mapped to a monomial in the resolved equation $f_j'=0$. After the push-forward to the singular equation $f_j=0$, the lattice point $u\in P_j$ corresponds to the following monomial in the original singular equation $f_j=0$:
\be
m_u=z_1^{u_1} z_2^{u_2} z_3^{u_3} z_4^{u_4} z_5^{u_5}\,.
\ee
As a consistency check, the lowest order terms $m_u$ ($u\in P_j$) should match the monomials in the original equation $f_j=0$.

The question is then how to compute $\{v_i\}$ and $F_j'$ ($j=1,2$) from the singular equation $f_1=f_2=0$, such that the lowest order monomials in $P_j$ match the ones in $f_j=0$ $(j=1,2)$, and the following equation holds:
\be
-K_{\Sigma'}=F_1'+F_2'\,.
\ee

The key is to solve the following set of equations which $v_i=(v_{i,1},v_{i,2},v_{i,3},v_{i,4},v_{i,5})$ and the integer coefficients $a_{i,1}$, $a_{i,2}$ in (\ref{Fj-decomp}) satisfy:
\be
\boxed{
\ba
\label{solve-v-eq}
&v_{i,k}>0\quad (k=1,2,\dots,5)\,,\cr
 &a_{i,j}>0\quad (j=1,2)\,,\cr
&\sum_{k=1}^5 v_{i,k}-1=a_{i,1}+a_{i,2}\,,\cr
&\forall m_u\equiv z_1^{u_1} z_2^{u_2} z_3^{u_3} z_4^{u_4} z_5^{u_5}\in f_j\ ,\ \la u,v_i\ra\geqslant a_{i,j}\quad (j=1,2)\,.
\ea
}
\ee
In (\ref{solve-v-eq}), the third equation is the condition for crepant resolution, and the fourth equation guarantees that the monomials in $f_j$ are indeed the generators for $\mc{O}(F_j)$.

One takes all the solutions $(v_i, a_{i,1}, a_{i,2})$ to (\ref{solve-v-eq}), and obtain the list of $\{v_i\}$ as well as
\be
F_j'=\sum_i a_{i,j}E_i\,.
\ee

\paragraph{Step 2} Now from the list of $\{v_i\}$, we need to construct the resolution sequence, which generates $\{v_i\}$ via a number of weighted blow-ups of the toric ambient fivefold $T_\Sigma$.

To generate the full list of exceptional toric rays for all the cases of ICIS, we need the following types of weighted blow-ups on $T_\Sigma$ (here $x_i$ denotes any global toric coordinate):
\begin{enumerate}
\item{Blow up a codimension-two locus $x_1=x_2=0$: $(x_1,x_2;\delta)$.}
\item{Blow up a codimension-three locus $x_1=x_2=x_3=0$: $(x_1,x_2,x_3;\delta)$.}
\item{Blow up a codimension-four locus $x_1=x_2=x_3=x_4=0$:
$(x_1,x_2,x_3,x_4;\delta)$, 
$(x_1^{(1)},x_2^{(1)},x_3^{(1)},x_4^{(2)};\delta)$, $(x_1^{(1)},x_2^{(1)},x_3^{(2)},x_4^{(3)};\delta)$. These weights are the ones used in \cite{caibar1999minimal,Closset:2021lwy}.
}

\item{Blow up codimension-five locus $x_1=x_2=x_3=x_4=x_5=0$. In this  case, the full list of blow-up weights is undiscovered in the literature. In this paper, we observe the following list of blow-up weights that need to be used in the resolution of ICIS examples, in order to generate all the rays $v_i$ in the toric fan of $T_\Sigma'$:
\be
\ba
\label{blowup-weights}
&(1,1,1,1,1)\ ,\ (2,1,1,1,1)\ ,\ (2,2,1,1,1)\ ,\ (3,1,1,1,1)\ ,\ (3,2,1,1,1)\cr
&(3,2,2,1,1)\ ,\ (3,3,2,1,1)\ ,\ (4,3,2,1,1)\ ,\ (5,3,2,1,1)\,.
\ea
\ee}

\end{enumerate}

In order to check the smoothness of the resolved CY3 $\widetilde{X}$, one need to check the absence of singularity on each coordinate patch. In practice we take each coordinate patch of the toric fivefold ambient space $T'_\Sigma$, which one-to-one corresponds to the 5d cones of $\Sigma'$. Denote such a coordinate patch by $\{x_1 x_2 x_3 x_4 x_5\}$, where we allow $x_1=x_2=x_3=x_4=x_5=0$ simultaneously. One need to check the following two things to make sure that there is no singularity on this patch:
\begin{enumerate}
\item{The Jacobian matrix $\{\ptl f_i/\ptl x_j\}$ has rank two.
}
\item{If the ambient space $T'_\Sigma$ is singular on a locus $\mc{S}$, $\mc{S}$ cannot intersect $\widetilde{X}:f_1=f_2=0$. Note that $\mc{S}$ can have different codimensions. For instance, if the there is a set of rays  $v_{i_1}$, $v_{i_2}$, $\dots$, $v_{i_j}$ such that
for all possible 5d cones $\sigma_5$ that contain $v_{i_1}v_{i_2}\dots v_{i_j}$ as a sub-cone, 
\be
\mathrm{gcd}(\{|\mathrm{det}(\sigma_5)|\})>1\,,
\ee
then there is a singularity $\mc{S}:x_1=x_2=\dots=x_j=0$ on the ambient space $T'_\Sigma$. One should check if such $\mc{S}$ intersects the resolved $\widetilde{X}$ or not.}
\end{enumerate}

In order to obtain a smooth resolved CY3 $\widetilde{X}$ with smooth exceptional divisors, one should carefully choose which weights to use in (\ref{blowup-weights})\footnote{The smoothness properties here are not birational invariants.}. We will list the blow-up sequence for the models in Section~\ref{sec:smooth-zerob3} on a case-by-case basis, which leads to a smooth $\widetilde{X}$ with smooth divisors.

For the compact exceptional divisors $S_i$ of $\widetilde{X}$ from the resolution sequence, one subtlety here is that the exceptional divisor $\delta_i=f_1'=f_2'=0$ is often reducible in $\widetilde{X}$. To compute the correct 5d rank $r\equiv h^{1,1}(\widetilde{X})$ in the 5d SCFT $\FT$ constructed from M-theory on $X$, one needs to correctly count all the irreducible components in each $\delta_i=0$. To achieve this, we carry out prime decomposition to the defining ideal of each $\delta_i=0$. Nevertheless, due to the complexity of the ambient space $T_{\Sigma'}$ after the blow-ups, in practice we carry out prime decomposition on each coordinate patch, and then use the transition function to check if the irreducible divisors on different patches are indeed the same. Gather all independent irreducible divisors and we will get the right $r$. 

There is also a subtlety that general algorithms performing prime decomposition to ideals on the polynomial ring $\mb{C}[z_1,z_2,z_3,z_4,z_5]$ is difficult or even impossible to find. Instead we would use algorithms on rings such as $\mb{Q}[i,\sqrt{2}][z_1,z_2,z_3,z_4,z_5]$ with carefully chosen coefficients on a case-by-case basis, to get the correct answer.

After correctly writing down all the irreducible divisors $S_i$, we can check the smoothness of $S_i$ using the same method as checking the smoothness of $\widetilde{X}$. Next, we can calculate the triple intersection numbers $S_i\cdot S_j\cdot S_k (i\neq j \neq k)$. The general principle here is quite the same as the case calculating $r$, but one difference is that we need to carry out primary decomposition, instead of prime decomposition, to the defining ideal of $S_i \cdot S_j \cdot S_k$, because here the multiplicity of zero points contributes to the intersection numbers. Once we have all those $S_i\cdot S_j\cdot S_k (i\neq j \neq k)$, we can solve the linear equivalence equations to get other triple intersection numbers. 

\subsection{An example of the resolution algorithm}
\label{sec:example-resol}

In this section, we apply the resolution algorithm to an example (Class (2) in \cite{Chen:2016bzh}):
\be
\begin{cases}
&z_1^2+z_2^2+z_3^2+z_4^3+z_5^3=0\\
&z_1^2+2z_2^2+3z_3^2+4z_4^3+5z_5^3=0\,.
\end{cases}
\ee 
Note that the two equations $f_1$ and $f_2$ have the same set of monomials, which are mapped to points in the $M$-lattice of $T_{\Sigma'}$:
\be
\ba
z_1^2: (2,0,0,0,0)\ ,\ z_2^2: (0,2,0,0,0)\ ,\ z_3^2: (0,0,2,0,0)\ ,\ z_4^3: (0,0,0,3,0)\ ,\ z_5^3: (0,0,0,0,3)\,.
\ea
\ee
From these monomials, we can obtain the solutions to the set of constraints (\ref{solve-v-eq}):
\be
\ba
\label{ex-2-v}
&v_1=(1,1,1,1,1)\ ,\ a_{1,1}=a_{1,2}=2\,,\cr
&v_2=(3,3,3,2,2)\ ,\ a_{2,1}=a_{2,2}=6\,.
\ea
\ee
From the solved $v_1$, $v_2$, we obtain the resolution sequence:
\be
\ba
\label{Class2-resol}
&(z_1,z_2,z_3,z_4,z_5;\delta_1)\cr
&(z_1^{(1)},z_2^{(1)},z_3^{(1)},\delta_1^{(2)};\delta_2)\,.
\ea
\ee

It is easy to check again that the $v_i$ for the exceptional divisors $E_i:\delta_i=0$ are given as in (\ref{ex-2-v}). From the linear equivalence relations, we get 
\be
D_1=D_2=D_3=-E_1-3E_2\ ,\ D_4=D_5=-E_1-2E_2\,,
\ee
hence
\be
\ba
-K_{\Sigma'}&=D_1+D_2+D_3+D_4+D_5+E_1+E_2\cr
&=-4E_1-12E_2\,.
\ea
\ee
From the values of $a_{i,j}$ in (\ref{ex-2-v}), we read off $F'_1=F'_2=-2E_1-6E_2$, which satisfies $F'_1+F'_2=-K_{\Sigma'}$.

Now let us check the lattice points in the polytope
\be
\ba
P_1=P_2=&\{u=(u_1,u_2,u_3,u_4,u_5)|\la u,v_1\ra\geqslant 2\ ,\ \la u,v_2\ra\geqslant 6\ ,\ u_k\geqslant 0\ (k=1,2,3,4,5)\}\cr
=&\{(u_1,u_2,u_3,u_4,u_5)|u_1+u_2+u_3+u_4+u_5\geqslant 2\ ,\ 3u_1+3u_2+3u_3+2u_4+2u_5\geqslant 6\ ,\cr
&u_k\geqslant 0\ (k=1,2,3,4,5)\}\,.
\ea
\ee
Indeed the solutions corresponding to the lowest order monomials are
\be
\ba
z_1^2: (2,0,0,0,0)\ ,\ z_2^2: (0,2,0,0,0)\ ,\ z_3^2: (0,0,2,0,0)\ ,\ z_4^3: (0,0,0,3,0)\ ,\ z_5^3: (0,0,0,0,3)\,,
\ea
\ee
which match the monomials in the original equations $f_1=0$ and $f_2=0$.

The resolved CY3 $\widetilde{X}$ is given by the set of equations
\be
\begin{cases}
&z_1^2+z_2^2+z_3^2+(z_4^3+z_5^3)\delta_1=0\\
&z_1^2+2z_2^2+3z_3^2+(4z_4^3+5z_5^3)\delta_1=0\,.
\end{cases}
\ee 

We consider the divisor $\delta_1=0$ on coordinate patch $\{z_1, z_2, \delta_1, z_4, \delta_2 \}$.
The defining ideal is $(z_1^2+z_2^2+1,z_1^2+2z_2^2+3,\delta_1)$ .
Carrying out prime decomposition on this ideal, we will get 4 prime ideals:
\be
(\sqrt{2}i+z_2,-1+z_1,\delta_1)\ ,\ (\sqrt{2}i+z_2,1+z_1,\delta_1)\ ,\ (-\sqrt{2}i+z_2,-1+z_1,\delta_1)\ ,\ (-\sqrt{2}i+z_2,1+z_1,\delta_1)\,.
\ee
We repeat this procedure on other patches and identify prime ideals on different patches using the transition function. We are able to find that the reducible divisor $\delta_1=0$ has 4 irreducible components:
\be
\ba
&S_1^{(1)}:\ \delta_1=\sqrt{2}i+z_2=-1+z_1=0\cr
&S_1^{(2)}:\ \delta_1=\sqrt{2}i+z_2=1+z_1=0\cr
&S_1^{(3)}:\ \delta_1=-\sqrt{2}i+z_2=-1+z_1=0\cr
&S_1^{(4)}:\ \delta_1=-\sqrt{2}i+z_2=1+z_1=0\,.
\ea
\ee

Now we consider the triple intersection numbers. As an example we set the goal to calculate $S_1^{(1)} \cdot S_2 \cdot S_2$. We define $S_2:\delta_2=0$ and $D_i:z_i=0$. First we use the linear equivalence relations to get
\be
\begin{cases}
\label{linear-equival}
&S_1^{(1)} \cdot S_2 \cdot (D_1+\sum_{i=1}^{4} S_1^{(i)}+3 S_2)=0\\
&S_1^{(1)} \cdot S_2 \cdot (D_4+\sum_{i=1}^{4} S_1^{(i)}+2 S_2)=0\,.
\end{cases}
\ee

In this equation, we take $S_1^{(1)} \cdot S_2 \cdot D_4$ as an example:
on coordinate patch $\{z_1, z_2, \delta_1, z_4, \delta_2 \}$, the defining ideal is $(\sqrt{2}i+z_2,-1+z_1,\delta_1,\delta_2)$, which is already prime. We can verify that on other coordinate patches the defining ideals are all prime and compatible with this one. So we can conclude $S_1^{(1)} \cdot S_2 \cdot D_4=1$.

Repeat this process, we can see that 
\be
S_1^{(1)} \cdot S_2 \cdot D_1=S_1^{(1)} \cdot S_2 \cdot S_1^{(j)}=0\quad(j=2,3,4)
\ee
Thus, solve the linear equations (\ref{linear-equival}) and we see that $S_1^{(1)} \cdot S_2 \cdot S_2=1$ 
and $S_1^{(1)} \cdot S_1^{(1)}  \cdot S_2=-3$.

Following this procedure,we can deduce the full intersection diagram. See (\ref{Class-2-int}).

\subsection{Topology of the link $\Sigma_5$}
\label{sec:linktop}

For an ICIS threefold singularity $X_3$, we define its link fivefold $\Sigma_5$ as following. We first define a 9-dimensional sphere $S_\epsilon$ with radius $\epsilon$ in the ambient space $\mb{C}_5$:
\be
S_\epsilon:=|z_1|^2+|z_2|^2+|z_3|^2+|z_4|^2+|z_5|^2=\epsilon^2\,.
\ee
Then we define the link $\Sigma_5$ as
\be
\Sigma_5=S_\epsilon\cap \{f_1(z_1,z_2,z_3,z_4,z_5)=0\}\cap\{f_2(z_1,z_2,z_3,z_4,z_5)=0\}\,.
\ee

Following \cite{RANDELL1975347}, we compute the homology classes of $\Sigma_5$, which has the general form
\be
\label{link-top}
\ba
H_0(\Sigma_5,\mb{Z})=\mb{Z}\ ,\  H_1(\Sigma_5,\mb{Z})=0\ ,\  H_2(\Sigma_5,\mb{Z})=\mb{Z}^f\oplus(\frak{f})^2\cr
H_3(\Sigma_5,\mb{Z})=\mb{Z}^f\ ,\  H_4(\Sigma_5,\mb{Z})=0\ ,\  H_5(\Sigma_5,\mb{Z})=\mb{Z}\,.
\ea
\ee
$f$ here is the flavor rank of $\FTfour$. $\frak{f}$ is a finite group (the defect group as in \cite{DelZotto:2015isa,Albertini:2020mdx}), which represents the torsional cycles of $\Sigma_5$, which in turn generate torsional non-compact 3-cycles in $X_3$ as an element of relative homology group $H_3(X_3,\Sigma_5;\mb{Z})$. We describe the detailed algorithm in Appendix~\ref{app:link-top}.

In the M-theory description of $\FT$, M2 or M5-brane wrapping these 3-cycles lead to either an electric 0-form symmetry or a magnetic 3-form symmetry $\frak{f}$. In the IIB description of $\FTfour$, D3-brane wrapping the 3-cycles lead to either an electric or magnetic 1-form symmetry $\frak{f}$\cite{Closset:2020scj}.

Below we list all the cases that can be computed by our method.
\be
\begin{array}{||c|c|c||c|c||}
\hline
 \text{Class}& \text{Equations}&\{n_i\} & f & \mathfrak{f}\\
\hline
\hline
1 &\begin{cases}
&z_1^2+ z_2^2+z_3^2+z_4^ 2+ z_5^n=0 \\
&z_1^2+2 z_2^2+3z_3^2+4z_4^ 2+ 5z_5^n=0\,.
\end{cases}  &n=2 & 5 & 0 \\
\cline{3-5}
 & &n=2k(k>1)  & 5&\mb{Z}_k\\
\cline{3-5}
& &n=2k+1  & 1&\mb{Z}_n\\
\hline
3&\begin{cases}
& z_1^2+z_2^2+z_3^2+z_4^3+z_5^4=0 \\
& z_1^2+2 z_2^2+3 z_3^2+4 z_4^3+5 z_5^4=0\,.
\end{cases} & - & 1 & \mb{Z}_3 \\
\hline
4&\begin{cases}
& z_1^2+z_2^2+z_3^2+z_4^3+z_5^5=0 \\
& z_1^2+2 z_2^2+3 z_3^2+4 z_4^3+5 z_5^5=0\,.
\end{cases} & - & 0 & 0 \\
\hline
\end{array}
\ee
\section{4d and 5d physics from ICIS}
\label{sec:4d-5d-phys}

\subsection{4d $\mc{N}=2$ SCFTs from ICIS}
\label{sec:4d-def}

The 4d $\mc{N}=2$ SCFT $\FTfour$ constructed from IIB superstring on the ICIS singularity $X$ was studied in \cite{Chen:2016bzh,Wang:2016yha}, and we will briefly summarize the physical dictionary. We denote the CB rank, HB dimension and flavor rank of $\FTfour$ as $\h r$, $\h{d}_H$ and $f$.

Starting from an ICIS equation,
\be
\label{ICIS-eq}
\begin{cases}
&f_1(z_1,z_2,z_3,z_4,z_5)=0\\
&f_2(z_1,z_2,z_3,z_4,z_5)=0\,,
\end{cases}
\ee
we assign weights $(q_1,q_2,q_3,q_4,q_5;d_1=1,d_2=d)$ to $(z_1,z_2,z_3,z_4,z_5;f_1,f_2)$, such that $d_2=d\geqslant 1$.

Now let us consider the mini-versal deformations of (\ref{ICIS-eq}), which form the Milnor ring
\be
\mc{J}=\frac{\mb{C}[z_1,z_2,z_3,z_4,z_5]}{\frac{\partial f_i}{\partial z_j}}\,.
\ee
The generators of the Milnor ring take the form of either
\be
f_1\rightarrow f_1+\lambda_\alpha\phi^1_\alpha
\ee
or
\be
f_2\rightarrow f_2+\lambda_\alpha\phi^2_\alpha\,.
\ee
For the first case, the generator $\phi^1_\alpha$ gives rise to a parameter $\lambda_\alpha$ on the 4d $\mc{N}=2$ Coulomb branch with scaling dimension
\be
\Delta[\lambda]=\frac{1-Q[\phi^1_\alpha]}{\sum q_i-1-d}\,.
\ee
For the second case, the generator $\phi^2_\alpha$ gives rise to a parameter $\lambda_\alpha$ with scaling dimension
\be
\Delta[\lambda]=\frac{d-Q[\phi^2_\alpha]}{\sum q_i-1-d}\,.
\ee

These parameters can be separated into the following categories:
\begin{enumerate}
\item $\Delta[\lambda]>1$, $\#=\h r$; they one-to-one correspond to the 4d $\mc{N}=2$ CB operators, and their total number $\hat r$ equals to the CB dimension of $\FTfour$.
\item $\Delta[\lambda]=1$, $\#=f$; they correspond to the mass parameters on the 4d $\mc{N}=2$ CB, and their total number $f$ equals to the flavor rank of $\FTfour$.
\item $0<\Delta[\lambda]<1$; they correspond to the relevant deformations of the CB.
\item $\Delta[\lambda]=0$; they correspond to the exact marginal deformation of the CB, and the moduli of the singularity $X$.
\item $\Delta[\lambda]<0$; they correspond to the irrelevant deformations.
\end{enumerate}

From the 4d CB spectrum, one can read off the generalized quiver description of the 4d $\mc{N}=2$ SCFT $\FTfour$, potentially with non-Lagrangian matter sectors, see the examples in \cite{Wang:2016yha}.

The 4d central charges $a$ and $c$ can also be computed from the Coulomb branch spectrum:
\be
a=\frac{R(A)}{4}+\frac{R(B)}{6}+\frac{5\h r}{24}\ ,\ c=\frac{R(B)}{3}+\frac{\h r}{6}\,.
\ee
$R(A)$ and $R(B)$ can be computed as
\be
R(A)=\sum_{\Delta[\lambda_\alpha]>1}(\lambda_\alpha-1)\ ,\ R(B)=\frac{\mu'}{4(\sum q_i-1-d)}\,.
\ee
Here $\mu'=\mu+\mu_1$, where $\mu$ is the Milnor number of $f_1=f_2=0$ and $\mu_1$ is the Milnor number of $f_1$=0.

From $a$ and $c$, we can also compute the effective number of vector multiplets and hypermultiplets:
\be
n_v=8a-4c\ ,\ n_h=-16a+20c\,.
\ee
The 4d $\mc{N}=2$ Higgs branch dimension can be computed as
\be
\label{4dHB-dim}
\h{d}_H=n_h-n_v+\frac{1}{2}b_3\,.
\ee
Here the correction term $\frac{1}{2}b_3\equiv \frac{1}{2}b_3(\widetilde{X})$ arises from the free vector multiplets from D3-brane wrapping 3-cycles in $\widetilde{X}$~\cite{Closset:2020scj,Closset:2021lwy}.

\subsection{5d $\mc{N}=1$ SCFT and the 5d/4d correspondence}

Now let us define a 5d $\mc{N}=1$ SCFT $\FT$ from M-theory on $X$. The basics of geometric engineering can be found in \cite{Intriligator:1997pq,Jefferson:2018irk,Apruzzi:2019opn,Eckhard:2020jyr}. Let us denote the (real) Coulomb branch rank, (quarternionic) Higgs branch dimension, the rank of flavor symmetry visible on the Coulomb branch\footnote{In the cases with $b_3>0$, the actual flavor rank of the 5d SCFT $\FT$ can be higher than the number $f$. One can only see the correct flavor symmetry group after a geometric transition, see~\cite{Jefferson:2018irk,Closset:2020scj}.} and flavor symmetry algebra of $\FT$ by $r$, $d_H$, $f$ and $G_F$.

The Coulomb branch information of $\FT$ can be read off from the resolution $\t X$. In particular when $\t X$ and the exceptional divisors $S_i$ are smooth, we can read off the prepotential $\mc{F(\phi)}$ from the triple intersection numbers $S_i\cdot S_j\cdot S_k$, as well as the IR gauge theory descriptions and the flavor rank $f$.

In principle, the Higgs branch information of $\FT$ can be read off from the deformation $\widehat{X}$. In this paper, we employ the framework of 5d/4d correspondence, proposed in \cite{Closset:2020scj}. Starting from the 4d $\mc{N}=2$ theory $\FTfour$, we reduce it on a circle $S^1$ and flow to the IR, to get the 3d $\mc{N}=4$ quiver gauge theory $\EQfour$. Then we gauge the flavor symmetry $U(1)^f$ of $\EQfour$, and end up with the Magnetic quiver $\MQfive$ for the 5d $\mc{N}=1$ theory $\FT$. $\MQfive$ contains the following information about the HB and CB of $\FT$:
\begin{enumerate}
\item{The Higgs branch dimension $d_H$ is given by the total rank of gauge groups in $\MQfive$. Hence we also have a non-trivial relation between $d_H$, $\h{r}$ and $f$
\be
d_H=\h r+f\,.
\ee}
\item{From $\MQfive$, one can compute the Hasse diagram using the quiver subtraction techniques in \cite{Bourget:2019aer}. The number of layers of the Hasse diagram is equal to the 5d CB rank $r$.}
\item{From the bottom layer of the Hasse diagram, one can read off the non-abelian part of the flavor symmetry algebra $G_{F,nA}$~\cite{Bourget:2019aer}. For each layer labeled by minimal nilpotent orbit of the Lie algebra $\mk{g}$, it contributes to a factor $\mk{g}$ to $G_{F,nA}$. For the Klein singularity $A_N(N>1)$, it does not contribute to $G_{F,nA}$. Finally, we can write down the full flavor symmetry $G_F=G_{F,nA}\times U(1)^{r-\text{rk}(G_{F,nA})}$.}
\end{enumerate}

Apart from the relation between $\EQfour$ and $\MQfive$, there is also a conjectured relation between the HB of $\FTfour$ and the CB of $\FT$\footnote{When the crepant resolution $\widetilde{X}$ has Q-factorial terminal singularities, the 4d Higgs branch is said to be non-existing~\cite{Xie:2019vzr}.}. Namely, we define the 3d mirror of $\EQfour$ as $\MQfour$, and the 3d mirror of $\MQfive$ as $\EQfive$. The conjecture in \cite{Closset:2020scj} states that we can obtain $\MQfour$ by gauging the flavor symmetry $U(1)^f$ of $\EQfive$. Because the HB information of $\FTfour$ is encoded in $\MQfour$, we also have a non-trivial relation:
\be
\label{4dHB-5dCB}
\h d_H=r+f\,.
\ee

Now using (\ref{4dHB-5dCB}) and (\ref{4dHB-dim}), we can compute the topological number
\be
b_3(\widetilde{X})=2(r+f-n_h+n_v)\,.
\ee
On the r.h.s., $r$ can be computed from the resolution procedure in section~\ref{sec:resol}, while $f$, $n_h$ and $n_v$ can be computed from the deformation of $X$ in section~\ref{sec:4d-def}.

If $b_3(\widetilde{X})>0$, then in the M-theory description of $\FT$, Euclidean M2-branes wrapping 3-cycles give rise to quantum corrections to the 5d Coulomb branch. However, one can remove these 3-cycles by a geometric transition\cite{Jefferson:2018irk,Apruzzi:2019opn,Apruzzi:2019enx,Closset:2020scj}.

\subsection{Flavor symmetry from resolution}
\label{sec:flavor}

In many cases of $\FT$, there is no good Lagrangian description of the magnetic quiver $\MQfive$. Nonetheless, we can often speculate the non-abelian flavor symmetry factors $G_{F,nA}$ from the fully resolved CY3, if $\widetilde{X}$ is smooth and $b_3(\widetilde{X})=0$\footnote{If $b_3(\widetilde{X})>0$, one should first perform geometric transitions on $\widetilde{X}$ to get a new resolved CY3 $\widetilde{X}'$ with $b_3(\widetilde{X})=0$}. The method was carried out in \cite{Closset:2021lwy}, generalizing the approaches in \cite{Apruzzi:2019opn,Apruzzi:2019enx,Apruzzi:2019kgb,Bhardwaj:2020ruf,Bhardwaj:2020avz}. Note that on $\widetilde{X}$, all compact surfaces $S_i$ are rational surfaces.

The idea is to perform a marginal deformation of the singularity $X\rightarrow X'$, where $X'$ is non-isolated and its resolution $\widetilde{X}'$ contains a number of non-compact divisors that lead to non-abelian flavor symmetry $G_{F,nA}$. More precisely speaking, $\widetilde{X}'$ should be diffeomorphic to $\widetilde{X}$. For the compact surfaces $S_i$, the intersection numbers $S_i\cdot S_j\cdot S_k$ are the same in $\widetilde{X}$ and $\widetilde{X}'$, but the compact surfaces in $\widetilde{X}'$ would contain a number of flavor curves~\cite{Apruzzi:2019opn} $C_\alpha$ with normal bundle $N_{C_\alpha|\widetilde{X}'}=\mc{O}\oplus\mc{O}(-2)$ that lead to flavor W-bosons on the 5d Coulomb branch. In the rank-1 example, $\widetilde{X}$ is the isolated local dP$_n$ singularity, while $\widetilde{X}'$ is the non-isolated local gdP$_n$ singularity\footnote{For the case of $n=1$, we either have $\widetilde{X}=\widetilde{X}'=$ local dP$_1$ or $\widetilde{X}=$ local $\mb{F}_0$, $\widetilde{X}'=$ local $\mb{F}_2$. These two cases give different rank-1 theories, as $\mb{F}_0$ and dP$_1$ are not diffeomorphic.}. The 5d SCFT obtained from M-theory on $\widetilde{X}'$ and $\widetilde{X}$ are equivalent, while M-theory on the non-isolated $\widetilde{X}'$ describes a 5d theory coupled to the background $G_{F,nA}$ gauge field, where we can read off $G_{F,nA}$. We conjecture that this methodology also apply to $r>1$ cases as well.

In practice, one first performs a number of flops on $\widetilde{X}'$, such that except for one compact divisor $\mbf{S}$, the negative curves of other compact divisors $S_i$ are all of the form $S_i\cdot S_j$. That is, one try to flop as many as possible compact curves into $\mbf{S}$. 

On $\mbf{S}$, besides the intersection curves $\mbf{S}\cdot S_i$, there can be other negative rational curves $C_\alpha=\mbf{S}\cdot D_\alpha$ as well. The non-compact divisors $D_\alpha$ generate the flavor Cartan $U(1)^f\subset G_F$. For simplicity, we assume that $D_\alpha$ only intersects the compact divisor $\mbf{S}$. $C_\alpha$ gives rise to a flavor W-boson, which corresponds to a simple root in the non-abelian part $G_{F,nA}$ if it satisfy one of the following conditions:

\begin{enumerate}
\item $C_\alpha^2|_\mbf{S}\equiv \mbf{S}\cdot D_\alpha^2=-2$.
\item $C_\alpha^2|_\mbf{S}\equiv \mbf{S}\cdot D_\alpha^2=-1$, and $C_\alpha$ intersects other compact divisors $S_i$ other than $\mbf{S}$: $C_\alpha\cdot \sum_i S_i=2$. The reason is that if one do a flop operation by blowing down $C_\alpha$ on $\mbf{S}$, it would split into two $\mc{O}(-1)\oplus\mc{O}(-1)$ curves $C_\alpha^{(1)}$ and $C_\alpha^{(2)}$, and $C_\alpha^{(1)}+C_\alpha^{(2)}$ gives a flavor W-boson, see Figure 1 of \cite{Apruzzi:2019kgb}.
\end{enumerate}

Then we connect the simple roots into a Dynkin diagram of $G_{F,nA}$, based on the intersection numbers $C_\alpha\cdot D_\beta$ $(\alpha\neq\beta)$. 

Finally note that there are multiple ways of choosing the flavor curves on $\mbf{S}$. One should choose the version which gives rise to the largest $G_{F,nA}$\footnote{An example would be the rank-1 example $\mbf{S}=$dP$_n$ versus gdP$_n$. If one choose the type $E_n$ gdP$_n$~\cite{derenthal2014singular}, it maximally gives rise to $G_{F,nA}=E_n$.}. When choosing the flavor curves, one should also make sure that all the intersection numbers between different negative curves are non-negative.

For the examples with smooth resolution $\widetilde{X}$ and $b_3(\widetilde{X})=0$ in this paper, we use this method to read off a conjectured $G_{F,nA}$ when there is not an available magnetic quiver $\MQfive$. Such conjectured $G_{F,nA}$ will be labeled with blue color.



\subsection{The example of Class (1)}
\label{sec:class1}

Here we give a set of simple examples to support the previous setups and conjectures, which are the Class (1) of \cite{Chen:2016bzh}:
\be
\begin{cases}
&z_1^2+z_2^2+z_3^2+z_4^2+z_5^n=0\\
&z_1^2+2z_2^2+3z_3^2+4z_4^2+5z_5^n=0\,.
\end{cases}
\ee 
We list the physical quantities for the cases of different $n$ in the table below:

\be
\begin{array}{||c||c|c||c|c|c||c|c||c|c||}
\hline
 & r & d_H &  \h r & \h d_H & f & b_3 & \mathfrak{f} & a & c\\
\hline
n=2 & 1 & 7 &  2 & 6 & 5 & 0 & 0 & \frac{7}{4} & 2\\
\hline
n=2k\ (k>1) & k & 8k-1 & 8k-6 & k+5 & 5 & 2k-2 & \mb{Z}_k & 3k^2-\frac{5}{4} & 3k^2-1\\
\hline
n=2k+1 & k & 8k+1 &  8k & k+1 & 1 & 2k & \mb{Z}_{2k+1} & 3k^2+3k+\frac{1}{24} & 3k^2+3k+\frac{1}{12}\\
\hline
\end{array}
\ee

Let $k=\lfloor\frac{n}{2}\rfloor$, the resolution sequence is
\be
\ba
&(z_1,z_2,z_3,z_4,z_5;\delta_1)\cr
&(z_1,z_2,z_3,z_4,\delta_j;\delta_{j+1})\quad (j=1,\dots,k-1)\,.
\ea
\ee
When $n=2k$, the resolved equation is
\be
\begin{cases}
&z_1^2+z_2^2+z_3^2+z_4^2+z_5^n\prod_{j=1}^{k-1}\delta_j^{2k-2j}=0\\
&z_1^2+2z_2^2+3z_3^2+4z_4^2+5z_5^n\prod_{j=1}^{k-1}\delta_j^{2k-2j}=0\,.
\end{cases}
\ee
There is no terminal singularity and all the exceptional divisors $\delta_j=0$ are smooth. The 5d theory is actually the rank-$k$ Seiberg $E_5$ theory. 

The triple intersection numbers are
\be
\includegraphics[height=2cm]{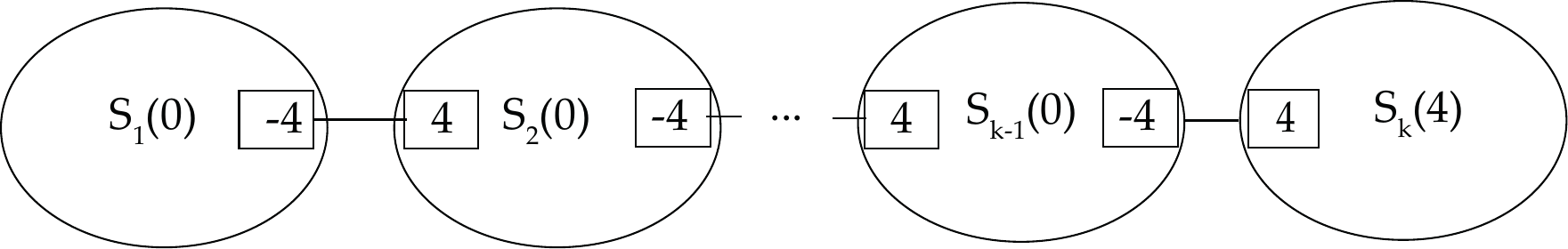}
\ee

Note that $S_1$, $\dots$, $S_{k-1}$ are $\mb{P}^1$ fibration over genus-1 curves, while $S_k$ is a dP$_5$ (gdP$_5$). We have
\be
b_3(\widetilde{X})=2k-2\,,
\ee
which matches the total number of compact 3-cycles on $S_1$, $\dots$, $S_{k-1}$ (each of these surfaces have two 3-cycles).

In particular when $k=1$, the 5d theory is the rank-1 Seiberg $E_5$ theory with $G_F=SO(10)$, which can be seen from the compact divisor $S_1=$gdP$_5$ of type $\mbf{D}_5$.

When $k>1$, the resolved geometry has a non-zero $b_3$, and one needs to do a geometric transition by blowing up $S_k$ at a double point on the intersection curve $S_{k-1}\cdot S_k$~\cite{Jefferson:2018irk,Closset:2020scj}. After the geometric transition, the new triple intersection numbers are
\be
\includegraphics[height=2cm]{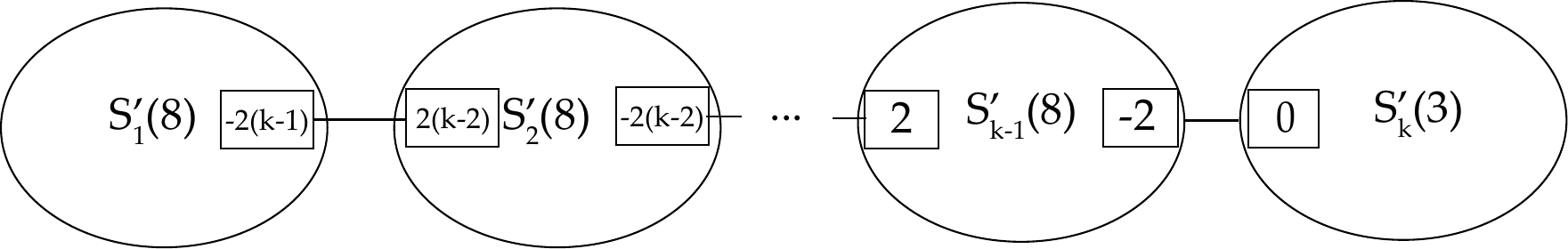}
\ee
The new surfaces $S_1'$, $\dots$, $S_{k-1}'$ are Hirzebruch surfaces, while $S_k'$ is a dP$_6$. Now we argue that one can read off the correct flavor symmetry $G_F=SO(10)\times SU(2)$ from the compact divisor $S_k'$ which is a type $\mbf{D}_5$ gdP$_6$, with the curve configuration
\be
\includegraphics[height=4cm]{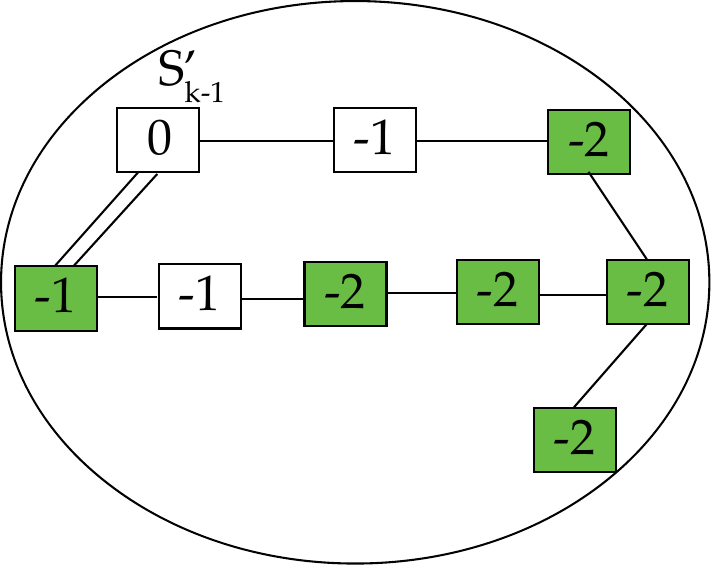}
\ee
$S_{k-1}'$ in the above figure denotes the intersection curve $S_{k-1}'\cdot S_k'$ on $S_k'$. As one can see, there is a flavor $(-1)$-curve that intersects another compact surface $S_{k-1}'$ at two points. Following the discussions in section~\ref{sec:flavor}, we indeed observe that the flavor curves form the Dynkin diagram of $G_F=SO(10)\times SU(2)$.

The 4d $\mc{N}=2$ quiver description is
\be
   \begin{tikzpicture}[x=.5cm,y=.5cm]
\draw[ligne, black](-0.7,-1.1)--(-0.3,-0.4);
\draw[ligne, black](-0.7,1.1)--(-0.3,0.4);
\draw[ligne, black](1.2,0)--(1.8,0);
\draw[ligne, black](3.3,0.4)--(3.7,1.1);
\draw[ligne, black](3.3,-0.4)--(3.7,-1.1);
\node[] at (-1,-1.5)  {\scriptsize$SU(k)$};
\node[] at (-1,1.5)  {\scriptsize$SU(k)$};
\node[] at (0,0) {\scriptsize$SU(2k)$};
\node[] at (3,0) {\scriptsize$SU(2k)$};
\node[] at (4,1.5) {\scriptsize$SU(k)$};
\node[] at (4,-1.5) {\scriptsize$SU(k)$};
\end{tikzpicture}
\ee

Hence after gauging $U(1)^f$, $f=5$, the Magnetic quiver $\MQfive$ for $\FT$ is that of the rank-$k$ Seiberg $E_5$ theory:
\be
 \begin{tikzpicture}[x=.5cm,y=.5cm]
\draw[ligne, black](0,0)--(-1,1);
\draw[ligne, black](0,0)--(-1,-1);
\draw[ligne, black](0,0)--(1,0);
\draw[ligne, black](1,0)--(2,1);
\draw[ligne, black](1,0)--(2,-1);
\node[] at (-3,0) {$\MQfive= $};
\node[bd] at (-1,1) [label=below:{{\scriptsize$k$}}] {};
\node[bd] at (-1,-1) [label=below:{{\scriptsize$k$}}] {};
\node[bd] at (0,0) [label=below:{{\scriptsize$2k$}}] {};
\node[bd] at (1,0) [label=below:{{\scriptsize$2k$}}] {};
\node[bd] at (2,1) [label=below:{{\scriptsize$k$}}] {};
\node[bd] at (2,-1) [label=below:{{\scriptsize$k$}}] {};
\end{tikzpicture} 
\ee
Here each node $k$ denotes $U(k)$, and a diagonal $U(1)$ is modded out from the quiver. Thus we can compute the 5d Higgs branch dimension
\be
d_H=8k-1\,.
\ee

When $n=2k+1$, the resolved equation is
\be
\begin{cases}
&z_1^2+z_2^2+z_3^2+z_4^2+z_5^n\prod_{j=1}^{k}\delta_j^{2k-2j+1}=0\\
&z_1^2+2z_2^2+3z_3^2+4z_4^2+5z_5^n\prod_{j=1}^{k}\delta_j^{2k-2j+1}=0\,.
\end{cases}
\ee
$\widetilde{X}$ is singular at $z_1=z_2=z_3=z_4=\delta_k=0$. In the local patch $(z_1,z_2,z_3,z_4,\delta_k)$, the singularity has the form
\be
\begin{cases}
&z_1^2+z_2^2+z_3^2+z_4^2+\delta_k=0\\
&z_1^2+2z_2^2+3z_3^2+4z_4^2+5\delta_k=0\,,
\end{cases}
\ee
which can be put into the form of a hypersurface equation
\be
4z_1^2+3z_2^2+2z_3^2+z_4^2=0\,.
\ee
As this residual terminal singularity is a conifold, the 5d SCFT $\FT$ in can be interpreted as the rank-$k$ Seiberg $E_5$ theory coupled to an additional  hypermultiplet.

The last exceptional divisor $S_k:\delta_k=0$ has the form of
\be
\begin{cases}
&z_1^2+z_2^2+z_3^2+z_4^2=0\\
&z_1^2+2z_2^2+3z_3^2+4z_4^2=0\,,
\end{cases}
\ee
which is also singular at
\be
\delta_k=z_1=z_2=z_3=z_4=0\,.
\ee
In this case, the 4d theory does not have an integral Coulomb branch, and the 4d $\mc{N}=2$ quiver description is
\be
  \begin{tikzpicture}[x=.5cm,y=.5cm]
\draw[ligne, black](-1.2,-1.1)--(-0.8,-0.4);
\draw[ligne, black](-1.7,1.1)--(-0.8,0.4);
\draw[ligne, black](1.2,0)--(1.8,0);
\draw[ligne, black](3.8,0.4)--(4.2,1.1);
\draw[ligne, black](3.8,-0.4)--(4.2,-1.1);
\node[] at (-1.5,-1.5)  {\scriptsize$D_2(SU(2k+1))$};
\node[] at (-1.5,1.5)  {\scriptsize$D_2(SU(2k+1))$};
\node[] at (-0.5,0) {\scriptsize$SU(2k+1)$};
\node[] at (3.5,0) {\scriptsize$SU(2k+1)$};
\node[] at (4.5,1.5) {\scriptsize$D_2(SU(2k+1))$};
\node[] at (4.5,-1.5) {\scriptsize$D_2(SU(2k+1))$};
\end{tikzpicture}
\ee
From $\FTfour$, we can derive $\MQfive$ using the methods of \cite{Closset:2020afy}, which turns out to be 
\be
 \begin{tikzpicture}[x=.5cm,y=.5cm]
\draw[ligne, black](0,0)--(-1.5,1);
\draw[ligne, black](0,0)--(-1.5,-1);
\draw[ligne, black](0,0)--(1.5,0);
\draw[ligne, black](1.5,0)--(3,1);
\draw[ligne, black](1.5,0)--(3,-1);
\node[] at (-3,0) {$\MQfive= $};
\node[bd] at (-1.5,1) [label=below:{{\scriptsize$k$}}] {};
\node[bd] at (-1.5,-1) [label=below:{{\scriptsize$k$}}] {};
\node[bd] at (0,0) [label=below:{{\scriptsize$2k+1$}}] {};
\node[bd] at (1.5,0) [label=above:{{\scriptsize$2k+1$}}] {};
\node[bd] at (3,1) [label=below:{{\scriptsize$k$}}] {};
\node[bd] at (3,-1) [label=below:{{\scriptsize$k$}}] {};
\end{tikzpicture} 
\ee
The quiver is apparently a bad quiver.

\section{Fully smooth models with $b_3=0$}
\label{sec:smooth-zerob3}

In this section, we present the examples where both the resolved CY3 $\widetilde{X}$ and the compact exceptional divisors $S_i$ are smooth. We also require that $b_3(\widetilde{X})=0$, such that there is no further quantum corrections from Euclidean M2-branes to $\widetilde{X}$. These cases are the most well-understood cases in the 5d/4d correspondence, as the $\MQfive$ for the 5d theory is a good quiver (if it has a Lagrangian description).

We list the infinite sequences of such models in section~\ref{sec:infiniteseq} and collect other sporadic models according to their rank in the later subsections.

\subsection{Infinite sequences}
\label{sec:infiniteseq}
\subsubsection{Class (209)}
\label{sec:209}
\be
\begin{cases}
&z_1 z_2+z_3 z_4=0 \\
&z_1 z_3+z_2^n z_3+z_4^3+z_5^3=0\,.
\end{cases}
\ee
$n\geqslant2$

The resolution is smooth for all odd $n$.
\be
\begin{array}{||c||c|c||c|c|c||c|c|c||}
\hline
 &r & d_H &  \h r & \h d_H & f & b_3 & a&c\\
\hline
 n=2k+1&3n & 6n+9  & 6n+4& 3n+5 & 5 & 0 & 3n^2+\frac{47n}{8}+\frac{59}{24}&3n^2+6n+\frac{8}{3}\\
\hline
\end{array}
\ee
When $n=2k+1$,the Coulomb branch spectrum is:

$
1^5,\underbrace{\frac{3}{2}^2,2,\frac{5}{2}^2,3^2,......,\frac{2n-3}{2}^2,n-1,\frac{2n-1}{2}^2,n^2},\frac{2n+1}{2}^2,n+1,\frac{2n+3}{2}^3,$

$\underbrace{{n+2}^2,n+3,\frac{2n+7}{2},......,{2n+1}^2,2n+2,\frac{4n+5}{2}},
\underbrace{\frac{4n+9}{2},\frac{4n+13}{2},......,\frac{6n+3}{2}}
$

We can see that even though we have a smooth resolution, the spectrum is still not integral.

We take $n=3$ as an example, which is a rank-9 5d SCFT with $G_F=\color{blue}{SU(3)\times U(1)^3}$.

The resolution sequence is
\be
\ba
&(z_1,z_2,z_3,z_4,z_5;\delta_1)\cr
&(z_1,z_3,\delta_1;\delta_6)\cr
&(z_1,z_3,z_4,z_5,\delta_6;\delta_8)\cr
&(z_1,z_4,z_5,\delta_6;\delta_7)\cr
&(z_1,z_4,\delta_1;\delta_5)\cr
&(z_2,z_3,\delta_1;\delta_3)\cr
&(z_1,\delta_1;\delta_4)\cr
&(z_1,\delta_8;\delta_{10})\cr
&(z_3,\delta_1;\delta_2)\cr
&(\delta_1,\delta_6;\delta_9)
\ea
\ee
The resolved equation is
\be
\begin{cases}
&z_1 z_2\delta_4+z_3 z_4 \delta_2\delta_8=0\\
&z_5^3 \delta_1 \delta_7+z_4^3 \delta_1 \delta_5^3 \delta_7+
z_1 z_3 \delta_6 \delta_{10}+z_2^3 z_3 \delta_1^2\delta_2^2\delta_3^5\delta_4\delta_5\delta_6^2\delta_9^3\delta_{11}=0\,.
\end{cases}
\ee 
We can deduce the following intersection diagram:
\be
\includegraphics[height=7cm]{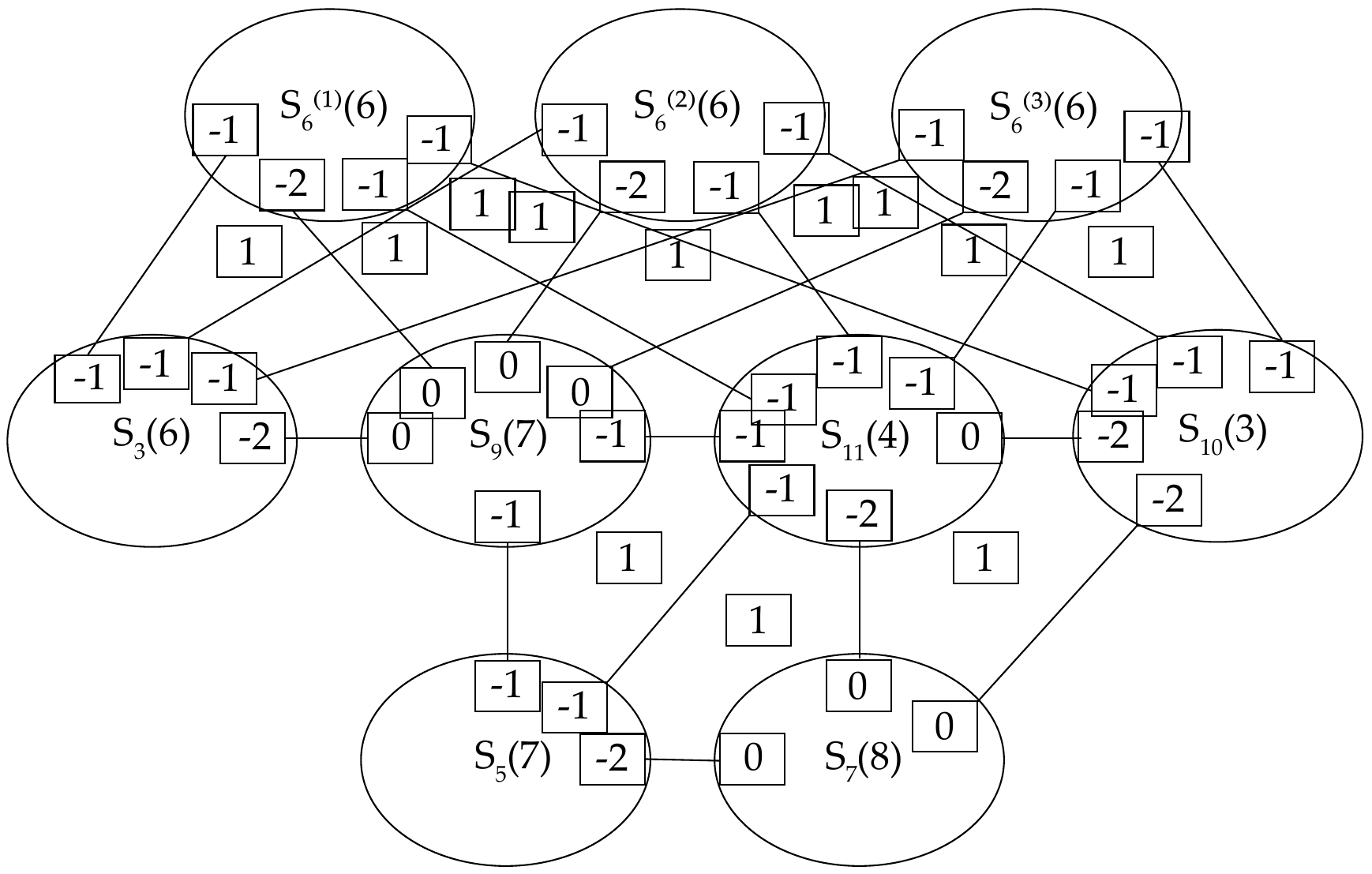}
\ee

For the non-abelian flavor symmetry, we consider the possible configuration of flavor curves on the compact divisors. From the intersection diagram, the only compact divisor with possible additional $(-2)$-curves is $S_{10}$. The gdP$_6$ with the most $(-2)$-curves that are compatible with the intersection diagram is of the type $2\mbf{A}_2$, see (\ref{fig:209-S10}). The additional $(-2)$-curves form a $A_2$ Dynkin diagram, hence the conjectured $G_{F,nA}=SU(3)$.
\be
\label{fig:209-S10}
\includegraphics[height=2cm]{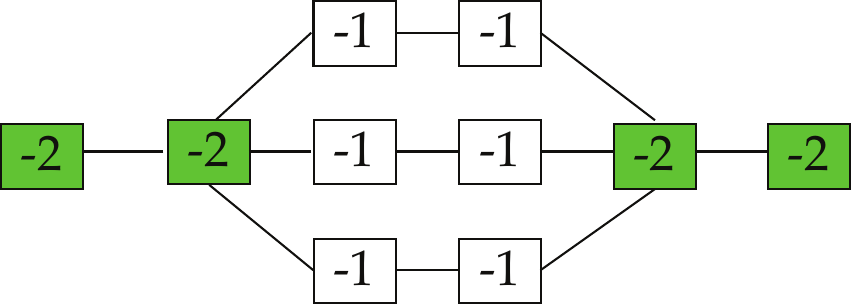}
\ee
Using the standard Picard group basis of gdP$_6$: $h^2=1$, $h\cdot e_i=0$, $e_i\cdot e_j=-\delta_{ij}$ (we will also use this type of Picard group basis in the later examples), the $(-2)$-curves on $S_{10}$ are
\be
\ba
&S_{10}\cdot S_{11}:\ h-e_1-e_2-e_3\cr
&S_{10}\cdot S_7:\ h-e_4-e_5-e_6\cr
&\text{Flavor\ curves}:\ e_4-e_5\ ,\ e_5-e_6\,.
\ea
\ee

\subsubsection{Class (7)}
\be
\begin{cases}
&z_1 z_3+z_4 z_5=0 \\
&z_1^2+z_2^2+z_3^n+z_4^2+2 z_5^n=0
\end{cases}
\ee
$n\geqslant3$

The resolution is smooth only for $n=2k$.
\be
\begin{array}{||c||c|c|c||c|c|c||c|c|c||}
\hline
 &r & d_H & G_F & \h r & \h d_H & f & b_3 & a&c\\
\hline
 n=2k& \frac{n}{2} & \frac{n^2+3n+4}{2} & SO(2n+4)\times U(1) & \frac{n(n+1)-2}{2}& \frac{3n}{2}+3 & n+3 & 0 & \frac{-6+5n+12n^2+4n^3}{48}&\frac{n(2+3n+n^2)}{12}\\
\hline
\end{array}
\ee
We have 4d $\mc{N}=2$ quiver description, which is the same as $\EQfour$: 
\be
   \begin{tikzpicture}[x=.5cm,y=.5cm]
\draw[ligne, black](2,0)--(3,0);
\draw[dotted, black](6,0)--(7,0);
\draw[ligne, black](-4,0)--(-5,0);
\draw[ligne, black](9,0)--(10,0);
\draw[ligne, black](-3,0.4)--(-3,1.6);
\draw[ligne, black](-3,2.4)--(-3,3.6);
\draw[ligne, black](-1.5,2)--(-0.5,2);
\draw[ligne, black](-2,0)--(-1,0);
\node[] at (-3,4) {\scriptsize$[1]$};
\node[] at (0.5,2) {\scriptsize$[1]$};
\node[] at (-3,2) {\scriptsize$SU(\frac{n}{2}+1)$};
\node[] at (0.5,0) {\scriptsize$SU(n-1)$};
\node[] at (4.5,0) {\scriptsize$SU(n-2)$};
\node[] at (8,0) {\scriptsize$SU(2)$};
\node[] at (11,0) {\scriptsize$[1]$};
\node[] at (-3,0) {\scriptsize$SU(n)$};
\node[] at (-6,0) {\scriptsize$SU(\frac{n}{2})$};
\end{tikzpicture}
\ee
After gauging $U(1)^f$ (replacing $SU(n)$ nodes with $U(n)$ nodes), we have the 5d magnetic quiver:
\be
 \begin{tikzpicture}[x=1cm,y=1cm]
\draw[ligne, black](-2,0)--(0,0);
\draw[ligne, black](0,0)--(1,0);
\draw[dotted, black](1,0)--(2,0);
\draw[ligne, black](2,0)--(3,0);
\draw[ligne, black](-1,1)--(-1,0);
\draw[ligne, black](-1,1)--(0,1);
\draw[ligne, black](-1,1)--(-1,2);
\node[] at (-4,0) {$\MQfive= $};
\node[bd] at (3,0) [label=below:{{\scriptsize$1$}}] {};
\node[bd] at (2,0) [label=below:{{\scriptsize$2$}}] {};
\node[bd] at (1,0) [label=below:{{\scriptsize$n-2$}}] {};
\node[bd] at (0,0) [label=below:{{\scriptsize$n-1$}}] {};
\node[bd] at (-1,0) [label=below:{{\scriptsize$n$}}] {};
\node[bd] at (-2,0) [label=below:{{\scriptsize$\frac{n}{2}$}}] {};
\node[bd] at (-1,1) [label=left:{{\scriptsize$\frac{n}{2}+1$}}] {};
\node[bd] at (0,1) [label=right:{{\scriptsize$1$}}] {};
\node[bd] at (-1,2) [label=right:{{\scriptsize$1$}}] {};
\end{tikzpicture} 
\ee
We can get the following Hasse diagram with $\frac{n}{2}$ layers:
\be
\begin{tikzpicture}
\node (1) [hasse] at (0,1) {};
\node (2) [hasse] at (0,0) {};
\node (3) [hasse] at (0,-1) {};
\node (4) [hasse] at (0,-2) {};
\node (5) [hasse] at (0,-3) {};
\node (6) [hasse] at (0,-4) {};
\draw (1) edge [] node[label=right:$\mathfrak{d}_5$] {} (2);
\draw (2) edge [] node[label=right:$\mathfrak{d}_6$] {} (3);
\draw (3) edge [] node[label=right:$\mathfrak{d}_8$] {} (4);
\draw (5) edge [] node[label=right:$\mathfrak{d}_{n+2}$] {} (6);
\draw[dotted, black](0,-2)--(0,-3);
\end{tikzpicture}
\ee

From the Hasse diagram, we found that the 5d SCFT is the one with IR gauge theory description $SU(\frac{n}{2}+1)_2+(n+2)\mbf{F}$~\cite{Bourget:2019aer}. The enhanced UV flavor symmetry is $G_F=SO(2n+4)\times U(1)$.

For $n=2k$, the resolution sequence is
\be
\ba
&(z_1,z_2,z_3,z_4,z_5;\delta_1)\cr
&(z_1,z_2,z_4,\delta_j;\delta_{j+1})\quad (j=1,\dots,k-1)\,.
\ea
\ee
The resolved equation is
\be
\begin{cases}
&z_1 z_3+z_4 z_5=0\\
&z_1^2+z_2^2+z_4^2+(z_3^n+2z_5^n)\prod_{j=1}^{k-1}\delta_j^{n-2j}=0\,.
\end{cases}
\ee

The triple intersection numbers are
\be
\includegraphics[height=2cm]{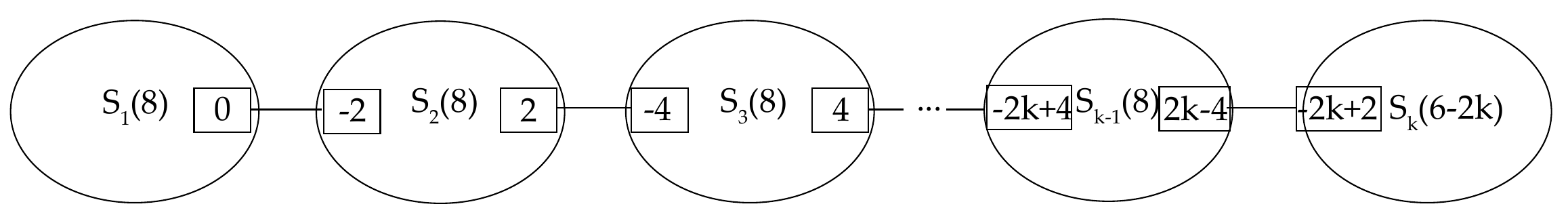}
\ee
Note that this case is also equivalent to the Class (182) with $n=2k$:
\be
\begin{cases}
& z_1 z_2+z_3 z_4=0 \\
& z_1 z_3+z_2^n+z_4^n+z_5^2=0\,.
\end{cases}
\ee 

\subsubsection{Class (59)}
\be
\begin{cases}
& z_1 z_2+z_3^2+z_4^2+z_5^n=0 \\
& z_1 z_5+2 z_3^2+z_4^2+3 z_2^n=0\,.
\end{cases}
\ee
$n\geqslant3$

The resolution is smooth only for $n=2k$.
\be
\begin{array}{||c||c|c|c||c|c|c||c|c|c||}
\hline
 &r & d_H & G_F & \h r & \h d_H & f & b_3 & a&c\\
\hline
 n=2k& \frac{n}{2} & \frac{n(n+5)}{2} & SO(2n+6) & \frac{n(n+3)-6}{2}& \frac{3n}{2}+3 & n+3 & 0 & \frac{-30-7n+24n^2+4n^3}{48}&\frac{-6-n+6n^2+n^3}{12}\\
\hline
\end{array}
\ee

We have the 5d magnetic quiver:
\be
 \begin{tikzpicture}[x=.5cm,y=.5cm]
\draw[ligne, black](-2,0)--(0,0);
\draw[ligne, black](0,0)--(1,0);
\draw[dotted, black](1,0)--(2,0);
\draw[ligne, black](2,0)--(3,0);
\draw[ligne, black](-1,1)--(-1,0);
\draw[ligne, black](0,0)--(0,1);
\node[] at (-4,0) {$\MQfive= $};
\node[bd] at (3,0) [label=below:{{\scriptsize$1$}}] {};
\node[bd] at (2,0) [label=below:{{\scriptsize$2$}}] {};
\node[bd] at (1,0) [label=below:{{\scriptsize$n-1$}}] {};
\node[bd] at (0,0) [label=below:{{\scriptsize$n$}}] {};
\node[bd] at (-1,0) [label=below:{{\scriptsize$n$}}] {};
\node[bd] at (-2,0) [label=below:{{\scriptsize$\frac{n}{2}$}}] {};
\node[bd] at (-1,1) [label=left:{{\scriptsize$\frac{n}{2}$}}] {};
\node[bd] at (0,1) [label=right:{{\scriptsize$1$}}] {};
\end{tikzpicture} 
\ee
We can get the following Hasse diagram with $\frac{n}{2}$ layers:
\be
\begin{tikzpicture}
\node (1) [hasse] at (0,1) {};
\node (2) [hasse] at (0,0) {};
\node (3) [hasse] at (0,-1) {};
\node (4) [hasse] at (0,-2) {};
\node (5) [hasse] at (0,-3) {};
\node (6) [hasse] at (0,-4) {};
\draw (1) edge [] node[label=right:$\mathfrak{d}_5$] {} (2);
\draw (2) edge [] node[label=right:$\mathfrak{d}_7$] {} (3);
\draw (3) edge [] node[label=right:$\mathfrak{d}_9$] {} (4);
\draw (5) edge [] node[label=right:$\mathfrak{d}_{n+3}$] {} (6);
\draw[dotted, black](0,-2)--(0,-3);
\end{tikzpicture}
\ee

From the Hasse diagram, we find out that the 5d SCFT is the one with quiver gauge theory description
\be
SU(2)_0-SU(2)-\dots-SU(2)-4\mbf{F}\,,
\ee
see (5.68) of \cite{VanBeest:2020kxw}.

For $n=2k$,the resolution sequence is
\be
\ba
&(z_1,z_2,z_3,z_4,z_5;\delta_1)\cr
&(z_1^{(2)},z_3^{(1)},z_4^{(1)},\delta_j^{(1)};\delta_{j+1})\quad (j=1,\dots,k-1)\,.
\ea
\ee
The resolved equation is
\be
\begin{cases}
&z_1 z_2+z_3^2+z_4^2+z_5^n \prod_{j=1}^{k-1}\delta_j^{n-2j}=0\\
&2z_3^2+z_4^2+z_1 z_5+3z_2^n \prod_{j=1}^{k-1}\delta_j^{n-2j}=0\,.
\end{cases}
\ee
The triple intersection numbers are
\be
\includegraphics[height=2cm]{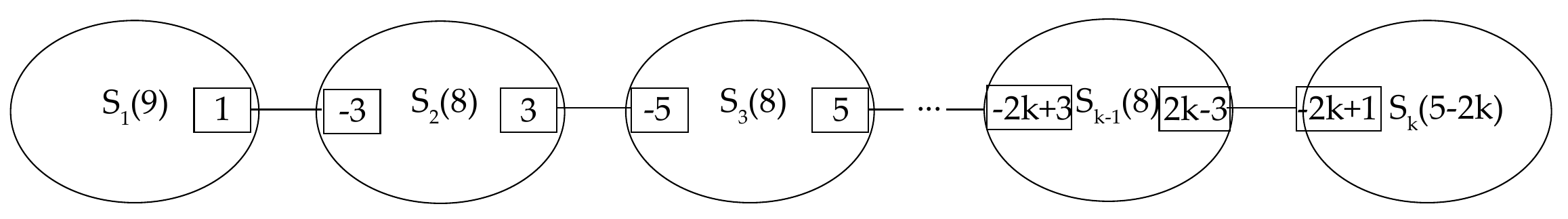}
\ee
Note that this case is also equivalent to the Class (143) with $n_2=n_5=2k$:
\be
\begin{cases}
& z_1 z_2+z_4^2+z_5^{n_5}=0 \\
& z_1 z_5+z_3^2+z_2^{n_2}=0\,,
\end{cases}
\ee 
the Class (144) with $2n_2=n_5=2k$:
\be
\begin{cases}
& z_1 z_2+z_4^2+z_5^{n_5}=0 \\
& z_1 z_5+z_3^2+z_2^{n_2} z_4=0\,,
\end{cases}
\ee 
 the Class (150) with $2n_5=n_2=2k$:
 \be
\begin{cases}
& z_1 z_2+z_4^2+z_3 z_5^{n_5}=0 \\
& z_1 z_5+z_3^2+z_2^{n_2}=0\,
\end{cases}
\ee 
and the Class (151) with $n_2=n_5=k$:
\be
\begin{cases}
& z_1 z_2+z_4^2+z_3 z_5^{n_5}=0 \\
& z_1 z_5+z_3^2+z_2^{n_2} z_4=0\,.
\end{cases}
\ee 

\subsubsection{Class (60)}
\be
\begin{cases}
& z_1 z_2+z_3^2+z_4^2+z_5^n=0 \\
& z_1 z_5+3 z_2^{1+2 n}+2 z_2 z_3^2+z_2 z_4^2=0\,.
\end{cases}
\ee
$n\geqslant3$

The resolution is smooth for arbitrary $n\geqslant3$.
\be
\begin{array}{||c|c|c||c|c|c||c|c|c||}
\hline
 r & d_H & G_F & \h r & \h d_H & f & b_3 & a&c\\
\hline
 2n-1 & n^2+5n+5 & SO(2n+6)\times U(1)^2 & n^2+4n & 3n+4 & n+5 & 0 &\frac{4+49n+48n^2+8n^3}{24}&\frac{2+13n+12n^2+2n^3}{6}\\
\hline
\end{array}
\ee

We have the 5d magnetic quiver:
\be
 \begin{tikzpicture}[x=1cm,y=1cm]
\draw[ligne, black](-2,0)--(0,0);
\draw[ligne, black](0,0)--(1,0);
\draw[ligne, black](1,0)--(2,0);
\draw[dotted, black](2,0)--(3,0);
\draw[ligne, black](1,1)--(1,0);
\draw[ligne, black](0,0)--(0,1);
\draw[ligne, black](0,2)--(0,1);
\node[] at (-4,0) {$\MQfive= $};
\node[bd] at (3,0) [label=below:{{\scriptsize$2$}}] {};
\node[bd] at (2,0) [label=below:{{\scriptsize$2n-2$}}] {};
\node[bd] at (1,0) [label=below:{{\scriptsize$2n$}}] {};
\node[bd] at (1,1) [label=right:{{\scriptsize$1$}}] {};
\node[bd] at (0,0) [label=below:{{\scriptsize$2n+1$}}] {};
\node[bd] at (0,1) [label=right:{{\scriptsize$n+1$}}] {};
\node[bd] at (0,2) [label=right:{{\scriptsize$1$}}] {};
\node[bd] at (-1,0) [label=below:{{\scriptsize$n+1$}}] {};
\node[bd] at (-2,0) [label=below:{{\scriptsize$1$}}] {};
\end{tikzpicture} 
\ee
Take $n=3$ as an example, we can get the following Hasse diagram with five layers:
\be
\begin{tikzpicture}
\node (1) [hasse] at (0,3) {};
\node (2) [hasse] at (-1.5,1.5) {};
\node (3) [hasse] at (0,1.5) {};
\node (4) [hasse] at (1.5,1.5) {};
\node (7) [hasse] at (0.75,0) {};
\node (6) [hasse] at (-0.75,0) {};
\node (8) [hasse] at (2.25,0) {};
\node (5) [hasse] at (-2.25,0) {};
\node (9) [hasse] at (-0.75,-1.5) {};
\node (10) [hasse] at (0.75,-1.5) {};
\node (11) [hasse] at (0,-3) {};
\node (12) [hasse] at (0,-4.5) {};
\draw (1) edge [] node[label=right:$\mathfrak{d}_5$] {} (4);
\draw (1) edge [] node[label=right:$\mathfrak{e}_6$] {} (3);
\draw (1) edge [] node[label=right:$\mathfrak{e}_6$] {} (2);
\draw (4) edge [] node[label=right:$\mathfrak{a}_5$] {} (6);
\draw (4) edge [] node[label=right:$\mathfrak{a}_6$] {} (7);
\draw (4) edge [] node[label=right:$\mathfrak{a}_6$] {} (8);
\draw (4) edge [] node[label=right:$\mathfrak{e}_6$] {} (5);
\draw (3) edge [] node[label=above right:$\mathfrak{a}_1$] {} (6);
\draw (2) edge [] node[label=right:$\mathfrak{d}_5$] {} (5);
\draw (2) edge [] node[label=right:$\mathfrak{a}_1$] {} (6);
\draw (5) edge [] node[label=right:$\mathfrak{a}_1$] {} (9);
\draw (6) edge [] node[label=right:$\mathfrak{d}_5$] {} (9);
\draw (6) edge [] node[label=right:$\mathfrak{a}_3$] {} (10);
\draw (7) edge [] node[label=right:$\mathfrak{a}_2$] {} (10);
\draw (8) edge [] node[label=right:$\mathfrak{a}_2$] {} (10);
\draw (9) edge [] node[label=right:$\mathfrak{a}_1$] {} (11);
\draw (10) edge [] node[label=right:$\mathfrak{d}_4$] {} (11);
\draw (11) edge [] node[label=right:$\mathfrak{d}_6$] {} (12);
\end{tikzpicture}
\ee
The resolution sequence is
\be
\ba
&(z_1,z_2,z_3,z_4,z_5;\delta_1)\cr
&(z_1,\delta_1;\delta_3)\cr
&(z_1,z_3,z_4,z_5,\delta_3;\delta_4)\cr
&(z_1^{(2)},z_3^{(1)},z_4^{(1)},\delta_4^{(1)};\delta_5)\cr
&(z_5,\delta_1;\delta_2)
\ea
\ee
The resolved equation is
\be
\begin{cases}
&z_3^2+z_4^2+z_1 z_2\delta_3+z_5^3\delta_1\delta_2^4\delta_3\delta_4^2=0\\
&z_1 z_5+2 z_2 z_3^2\delta_1+z_2 z_4^2 \delta_1+3z_2^7\delta_1^5\delta_2^4\delta_3^4\delta_4^2=0\,.
\end{cases}
\ee
We can deduce the following intersection diagram:
\be
\includegraphics[height=5cm]{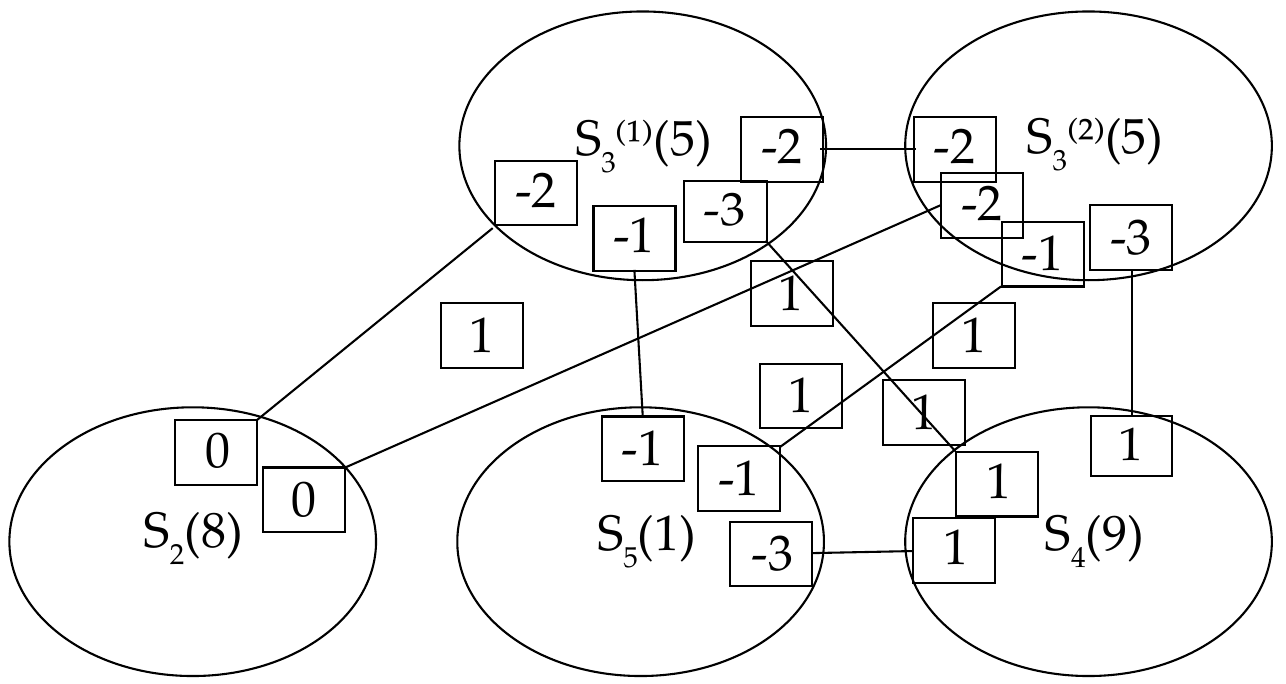}
\ee

\subsubsection{Class (113)}
\be
\begin{cases}
&z_1 z_2+z_4 z_5=0 \\
&z_1 z_4+z_3^2+z_2^{n_2}+z_5^{n_5}=0\,.
\end{cases}
\ee 
$n_2 \geqslant 3 \text { and } 2 \leqslant n_5 \leqslant n_2$

The resolution is smooth for $n_2=n_5=2n$.
\be
\begin{array}{||c|c|c||c|c|c||c|c|c||}
\hline
 r & d_H & G_F & \h r & \h d_H & f & b_3 & a&c\\
\hline
 n & 2(n+1)^2 & SO(4n+4)\times U(1) & 2n^2+2n-1 & 3n+3 & 2n+3 & 0 &-\frac{1}{8}+\frac{5n}{24}+n^2+\frac{2n^3}{3}&\frac{1}{3}n(1+3n+2n^2)\\
\hline
\end{array}
\ee

We have the 5d magnetic quiver:
\be
 \begin{tikzpicture}[x=1cm,y=1cm]
\draw[ligne, black](-2,0)--(0,0);
\draw[ligne, black](0,0)--(1,0);
\draw[dotted, black](1,0)--(2,0);
\draw[ligne, black](2,0)--(3,0);
\draw[ligne, black](-1,1)--(-1,0);
\node[] at (-4,0) {$\MQfive= $};
\node[bd] at (3,0) [label=below:{{\scriptsize$1$}}] {};
\node[bd] at (2,0) [label=below:{{\scriptsize$2$}}] {};
\node[bd] at (1,0) [label=below:{{\scriptsize$2n-2$}}] {};
\node[bd] at (0,0) [label=below:{{\scriptsize$2n-1$}}] {};
\node[bd] at (-1,1) [label=right:{{\scriptsize$n$}}] {};
\node[bd] at (-1,0) [label=below:{{\scriptsize$2n$}}] {};
\node[bd] at (-2,0) [label=below:{{\scriptsize$n+1$}}] {};
\end{tikzpicture} 
\ee
We can get the following Hasse diagram with $n$ layers:
\be
\begin{tikzpicture}
\node (1) [hasse] at (0,1) {};
\node (2) [hasse] at (0,0) {};
\node (3) [hasse] at (0,-1) {};
\node (4) [hasse] at (0,-2) {};
\node (5) [hasse] at (0,-3) {};
\node (6) [hasse] at (0,-4) {};
\draw (1) edge [] node[label=right:$\mathfrak{d}_5$] {} (2);
\draw (2) edge [] node[label=right:$\mathfrak{d}_6$] {} (3);
\draw (3) edge [] node[label=right:$\mathfrak{d}_8$] {} (4);
\draw (5) edge [] node[label=right:$\mathfrak{d}_{2n+2}$] {} (6);
\draw[dotted, black](0,-2)--(0,-3);
\end{tikzpicture}
\ee
The Hasse diagram correponds to the 5d SCFT with IR gauge description $SU(n+1)_2+(2n+2)\mbf{F}$, see table 27 of \cite{Bourget:2019aer}. The UV flavor symmetry is $G_F=SO(4n+4)\times U(1)$.

For $n_2=n_5=2n$, the resolution sequence is
\be
\ba
&(z_1,z_2,z_3,z_4,z_5;\delta_1)\cr
&(z_1,z_3,z_4,\delta_j;\delta_{j+1})\quad (j=1,\dots,n-1)\,.
\ea
\ee
The resolved equation is
\be
\begin{cases}
&z_1 z_2+z_4 z_5=0\\
&z_3^2+z_1 z_4+(z_2^{2n}+z_5^{2n})\prod_{j=1}^{n-1}\delta_j^{2n-2j}=0\,.
\end{cases}
\ee
The triple intersection numbers are
\be
\includegraphics[height=2cm]{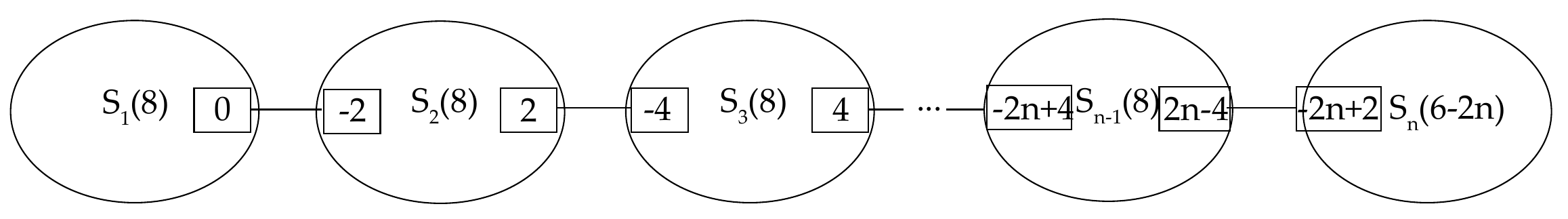}
\ee

Note that this model is also equivalent to the Class (114) with $n_2=2n_5=2n$:
\be
\begin{cases}
& z_1 z_2+z_4 z_5=0 \\
& z_1 z_4+z_3^2+z_2^{n_2}+z_1 z_5^{n_5}=0\,,
\end{cases}
\ee 
the Class (115) with $2n_2=n_5=2n$
\be
\begin{cases}
& z_1 z_2+z_4 z_5=0 \\
& z_1 z_4+z_3^2+z_2^{n_2} z_4+z_5^{n_5}=0\,
\end{cases}
\ee 
and the the Class (116) with $n_2=n_5=n$
\be
\begin{cases}
& z_1 z_2+z_4 z_5=0 \\
& z_1 z_4+2 z_3^2+3 z_2^{n_2} z_4+4 z_1 z_5^{n_5}+5 z_2^{n_2} z_5^{n_5}=0\,.
\end{cases}
\ee 
\subsubsection{Class (152)}
\be
\begin{cases}
&z_1 z_2+z_4^3+z_5^3=0 \\
&z_1 z_4+z_3^2+z_2^n=0\,.
\end{cases}
\ee
$n\geqslant3$

We have a smooth resolution for all odd $n$, but only cases with $n=4k+1,k\geqslant1$ satisfy $b_3=0$.
\be
\begin{array}{||c||c|c||c|c|c||c|c|c||}
\hline
 &r & d_H & \h r & \h d_H & f & b_3 & a&c\\
\hline
 n=4k+1&6k+1& 18k+5 & 6n& 3n+6 & 6 & 0 & 24k^2+\frac{83}{4}k+\frac{109}{24}&24k^2+21k+\frac{29}{6}\\
\hline
\end{array}
\ee
Take $n=5$ as an example, which is a rank-7 5d SCFT with $G_F=\color{blue}{SU(2)^4\times U(1)^2}$.

The resolution sequence is
\be
\ba
&(z_1,z_2,z_3,z_4,z_5;\delta_1)\cr
&(z_1^{(5)},z_3^{(3)},z_4^{(1)},z_5^{(1)},\delta_1^{(2)};\delta_8)\cr
&(z_1,z_3,z_4,\delta_1;\delta_5)\cr
&(z_1,z_3,\delta_1;\delta_4)\cr
&(z_1,\delta_1,\delta_4;\delta_7)\cr
&(z_1,\delta_1;\delta_3)\cr
&(z_1,\delta_4;\delta_6)\cr
&(z_2,\delta_1;\delta_2)\cr
\ea
\ee
The resolved equation is
\be
\begin{cases}
&z_5^3\delta_1+z_4^3\delta_1\delta_5^3+z_1 z_2 \delta_6=0\\
&z_1 z_4\delta_3+z_3^2\delta_4+z_2^5\delta_1^3\delta_2^8\delta_3^3\delta_4^2\delta_5\delta_6\delta_7^4=0\,.
\end{cases}
\ee
We can deduce the following intersection diagram:
\be
\includegraphics[height=7cm]{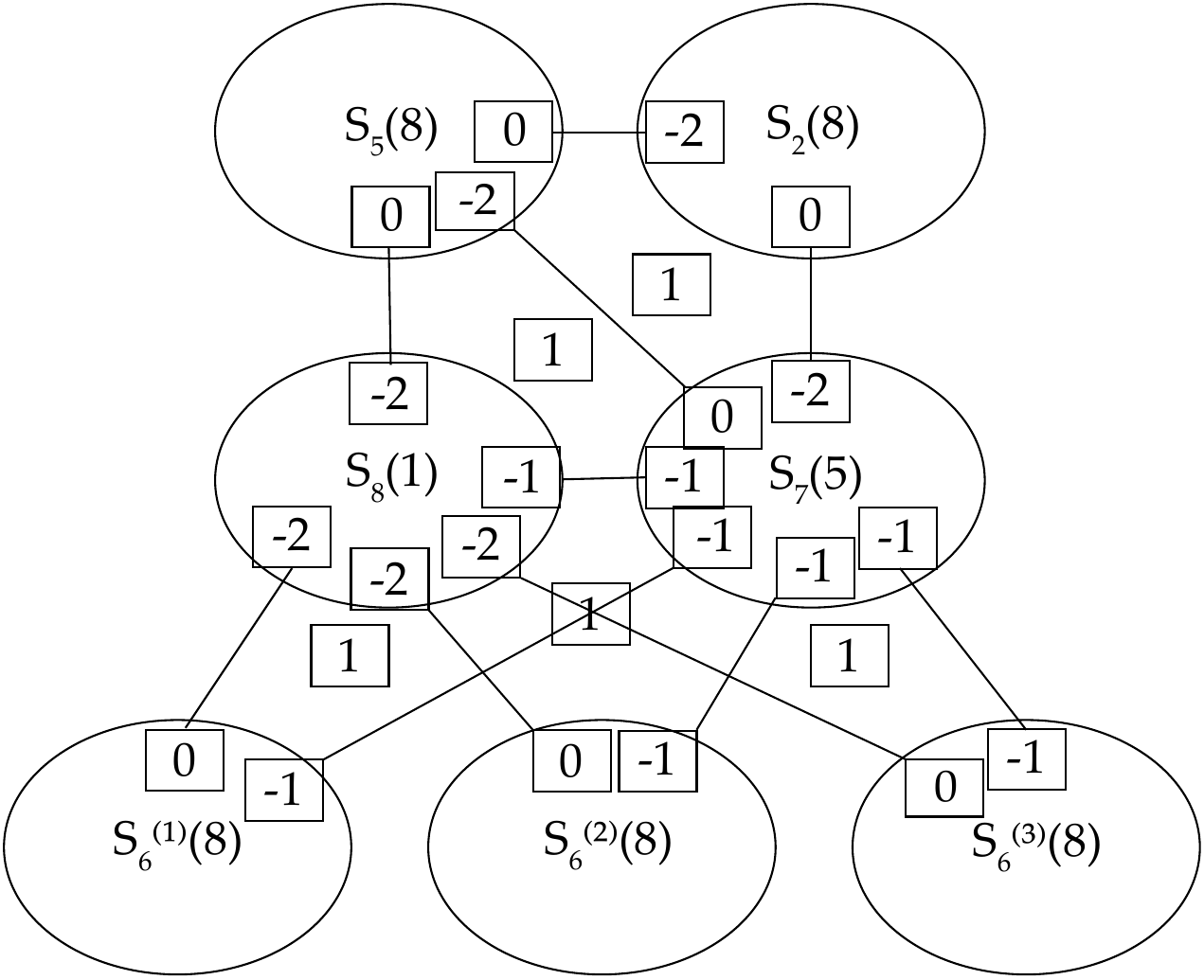}
\ee
For $G_{F,nA}$, we observe that the only compact divisor with additional flavor $(-2)$-curves is $S_8$. We take it to be gdP$_8$ of type $8\mbf{A}_1$. The $(-2)$-curves on $S_8$ are
\be
\ba
&S_8\cdot S_6^{(1)}:\ h-e_1-e_2-e_3\cr
&S_8\cdot S_6^{(2)}:\ h-e_1-e_4-e_5\cr
&S_8\cdot S_6^{(3)}:\ h-e_1-e_6-e_7\cr
&S_8\cdot S_5:\ 3h-e_1-e_2-e_3-e_4-e_5-e_6-e_7-2e_8\cr
&\text{Flavor\ curves}:\ e_2-e_3\ ,\ e_4-e_5\ ,\ e_6-e_7\ ,\ 2h-e_1-e_2-e_3-e_4-e_5-e_6\,.
\ea
\ee
Hence from the flavor curves, we read off $G_{F,nA}=SU(2)^4$ and $G_F=SU(2)^4\times U(1)^2$.

\subsubsection{Class (202)}
\be
\begin{cases}
&z_1 z_2+z_3 z_4=0 \\
&z_1 z_3+z_2^n z_5+z_4^3+z_5^3=0\,.
\end{cases}
\ee
$n\geqslant2$

We have a smooth resolution for all $n\geqslant2$, but only cases with odd $n$ satisfy $b_3=0$.
\be
\begin{array}{||c||c|c||c|c|c||c|c|c||}
\hline
 &r & d_H &  \h r & \h d_H & f & b_3 & a&c\\
\hline
 n=2k+1&3n-1 & 6n+7 &  6n& 3n+6 & 7 & 0 & \frac{n(23+24n)}{8}&\frac{1}{4}+3n+3n^2\\
\hline
\end{array}
\ee

We take $n=3$ as an example, which is a rank-7 5d SCFT with $G_F=\color{blue}{SU(2)^3\times U(1)^4}$.

The resolution sequence is 
\be
\ba
&(z_1,z_2,z_3,z_4,z_5;\delta_1)\cr
&(z_1,z_3,\delta_1;\delta_6)\cr
&(z_1,z_4,z_5,\delta_6;\delta_7)\cr
&(z_1,z_4,\delta_1;\delta_5)\cr
&(z_2,z_3,\delta_1;\delta_3)\cr
&(z_1,\delta_1;\delta_4)\cr
&(z_3,\delta_1;\delta_2)\cr
&(\delta_1,\delta_6;\delta_8)\cr
&(\delta_6,\delta_7;\delta_9)
\ea
\ee

The resolved equation is
\be
\begin{cases}
&z_3 z_4\delta_2+z_1 z_2 \delta_4=0\\
&z_1 z_3 \delta_6+z_5^3\delta_1\delta_7+z_4^3\delta_1\delta_5^3\delta_7+z_2^3z_5\delta_1^2\delta_2\delta_3^4\delta_4\delta_5\delta_6\delta_8^2=0\,.
\end{cases}
\ee 

We can deduce the following intersection diagram:
\be
\includegraphics[height=7cm]{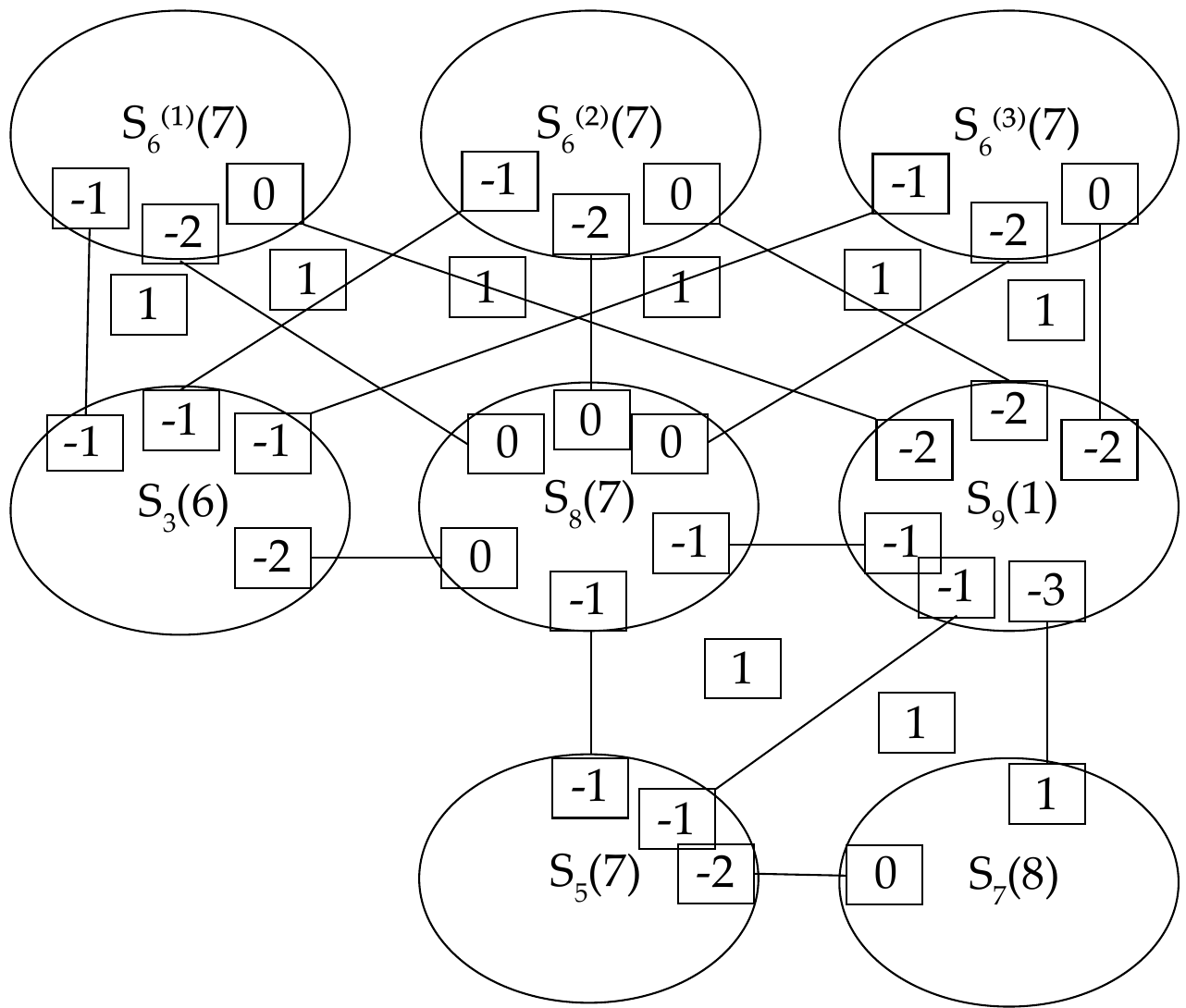}
\ee
For the $G_{F,nA}$, we observe that only $S_9$ can have additional flavor curves. The maximal set of flavor $(-2)$-curves give $G_{nF,A}=SU(2)^3$. We write the $(-2)$ and $(-3)$-curves on $S_9$ using the standard Picard group basis:
\be
\ba
&S_{9}\cdot S_{7}:\ 2h-\sum_{i=1}^7 e_i\cr
&S_{9}\cdot S_6^{(1)}:\ e_1-e_2\cr
&S_{9}\cdot S_6^{(2)}:\ e_3-e_4\cr
&S_{9}\cdot S_6^{(3)}:\ e_5-e_6\cr
&\text{Flavor\ curves}:\ h-e_1-e_2-e_8\ ,\ h-e_3-e_4-e_8\ ,\ h-e_5-e_6-e_8\,.
\ea
\ee

\subsubsection{Class (220)}
\be
\begin{cases}
&z_1 z_2+z_3 z_4+z_5^n=0 \\
&z_1 z_4+z_2 z_3+z_1 z_5+z_2^n+z_4^n=0\,.
\end{cases}
\ee
$n\geqslant3$

The resolution is smooth for arbitrary $n\geqslant3$.
\be
\begin{array}{||c|c|c||c|c|c||c|c|c||}
\hline
 r & d_H & G_F & \h r & \h d_H & f & b_3 & a&c\\
\hline
 n-1 & n^2+n+1 & SU(2n+1)\times U(1) & n^2-n& 3n & 2n+1 & 0 & \frac{n(-7+6n+4n^2)}{24}&\frac{n(-2+3n+2n^2)}{12}\\
\hline
\end{array}
\ee
We have the 5d magnetic quiver:
\be
 \begin{tikzpicture}[x=.5cm,y=.5cm]
\draw[ligne, black](-3,0)--(-1,0);
\draw[dotted, black](-1,0)--(0,0);
\draw[ligne, black](0,0)--(1,0);
\draw[dotted, black](1,0)--(2,0);
\draw[ligne, black](2,0)--(4,0);
\draw[ligne, black](0,1)--(0,0);
\draw[ligne, black](1,0)--(1,1);
\node[] at (-5,0) {$\MQfive= $};
\node[bd] at (4,0) [label=below:{{\scriptsize$1$}}] {};
\node[bd] at (3,0) [label=below:{{\scriptsize$2$}}] {};
\node[bd] at (2,0) [label=below:{{\scriptsize$3$}}] {};
\node[bd] at (1,0) [label=below:{{\scriptsize$n$}}] {};
\node[bd] at (0,0) [label=below:{{\scriptsize$n$}}] {};
\node[bd] at (-1,0) [label=below:{{\scriptsize$3$}}] {};
\node[bd] at (-2,0) [label=below:{{\scriptsize$2$}}] {};
\node[bd] at (-3,0) [label=below:{{\scriptsize$1$}}] {};
\node[bd] at (0,1) [label=right:{{\scriptsize$1$}}] {};
\node[bd] at (1,1) [label=right:{{\scriptsize$1$}}] {};
\end{tikzpicture} 
\ee
We can get the following Hasse diagram with $n-1$ layers:
\be
\begin{tikzpicture}
\node (1) [hasse] at (0,1) {};
\node (2) [hasse] at (0,0) {};
\node (3) [hasse] at (0,-1) {};
\node (4) [hasse] at (0,-2) {};
\node (5) [hasse] at (0,-3) {};
\draw (1) edge [] node[label=right:$\mathfrak{d}_5$] {} (2);
\draw (2) edge [] node[label=right:$\mathfrak{a}_6$] {} (3);
\draw (4) edge [] node[label=right:$\mathfrak{a}_{2n}$] {} (5);
\draw[dotted, black](0,-1)--(0,-2);
\end{tikzpicture}
\ee
From the Hasse diagram, we found that the 5d SCFT is the one with IR gauge theory description $SU(n)_1+2n\mbf{F}$~\cite{Bourget:2019aer}. The enhanced UV flavor symmetry is $G_F=SU(2n+1)\times U(1)$.

The resolution sequence is
\be
\ba
&(z_1,z_2,z_3,z_4,z_5;\delta_1)\cr
&(z_1,z_3,\delta_j;\delta_{j+1})\quad (j=1,\dots,n-2)\,.
\ea
\ee
The resolved equation is
\be
\begin{cases}
&z_1 z_2+z_3 z_4+z_5^n\prod_{j=1}^{n-2}\delta_j^{n-1-j}=0\\
&z_2 z_3+z_1 z_4+z_1 z_5+(z_2^n+z_4^n)\prod_{j=1}^{n-2}\delta_j^{n-1-j}=0\,.
\end{cases}
\ee
The triple intersection numbers are
\be
\includegraphics[height=2cm]{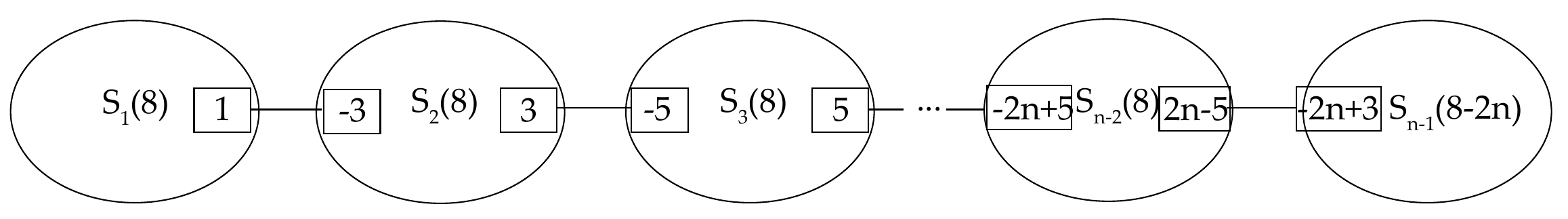}
\ee
Note that this case is also equivalent to the Class (234) with $n_2=n_5=n$:
\be
\begin{cases}
& z_1 z_2+z_3 z_4+z_5^{n_5}=0 \\
& z_1 z_4+z_3 z_5+z_2^{n_2}=0\,,
\end{cases}
\ee 
the Class (235) with $n_2=n_5=n$:
\be
\begin{cases}
& z_1 z_2+z_3 z_4=0 \\
& z_1 z_4+z_3 z_5+z_2^{n_2}+z_5^{n_5}=0\,,
\end{cases}
\ee 
the Class (259) with $n_2+1=n_5=n$:
\be
\begin{cases}
& z_1 z_2+z_3 z_4+z_5^{n_5}=0 \\
& z_1 z_4+z_3 z_5+z_2^{n_2} z_4=0\,
\end{cases}
\ee 
and the Class (260) with $n_2+1=n_5=n$:
\be
\begin{cases}
& z_1 z_2+z_3 z_4=0 \\
& z_1 z_4+z_3 z_5+z_2^{n_2} z_4+z_5^{n_5}=0\,.
\end{cases}
\ee 
\subsection{The only rank-1 example: Class (1)}

There is only one rank-1 example with smooth $\widetilde{X}_3$, smooth exceptional divisor and $b_3=0$, which is the case of Class (1): 
\be
\begin{cases}
&z_1^2+z_2^2+z_3^2+z_4^2+z_5^2=0\\
&z_1^2+2z_2^2+3z_3^2+4z_4^2+5z_5^2=0\,.
\end{cases}
\ee 
\be
\begin{array}{||c|c|c||c|c|c||c|c||c|c||}
\hline
r & d_H & G_F & \h r & \h d_H & f & b_3 & \mathfrak{f} & a & c\\
\hline
1 & 7 & SO(10) & 2 & 6 & 5 & 0 & 0 & \frac{7}{4} & 2\\
\hline
\end{array}
\ee

It describes a local $dP_5$ singularity, which corresponds to the 5d Seiberg rank-1 $E_5$ theory. The case was already discussed in section~\ref{sec:class1}, and we would not repeat here again.
 
\subsection{Rank-3 examples}
\subsubsection{Class (15)}
\be
\begin{cases}
&z_1 z_3+z_4 z_5=0\\
&z_1^2+z_2^2+z_3^6+z_4^3+z_5^3=0\,.
\end{cases}
\ee 

\be
\begin{array}{||c|c|c||c|c|c||c|c|c||}
\hline
 r & d_H & G_F & \h r & \h d_H & f & b_3 & a&c\\
\hline
 3 & 23 & E_6\times SU(2) & 16 & 10 & 7 & 0 & \frac{263}{12}&\frac{67}{3}\\
\hline
\end{array}
\ee

The resolution sequence is
\be
\ba
&(z_1,z_2,z_3,z_4,z_5;\delta_1)\cr
&(z_1^{(2)},z_2^{(2)},z_4^{(1)},z_5^{(1)},\delta_1^{(1)};\delta_2)\,.
\ea
\ee
The resolved equation is
\be
\begin{cases}
&z_1 z_3+z_4 z_5=0\\
&z_1^2+z_2^2+z_3^6\delta_1^4+z_4^3\delta_1+z_5^3\delta_1=0\,.
\end{cases}
\ee
Note that the exceptional divisor $\delta_1=0$ takes the form of
\be
\delta_1=z_1^2+z_2^2=z_1 z_3+z_4 z_5=0\,.
\ee
Hence $\delta_1=0$ is reducible, and we denote the two irreducible components as $S_1^{(1)}:\delta_1=z_1+iz_2=0$ and $S_1^{(2)}:\delta_1=z_1-iz_2=0$. Note that $S_1^{(1)}$ and $S_1^{(2)}$ intersects at
\be
\delta_1=z_1=z_2=z_4 z_5=0\,,
\ee
which is a disjoint union of two curves
\be
\ba
S_1^{(1)}\cdot S_2^{(2)}&=C_1+C_2\,,\cr
C_1:\ &\delta_1=z_1=z_2=z_4=0\cr
C_2:\ &\delta_1=z_1=z_2=z_5=0\,.
\ea
\ee

We can compute the triple intersection number between $S_1=S_1^{(1)}+S_1^{(2)}$ and $S_2$:
\be
S_1^3=S_2^3=2\ ,\ S_1\cdot S_2^2=-2\ ,\ S_2\cdot S_1^2=2\,.
\ee
We can deduce the following intersection diagram:
\be
\includegraphics[height=5cm]{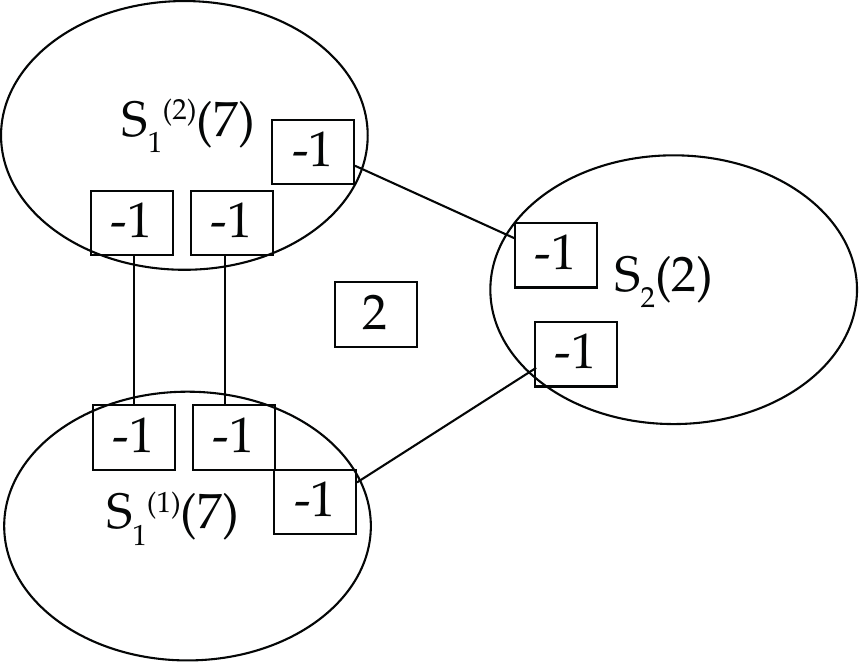}
\ee
From the resolved geometry, we can observe that the 5d SCFT has an IR quiver gauge theory description of
\be
SU(3)_0-SU(2)-[4\mbf{F}]\,,
\ee
where the $SU(3)$ factor comes from the irreducible components $S_1^{(1)}$ and $S_2^{(2)}$, and the $SU(2)$ factor correponds to $S_2$.

Now we study the 5d Higgs branch and magnetic quiver. First, one observes that the 4d $\mc{N}=2$ theory $\FTfour$ has a simple description:
\be
   \begin{tikzpicture}[x=.5cm,y=.5cm]
\draw[ligne, black](1,0)--(2,0);
\draw[ligne, black](4,0)--(5,0);
\draw[ligne, black](-1,0)--(-2,0);
\draw[ligne, black](-4,0)--(-5,0);
\draw[ligne, black](0,0.4)--(0,1.6);
\draw[ligne, black](0,2.4)--(0,3.6);
\node[] at (0,4) {\scriptsize$[2]$};
\node[] at (0,2) {\scriptsize$SU(4)$};
\node[] at (-3,0) {\scriptsize$SU(4)$};
\node[] at (-6,0) {\scriptsize$SU(2)$};
\node[] at (0,0) {\scriptsize$SU(6)$};
\node[] at (3,0) {\scriptsize$SU(4)$};
\node[] at (6,0) {\scriptsize$SU(2)$};
\end{tikzpicture}
\ee
Now let us gauge the $U(1)^f$, $f=7$ flavor symmetry. The question is what happens to the flavor node $[2]$. If it becomes $U(2)$, then the $\MQfive$ is identical to the Magnetic quiver of the rank-2 Seiberg $E_6$ theory. However, this does not match the rank $r=3$. Hence the only option to split $[2]$ into two $U(1)$ gauge nodes after the gauging, and we get
\be
 \begin{tikzpicture}[x=.5cm,y=.5cm]
\draw[ligne, black](2,0)--(-2,0);
\draw[ligne, black](0,0)--(0,1);
\draw[ligne, black](1,2)--(0,1);
\draw[ligne, black](-1,2)--(0,1);
\node[] at (-4,0) {$\MQfive= $};
\node[bd] at (2,0) [label=below:{{\scriptsize$2$}}] {};
\node[bd] at (1,0) [label=below:{{\scriptsize$4$}}] {};
\node[bd] at (0,0) [label=below:{{\scriptsize$6$}}] {};
\node[bd] at (-1,0) [label=below:{{\scriptsize$4$}}] {};
\node[bd] at (-2,0) [label=below:{{\scriptsize$2$}}] {};
\node[bd] at (0,1) [label=right:{{\scriptsize$4$}}] {};
\node[bd] at (1,2) [label=right:{{\scriptsize$1$}}] {};
\node[bd] at (-1,2) [label=right:{{\scriptsize$1$}}] {};
\end{tikzpicture} 
\ee

We can get the following Hasse diagram with three layers:
\be
\begin{tikzpicture}
\node (1) [hasse] at (0,4) {};
\node (2) [hasse] at (2,2) {};
\node (3) [hasse] at (0,2) {};
\node (4) [hasse] at (-2,2) {};
\node (5) [hasse] at (1,0) {};
\node (6) [hasse] at (-1,0) {};
\node (7) [hasse] at (0,-2) {};
\draw (1) edge [] node[label=right:$\mathfrak{e}_6$] {} (4);
\draw (1) edge [] node[label=right:$\mathfrak{e}_6$] {} (3);
\draw (1) edge [] node[label=right:$\mathfrak{d}_5$] {} (2);
\draw (4) edge [] node[label=left:$\mathfrak{e}_6$] {} (6);
\draw (4) edge [] node[label=right:$\mathfrak{a}_1$] {} (5);
\draw (3) edge [] node[label=left:$\mathfrak{e}_6$] {} (6);
\draw (3) edge [] node[label=right:$\mathfrak{a}_1$] {} (5);
\draw (2) edge [] node[label=right:$\mathfrak{a}_5$] {} (5);
\draw (5) edge [] node[label=right:$\mathfrak{e}_6$] {} (7);
\draw (6) edge [] node[label=right:$\mathfrak{a}_1$] {} (7);
\end{tikzpicture}
\ee
From the bottom layer, we can read off that $G_F=E_6\times SU(2)$.

\subsubsection{Class (55)}
\be
\begin{cases}
&z_1 z_2+z_3^2+z_4^2+ z_5^2=0\\
&z_1 z_3+2z_2^5+z_2 z_4^2=0\,.
\end{cases}
\ee 
\be
\begin{array}{||c|c|c||c|c|c||c|c|c||}
\hline
 r & d_H & G_F & \h r & \h d_H & f & b_3 & a&c\\
\hline
 3 & 19 & SO(10)\times SU(2)\times U(1) & 12 & 10 & 7 & 0&\frac{179}{12} & \frac{46}{3}\\
\hline
\end{array}
\ee
The resolution sequence is
\be
\ba
&(z_1,z_2,z_3,z_4,z_5;\delta_1)\cr
&(z_1^{(2)},z_3^{(1)},z_4^{(1)},z_5^{(1)},\delta_1^{(1)};\delta_4)\cr
&(z_1,\delta_1;\delta_3)\cr
&(z_3,\delta_1;\delta_2)
\ea
\ee
The resolved equation is
\be
\begin{cases}
&z_1 z_2 \delta_3+z_5^2 +z_4^2+z_3^2 \delta_2^2=0\\
&z_1z_3+z_2z_4^2\delta_1+2 z_2^5 \delta_1^3 \delta_2^2 \delta_3^2=0\,.
\end{cases}
\ee 
We have the 5d magnetic quiver:
\be
 \begin{tikzpicture}[x=.5cm,y=.5cm]
\draw[ligne, black](2,0)--(-2,0);
\draw[ligne, black](0,0)--(0,1);
\draw[ligne, black](0,2)--(0,1);
\draw[ligne, black](-1,0)--(-1,1);
\node[] at (-4,0) {$\MQfive= $};
\node[bd] at (2,0) [label=below:{{\scriptsize$1$}}] {};
\node[bd] at (1,0) [label=below:{{\scriptsize$3$}}] {};
\node[bd] at (0,0) [label=below:{{\scriptsize$5$}}] {};
\node[bd] at (-1,0) [label=below:{{\scriptsize$4$}}] {};
\node[bd] at (-2,0) [label=below:{{\scriptsize$2$}}] {};
\node[bd] at (0,1) [label=right:{{\scriptsize$3$}}] {};
\node[bd] at (0,2) [label=right:{{\scriptsize$1$}}] {};
\node[bd] at (-1,1) [label=right:{{\scriptsize$1$}}] {};
\end{tikzpicture} 
\ee
We can get the following Hasse diagram with three layers:
\be
\begin{tikzpicture}
\node (1) [hasse] at (0,4) {};
\node (2) [hasse] at (2,2) {};
\node (3) [hasse] at (0,2) {};
\node (4) [hasse] at (-2,2) {};
\node (5) [hasse] at (1,0) {};
\node (6) [hasse] at (-1,0) {};
\node (7) [hasse] at (0,-2) {};
\draw (1) edge [] node[label=right:$\mathfrak{d}_5$] {} (4);
\draw (1) edge [] node[label=right:$\mathfrak{e}_6$] {} (3);
\draw (1) edge [] node[label=right:$\mathfrak{e}_6$] {} (2);
\draw (4) edge [] node[label=left:$\mathfrak{e}_6$] {} (6);
\draw (4) edge [] node[label=right:$\mathfrak{a}_5$] {} (5);
\draw (3) edge [] node[label=left:$\mathfrak{d}_6$] {} (6);
\draw (3) edge [] node[label=right:$\mathfrak{a}_1$] {} (5);
\draw (2) edge [] node[label=right:$\mathfrak{a}_1$] {} (5);
\draw (5) edge [] node[label=right:$\mathfrak{d}_5$] {} (7);
\draw (6) edge [] node[label=right:$\mathfrak{a}_1$] {} (7);
\end{tikzpicture}
\ee

We can deduce the following intersection diagram:
\be
\includegraphics[height=5cm]{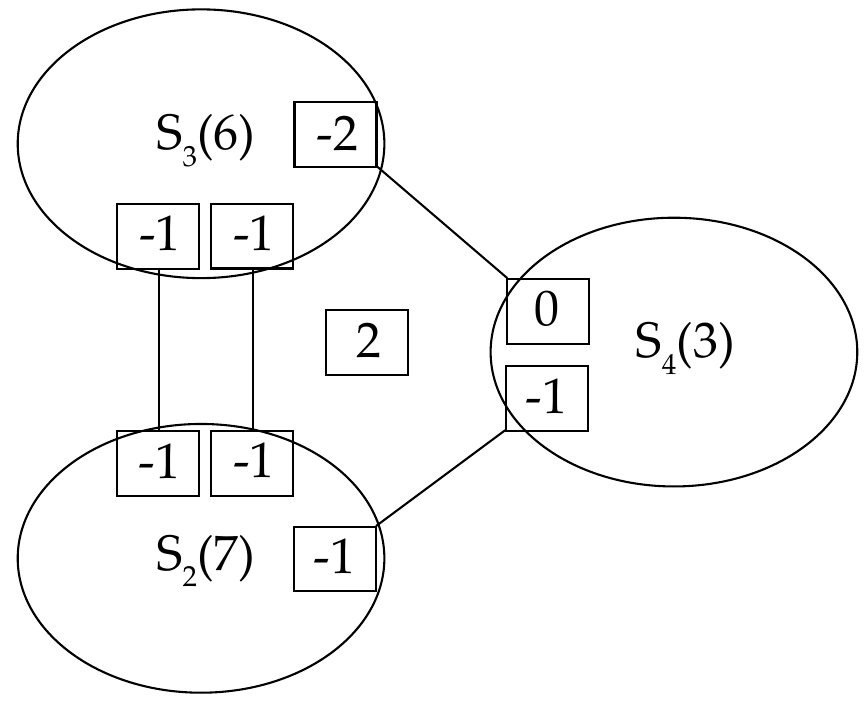}
\ee

\subsubsection{Class (228)}
\be
\begin{cases}
&z_1 z_2+z_3 z_4=0\\
&z_1 z_4+z_3^2+z_2^n+z_5^3=0\,.
\end{cases}
\ee
$n\geqslant3$

The resolution is smooth only for $n=6$.
\be
\begin{array}{||c|c|c||c|c|c||c|c|c||}
\hline
 r & d_H & G_F & \h r & \h d_H & f & b_3 & a&c\\
\hline
 3 & 24 & \color{blue}{E_7} & 17 & 10 & 7 & 0&\frac{71}{3} & \frac{289}{12}\\
\hline
\end{array}
\ee
The resolution sequence is
\be
\ba
&(z_1,z_2,z_3,z_4,z_5;\delta_1)\cr
&(z_1^{(3)},z_3^{(2)},z_4^{(1)},z_5^{(1)},\delta_1^{(1)};\delta_3)\cr
&(z_1,z_3,\delta_1;\delta_2)
\ea
\ee
The resolved equation is
\be
\begin{cases}
&z_1 z_2+z_3 z_4=0\\
&z_1 z_4+z_5^3 \delta_1+z_3^2 \delta_2+z_2^6\delta_1^4\delta_2^3=0\,.
\end{cases}
\ee 

We can deduce the following intersection diagram:
\be
\includegraphics[height=5cm]{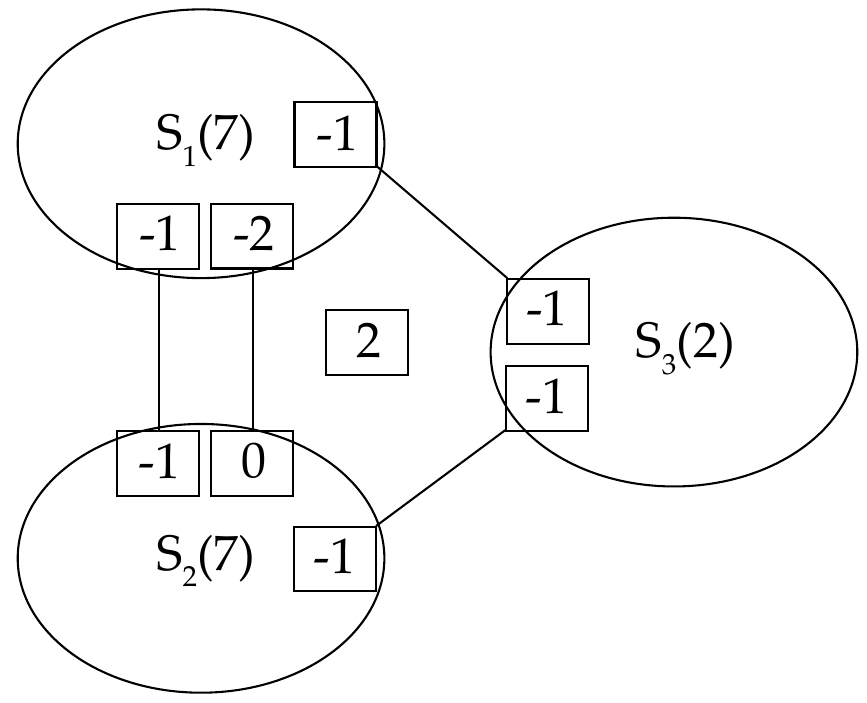}
\ee

For the $G_{F,nA}$, we observe that  $S_3$ is a gdP$_7$, which can contain flavor curves forming the Dynkin diagram of $E_6$. In addition, there is also another $(-1)$-curve $C$ on $S_2$ that intersects $S_2\cdot S_3$ at one point. After flopping $C$ into $S_3$, one get the flavor curves of $E_7$, hence $G_{F,nA}=E_7$

Class (244) and Class (254) have a smooth resolution only for $n=4$, and in that case they are the same theory as Class (228) (with $n=6$).

\subsubsection{Class (262)}
\be
\begin{cases}
& z_1 z_2+z_3 z_4+z_5^2=0 \\
& z_1 z_5+z_3^2+z_4^4+z_2^8=0\,.
\end{cases}
\ee
\be
\begin{array}{||c|c|c||c|c|c||c|c|c||}
\hline
 r & d_H & G_F & \h r & \h d_H & f & b_3 & a&c\\
\hline
 3 & 33 & E_7 & 26 & 10 & 7 & 0&\frac{527}{12} & \frac{133}{3}\\
\hline
\end{array}
\ee
The resolution sequence is
\be
\ba
&(z_1,z_2,z_3,z_4,z_5;\delta_1)\cr
&(z_1^{(4)},z_3^{(3)},z_4^{(1)},z_5^{(2)},\delta_1^{(1)};\delta_3)\cr
&(z_1,z_3,z_5,\delta_1;\delta_2)
\ea
\ee
The resolved equation is
\be
\begin{cases}
&z_1 z_2+z_3 z_4+z_5^2\delta_2=0\\
&z_3^2+z_1 z_5+z_4^4 \delta_1^2+z_2^8\delta_1^6\delta_2^4=0\,.
\end{cases}
\ee 
We have the 5d magnetic quiver:
\be
 \begin{tikzpicture}[x=.5cm,y=.5cm]
\draw[ligne, black](3,0)--(-2,0);
\draw[ligne, black](1,0)--(1,1);
\draw[ligne, black](2,0)--(2,1);
\node[] at (-4,0) {$\MQfive= $};
\node[bd] at (3,0) [label=below:{{\scriptsize$3$}}] {};
\node[bd] at (2,0) [label=below:{{\scriptsize$6$}}] {};
\node[bd] at (1,0) [label=below:{{\scriptsize$8$}}] {};
\node[bd] at (0,0) [label=below:{{\scriptsize$6$}}] {};
\node[bd] at (-1,0) [label=below:{{\scriptsize$4$}}] {};
\node[bd] at (-2,0) [label=below:{{\scriptsize$2$}}] {};
\node[bd] at (1,1) [label=right:{{\scriptsize$4$}}] {};
\node[bd] at (2,1) [label=right:{{\scriptsize$1$}}] {};
\end{tikzpicture} 
\ee
We can get the following Hasse diagram with three layers:
\be
\begin{tikzpicture}
\node (1) [hasse] at (0,1) {};
\node (2) [hasse] at (0,0) {};
\node (3) [hasse] at (0,-1) {};
\node (4) [hasse] at (0,-2) {};
\draw (1) edge [] node[label=right:$\mathfrak{d}_5$] {} (2);
\draw (2) edge [] node[label=right:$\mathfrak{d}_6$] {} (3);
\draw (3) edge [] node[label=right:$\mathfrak{e}_7$] {} (4);
\end{tikzpicture}
\ee

We can deduce the following intersection diagram:
\be
\includegraphics[height=5cm]{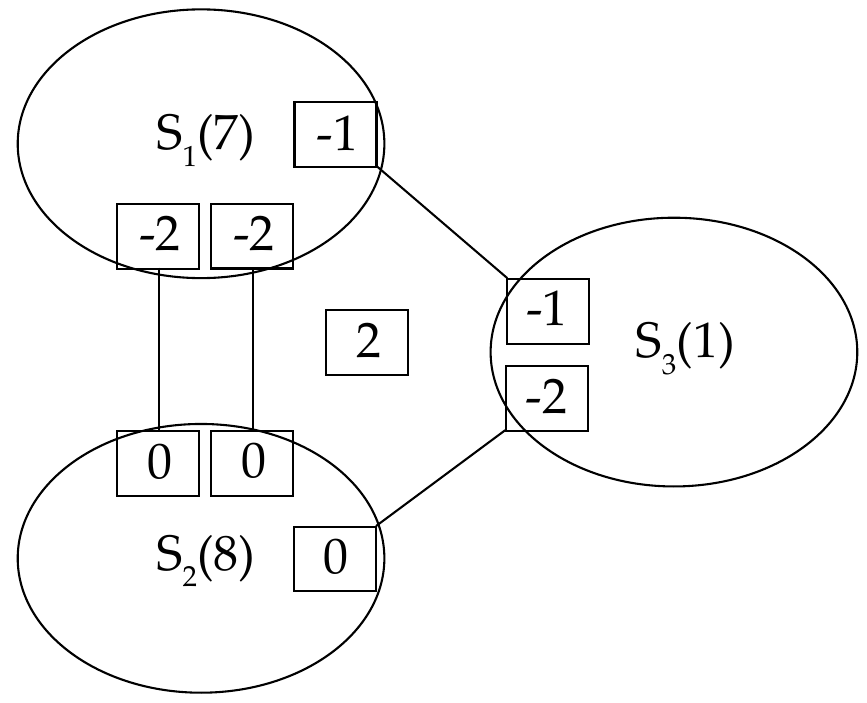}
\ee

Note that this is equivalent to the Class (118) with $n=8$:
\be
\begin{cases}
& z_1 z_2+z_4^2+z_5^3=0 \\
& z_1 z_4+z_3^2+z_2^n=0\,.
\end{cases}
\ee
and the Class (134) with $n=3$:
\be
\begin{cases}
& z_1 z_2+z_4^2+z_5^n=0 \\
& z_1 z_4+z_3^2+z_2^6 z_5=0\,.
\end{cases}
\ee

\subsection{Rank-5 examples}
\subsubsection{Class (2)}
\be
\begin{cases}
&z_1^2+z_2^2+z_3^2+z_4^3+z_5^3=0\\
&z_1^2+2z_2^2+3z_3^2+4z_4^3+5z_5^3=0\,.
\end{cases}
\ee 

\be
\begin{array}{||c|c|c||c|c|c||c|c|c||}
\hline
 r & d_H & G_F & \h r & \h d_H & f & b_3 & a&c\\
\hline
 5 & 19 & \color{blue}{SU(2)^6} & 13 & 11 & 6 & 0 &\frac{473}{24} & \frac{121}{6}\\
\hline
\end{array}
\ee

The resolution sequence is
\be
\ba
&(z_1,z_2,z_3,z_4,z_5;\delta_1)\cr
&(z_1^{(1)},z_2^{(1)},z_3^{(1)},\delta_1^{(2)};\delta_2)
\ea
\ee
The resolved equation is
\be
\begin{cases}
&z_1^2+z_2^2+z_3^2+(z_4^3+z_5^3)\delta_1=0\\
&z_1^2+2z_2^2+3z_3^2+(4z_4^3+5z_5^3)\delta_1=0\,.
\end{cases}
\ee 
In fact, the exceptional divisor $\delta_1=0$ has four irreducible components:
\be
\ba
S_1^{(1)}:\ &\delta_1=z_2+i\sqrt{2}z_1=z_3+z_1=0\cr
S_1^{(2)}:\ &\delta_1=z_2+i\sqrt{2}z_1=z_3-z_1=0\cr
S_1^{(3)}:\ &\delta_1=z_2-i\sqrt{2}z_1=z_3+z_1=0\cr
S_1^{(4)}:\ &\delta_1=z_2-i\sqrt{2}z_1=z_3-z_1=0\,.
\ea
\ee
Along with $S_2:\delta_2=0$, the theory has 5d rank $r=5$, and the triple intersection numbers are:
\be
\label{Class-2-int}
\includegraphics[height=7cm]{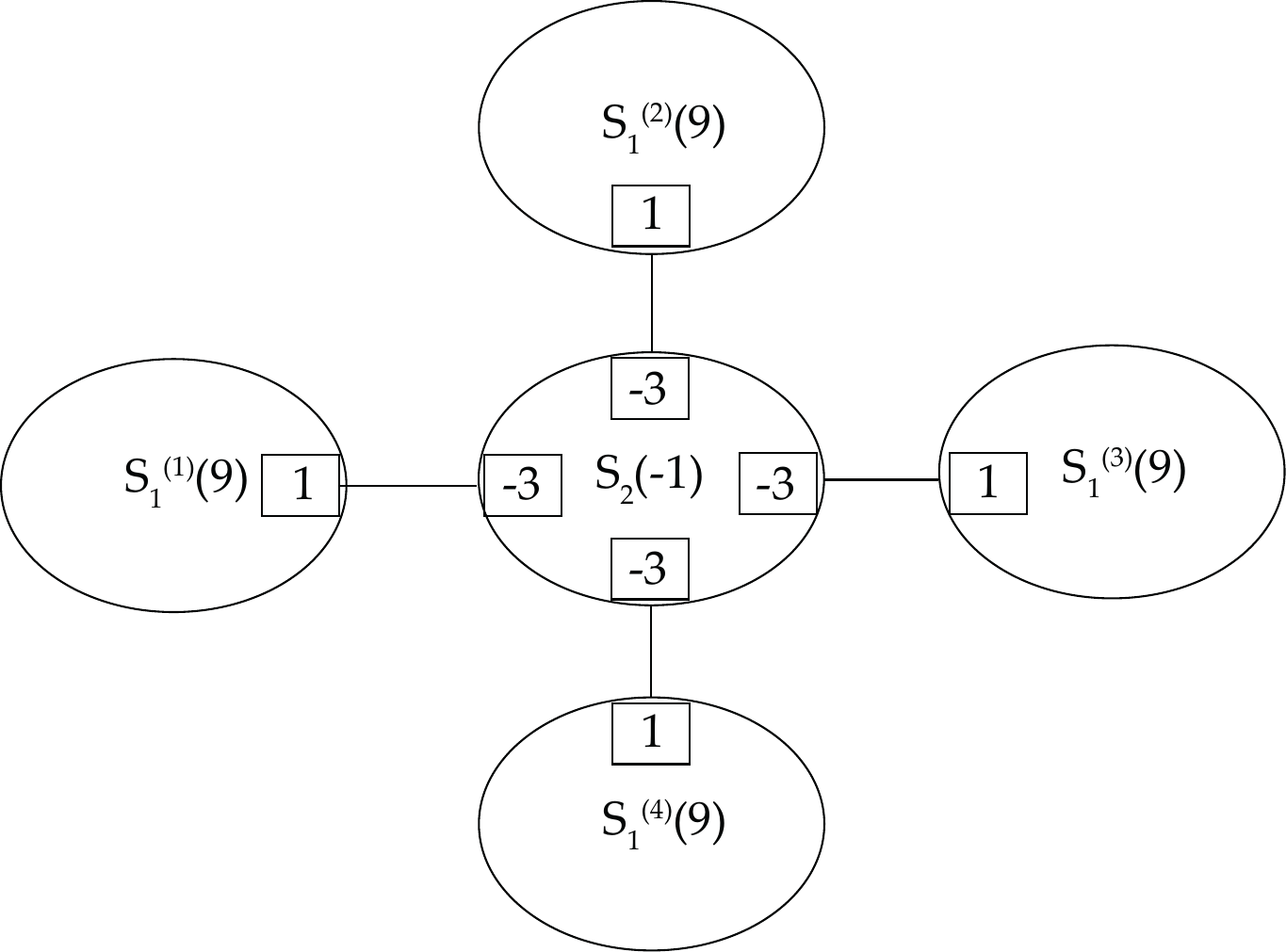}
\ee
The flavor rank can be counted from the number of linearly independent 2-cycles minus $r$:
\be
\ba
f&=b_2(\widetilde{X})-r\cr
&=11-5\cr
&=6\,.
\ea
\ee
Note we have $b_2=11$ from the divisor $S_2$, and the other $\mb{P}^2$s do not contribute to $b_2(\widetilde{X})$. This geometry has $b_3(\widetilde{X})=0$.

In fact, we observe that the 5d SCFT is a descendant of the $\mbf{AD}[E_7,E_7]$ theory, by comparing with Figure 1 of \cite{Closset:2021lwy}. Hence it has an IR quiver gauge theory description of
\be
 \begin{tikzpicture}[x=.6cm,y=.6cm]
\node at (-3.3,0) {$SU(2)_0$};
\node at (0,0) {$SU(4)_{0}$};
\node at (3.3,0) {$SU(2)_0$};
\node at (0,1.7) {$1\mathbf{AS}$};
\draw[ligne, black](-2,0)--(-1.3,0);
\draw[ligne, black](1.3,0)--(2,0);
\draw[ligne, black](0,0.5)--(0,1.3);
\end{tikzpicture}
\ee

The combined fiber diagram (CFD)~\cite{Apruzzi:2019vpe,Apruzzi:2019opn,Apruzzi:2019enx,Apruzzi:2019kgb} is read off by contracting the top $(-1)$-curve from (5.9) of \cite{Closset:2021lwy}. Hence we can see that the UV flavor symmetry is $G_F=SU(2)^6$.

\subsubsection{Class (58)}
\be
\begin{cases}
& z_1 z_2+z_3^2+z_4^2+z_5^3=0 \\
& z_1 z_3+z_2^2 z_4+z_4 z_5^2=0\,.
\end{cases}
\ee
\be
\begin{array}{||c|c|c||c|c|c||c|c|c||}
\hline
 r & d_H & G_F & \h r & \h d_H & f & b_3 & a&c\\
\hline
 5 & 21 & \color{blue}{SU(4)\times SU(2)^2\times U(1)} & 15& 10 & 6 & 0 & \frac{581}{24}&\frac{74}{3}\\
\hline
\end{array}
\ee
The resolution sequence is
\be
\ba
&(z_1,z_2,z_3,z_4,z_5;\delta_1)\cr
&(z_1,z_3,z_4,\delta_1;\delta_4)\cr
&(z_1,\delta_1,\delta_4;\delta_5)\cr
&(z_1,\delta_1;\delta_3)\cr
&(z_3,\delta_1;\delta_2)
\ea
\ee
The resolved equation is
\be
\begin{cases}
&z_1 z_2 \delta_3+z_5^3 \delta_1\delta_2\delta_3+z_4^2\delta_4+z_3^2\delta_2^2\delta_4=0\\
&z_1z_3+z_2^2z_4\delta_1+z_4z_5^2\delta_1=0\,.
\end{cases}
\ee 
The first equation for the divisor $\delta_3=0$ is 
\be
\ba
\delta_4(z_4^2+z_3^2\delta_2^2)=0
\ea
\ee
Thus we have two irreducible components with corresponding first equation:
\be
\ba
&S_3^{(1)}:\ z_4+i z_3 \delta_2=0\cr
&S_3^{(2)}:\ z_4-i z_3 \delta_2=0\,.
\ea
\ee

On the other hand, the exceptional divisor $\delta_1=0$ is an empty set, which means the theory has 5d rank $r=5$.

The triple intersection numbers are:
\be
\label{Class-58-int}
\includegraphics[height=7cm]{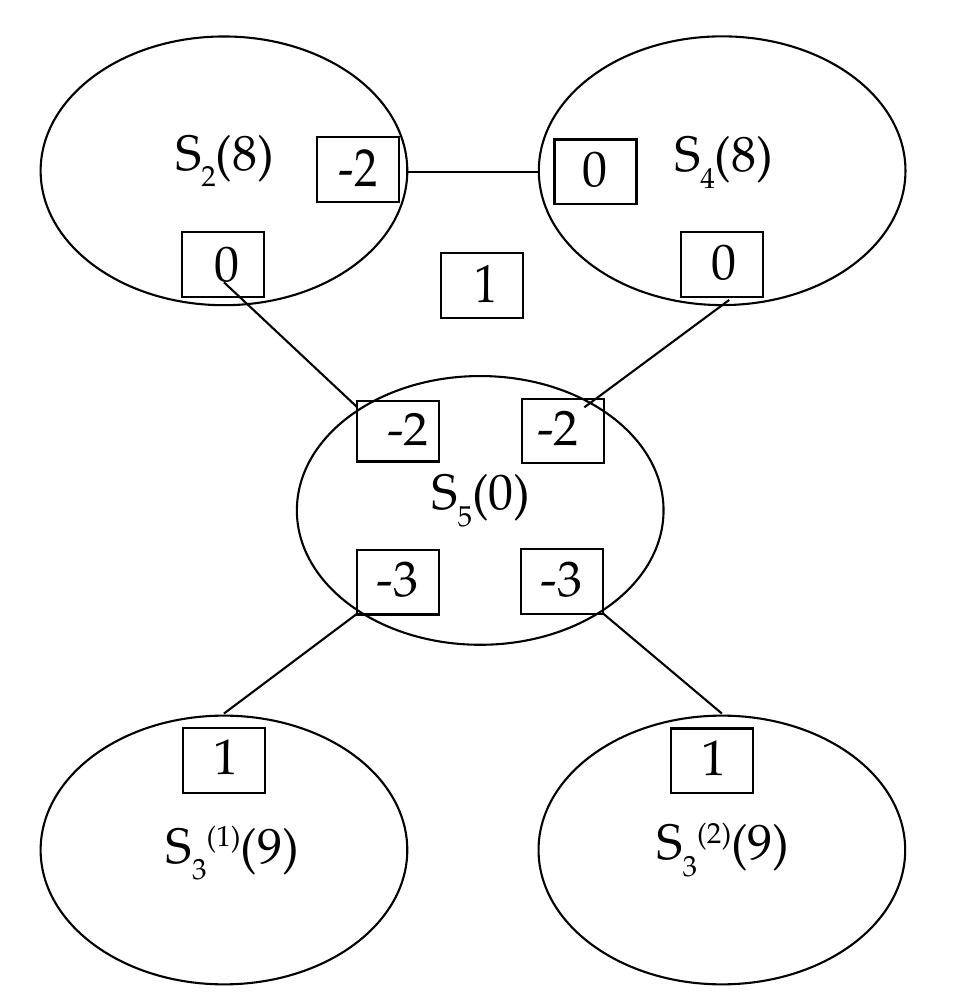}
\ee

Now let us discuss the field theory interpretations. We first perform two flops by blowing up $S_3^{(1)}$ and $S_3^{(2)}$, while blowing down $S_5$ twice. The resulting configuration is
\be
\includegraphics[height=7cm]{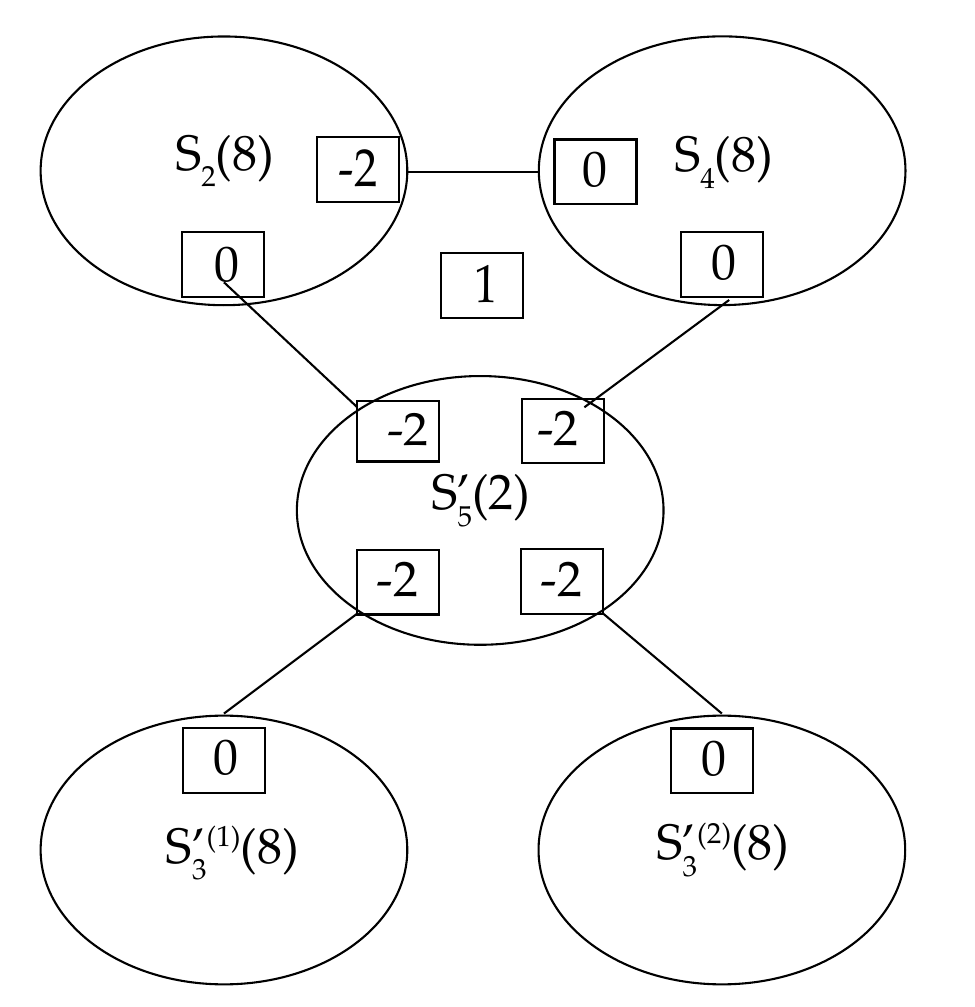}
\ee
Note that now $S_3^{\prime (1)}$ and $S_3^{\prime (2)}$ and Hirzebruch surfaces $\mb{F}_1$. Now we speculate the configuration of curves on $S'_5$, which takes the form of type $\mbf{A}_5+\mbf{A}_2$ $gdP_7$\cite{derenthal2014singular}:
\be
\includegraphics[height=4cm]{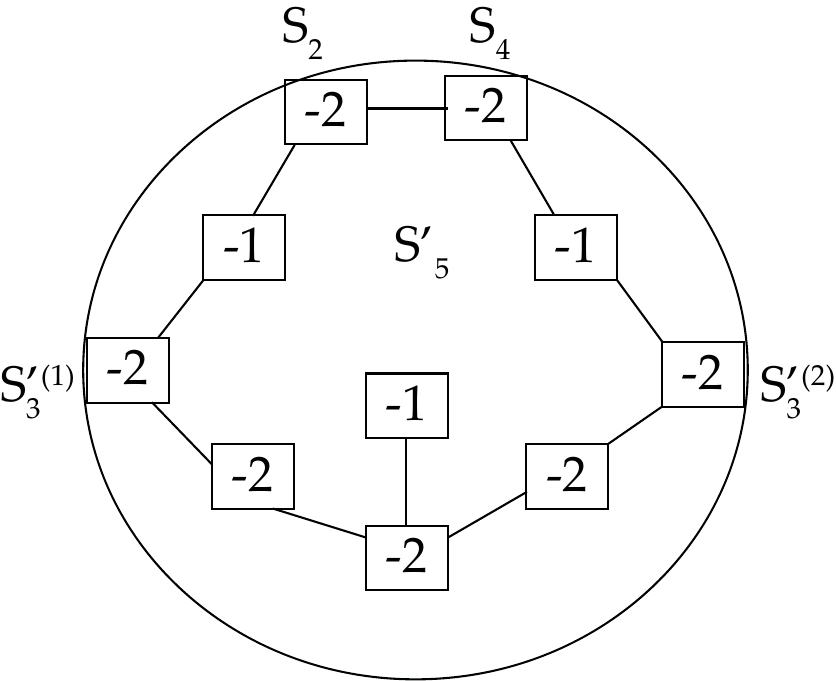}
\ee
The geometry does not have a well-defined ruling structure and hence we do not observe an IR quiver gauge theory description.

For the non-abelian flavor symmetry factors, let us go back to (\ref{Class-58-int}) and plot the curves on $S_5$:
\be
\includegraphics[height=4cm]{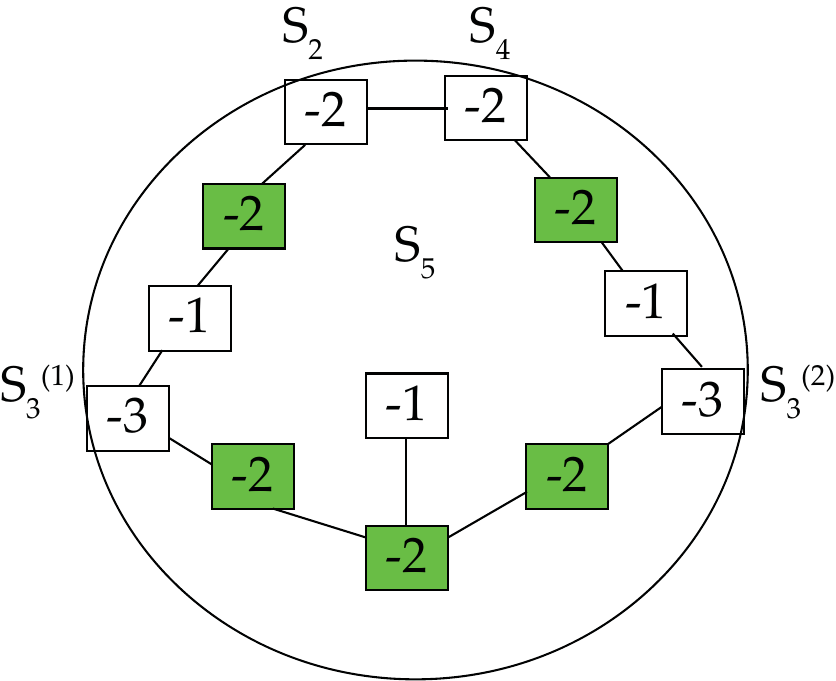}
\ee
The $(-2)$-curves in green on $S_5$ lead to $G_F=SU(4)\times SU(2)^2$ flavor symmetry factors.

\subsection{Rank-6 examples}
\subsubsection{Class (16)}
\be
\begin{cases}
&z_1 z_3+z_4 z_5=0\\
&z_1^2+z_2^2+z_3^{12}+z_4^3+z_5^4=0\,.
\end{cases}
\ee
\be
\begin{array}{||c|c|c||c|c|c||c|c|c||}
\hline
 r & d_H & G_F & \h r & \h d_H & f & b_3 & a&c\\
\hline
 6 & 50 & E_7\times U(1) & 42 & 14 & 8 & 0 & \frac{1183}{12}&\frac{595}{6}\\
\hline
\end{array}
\ee
We have the 5d magnetic quiver:
\be
 \begin{tikzpicture}[x=.5cm,y=.5cm]
\draw[ligne, black](2,0)--(-3,0);
\draw[ligne, black](0,0)--(0,1);
\draw[ligne, black](1,2)--(0,1);
\draw[ligne, black](-1,2)--(0,1);
\node[] at (-5,0) {$\MQfive= $};
\node[bd] at (2,0) [label=below:{{\scriptsize$4$}}] {};
\node[bd] at (1,0) [label=below:{{\scriptsize$8$}}] {};
\node[bd] at (0,0) [label=below:{{\scriptsize$12$}}] {};
\node[bd] at (-1,0) [label=below:{{\scriptsize$9$}}] {};
\node[bd] at (-2,0) [label=below:{{\scriptsize$6$}}] {};
\node[bd] at (-3,0) [label=below:{{\scriptsize$3$}}] {};
\node[bd] at (0,1) [label=right:{{\scriptsize$7$}}] {};
\node[bd] at (1,2) [label=right:{{\scriptsize$1$}}] {};
\node[bd] at (-1,2) [label=right:{{\scriptsize$1$}}] {};
\end{tikzpicture} 
\ee
We can get the hasse diagram with six layers:
\be
\begin{tikzpicture}
\node (1) [hasse] at (0,3) {};
\node (2) [hasse] at (-1.5,1.5) {};
\node (3) [hasse] at (0,1.5) {};
\node (4) [hasse] at (1.5,1.5) {};
\node (5) [hasse] at (-1.5,0) {};
\node (6) [hasse] at (0,0) {};
\node (7) [hasse] at (1.5,0) {};
\node (8) [hasse] at (-0.75,-1.5) {};
\node (9) [hasse] at (0.75,-1.5) {};
\node (10) [hasse] at (0,-3) {};
\node (11) [hasse] at (0,-4.5) {};
\node (12) [hasse] at (0,-6) {};
\draw (1) edge [] node[label=right:$\mathfrak{d}_5$] {} (2);
\draw (1) edge [] node[label=right:$\mathfrak{e}_6$] {} (3);
\draw (1) edge [] node[label=right:$\mathfrak{e}_6$] {} (4);
\draw (4) edge [] node[label=above right:$\mathfrak{a}_1$] {} (6);
\draw (4) edge [] node[label=right:$\mathfrak{e}_6$] {} (7);
\draw (3) edge [] node[label=right:$\mathfrak{a}_1$] {} (6);
\draw (3) edge [] node[label=right:$\mathfrak{e}_6$] {} (7);
\draw (2) edge [] node[label=right:$\mathfrak{d}_6$] {} (5);
\draw (2) edge [] node[label=right:$\mathfrak{a}_5$] {} (6);
\draw (5) edge [] node[label=right:$\mathfrak{a}_1$] {} (8);
\draw (6) edge [] node[label=right:$\mathfrak{a}_5$] {} (8);
\draw (6) edge [] node[label=right:$\mathfrak{e}_6$] {} (9);
\draw (7) edge [] node[label=right:$\mathfrak{a}_1$] {} (9);
\draw (8) edge [] node[label=right:$\mathfrak{d}_5$] {} (10);
\draw (9) edge [] node[label=right:$\mathfrak{a}_1$] {} (10);
\draw (10) edge [] node[label=right:$\mathfrak{d}_6$] {} (11);
\draw (11) edge [] node[label=right:$\mathfrak{e}_7$] {} (12);
\end{tikzpicture}
\ee
The resolution sequence is
\be
\ba
&(z_1,z_2,z_3,z_4,z_5;\delta_1)\cr
&(z_1^{(2)},z_2^{(2)},z_4^{(1)},z_5^{(1)},\delta_1^{(1)};\delta_3)\cr
&(z_1^{(3)},z_2^{(3)},z_4^{(2)},z_5^{(1)},\delta_3^{(1)};\delta_5)\cr
&(z_1,z_2,z_4,\delta_1;\delta_2)\cr
&(z_1,z_2,z_4,\delta_3;\delta_4)
\ea
\ee
The resolved equation is
\be
\begin{cases}
&z_1 z_3+z_4 z_5=0\\
&z_1^2+z_2^2+z_5^4\delta_1^2\delta_3^2+z_4^3\delta_1\delta_2^2\delta_4+z_3^{12}\delta_1^{10}\delta_2^8\delta_3^6\delta_4^4=0\,.
\end{cases}
\ee
We can deduce the following intersection diagram:
\be
\includegraphics[height=6cm]{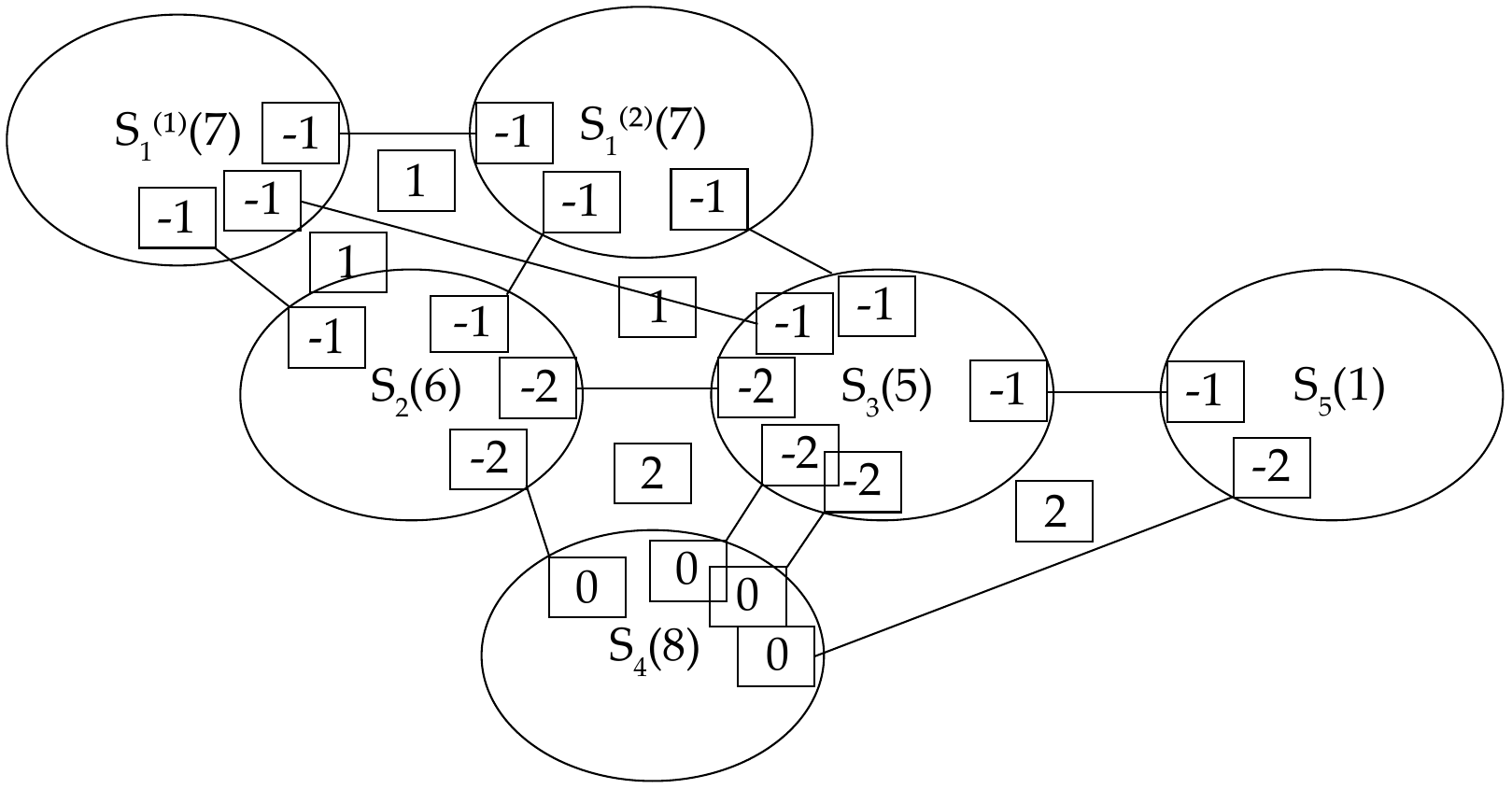}
\ee
\subsubsection{Class (31)}
\be
\begin{cases}
& z_1 z_3+z_4^2+z_5^4=0 \\
& z_1^2+z_2^2+z_3^2 z_4+z_5^5=0\,.
\end{cases}
\ee
\be
\begin{array}{||c|c|c||c|c|c||c|c|c||}
\hline
 r & d_H & G_F & \h r & \h d_H & f & b_3 & a&c\\
\hline
 6 & 26 & \color{blue}{SU(2)^4\times U(1)} & 21 & 11 & 5 & 0 & \frac{1049}{24}&\frac{265}{6}\\
\hline
\end{array}
\ee

The resolution sequence is
\be
\ba
&(z_1,z_2,z_3,z_4,z_5;\delta_1)\cr
&(z_1,z_2,z_3,z_4,\delta_1;\delta_3)\cr
&(z_1^{(2)},z_2^{(2)},z_4^{(1)},\delta_1^{(1)},\delta_3^{(1)};\delta_5)\cr
&(z_1,z_2,z_4,\delta_1;\delta_2)\cr
&(z_1,\delta_2;\delta_4)
\ea
\ee
The resolved equation is
\be
\begin{cases}
&z_1 z_3+z_4^2 \delta_2+z_5^4 \delta_1^2 \delta_2=0\\
&z_2^2+z_3^2 z_4 \delta_1 \delta_3^2+z_5^5 \delta_1^3 \delta_2 \delta_3 \delta_4+z_1^2 \delta_4^2=0\,.
\end{cases}
\ee 
We can deduce the following intersection diagram:
\be
\includegraphics[height=8cm]{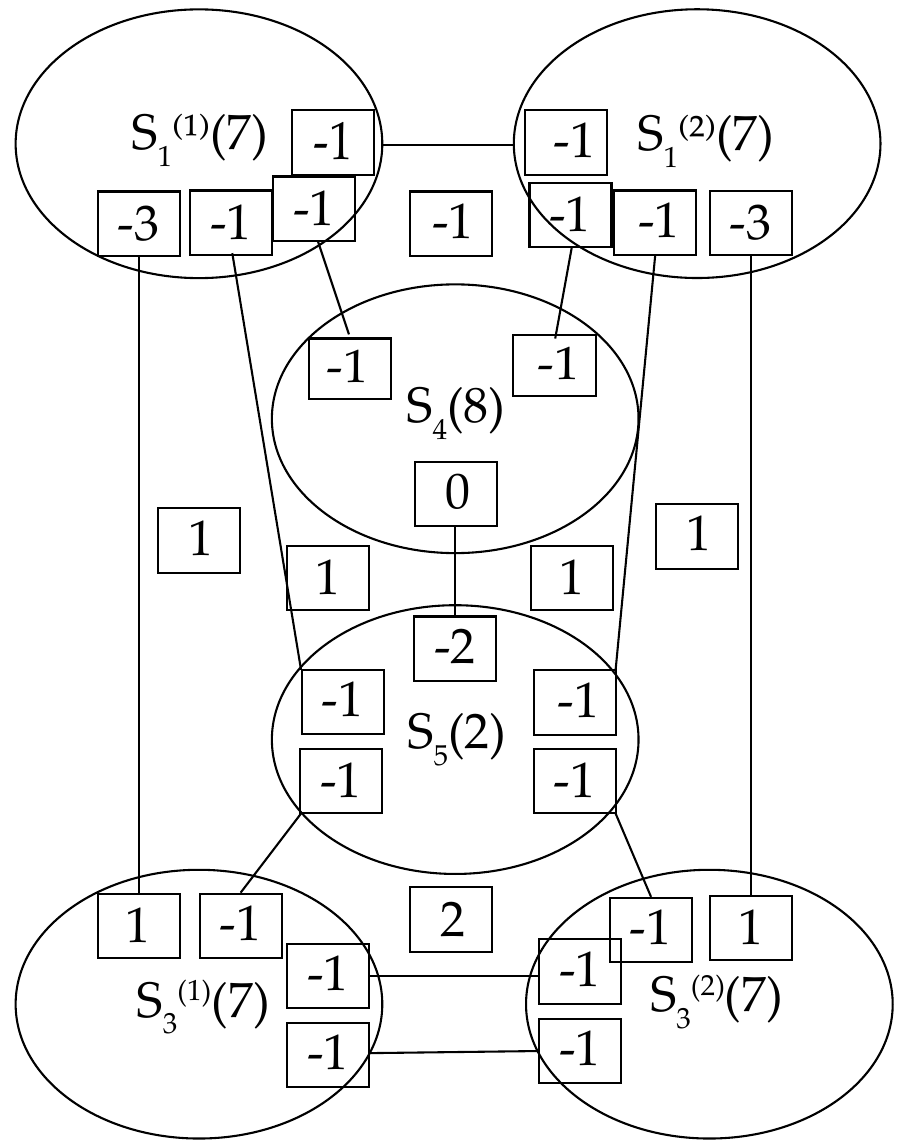}
\ee

Note that these curves coincide: $S_4\cdot S_1^{(1)}=S_4\cdot S_1^{(2)}=S_1^{(1)}\cdot S_1^{(2)}$.

The theory does not have a 5d IR non-abelian gauge theory description. 

For the $G_{F,nA}$, we focus on $S_5$, which is the only compact divisor with possibly extra $(-2)$-curves. We take $S_5$ to be gdP$_7$ of type $\mbf{A}_3+2\mbf{A}_1$, whose $(-2)$-curves in the standard Picard group generators are
\be
\ba
&S_5\cdot S_4:\ e_1-e_2\cr
&\textrm{Flavor\ curves}:\ h-e_1-e_4-e_5\ ,\ h-e_1-e_6-e_7\ ,\ e_4-e_5,\ e_6-e_7\,.
\ea
\ee
Hence the additional flavor curves give rise to $G_{F,nA}=SU(2)^4$.

\subsubsection{Class (57)}
\be
\begin{cases}
& z_1 z_2+z_3^2+z_4^2+z_5^4=0 \\
& z_1 z_3+z_2^3+z_2 z_5^3=0\,.
\end{cases}
\ee
\be
\begin{array}{||c|c|c||c|c|c||c|c|c||}
\hline
 r & d_H & G_F & \h r & \h d_H & f & b_3 & a&c\\
\hline
 6 & 24 & \color{blue}{SU(4)\times SU(2)\times U(1)} & 19 & 11 & 5 & 0 & \frac{893}{24}&\frac{113}{3}\\
\hline
\end{array}
\ee

The resolution sequence is
\be
\ba
&(z_1,z_2,z_3,z_4,z_5;\delta_1)\cr
&(z_1,z_2,z_3,z_4,\delta_1;\delta_4)\cr
&(z_1^{(2)},z_3^{(1)},z_4^{(1)},\delta_1^{(1)},\delta_4^{(1)};\delta_7)\cr
&(z_1,\delta_1;\delta_3)\cr
&(z_1,\delta_4;\delta_6)\cr
&(z_3,\delta_1;\delta_2)\cr
&(z_3,\delta_4;\delta_5)
\ea
\ee
The resolved equation is
\be
\begin{cases}
&z_4^2+z_5^4 \delta_1^2 \delta_2^2 \delta_3^2+z_3^2 \delta_2^2 \delta_5^2 +z_1 z_2 \delta_3 \delta_6=0\\
&z_1 z_3+z_2 z_5^3 \delta_1^2 \delta_2 \delta_3 \delta_4+z_2^3 \delta_1 \delta_4^2 \delta_5 \delta_6=0\,.
\end{cases}
\ee 
We can deduce the following intersection diagram:
\be
\includegraphics[height=7cm]{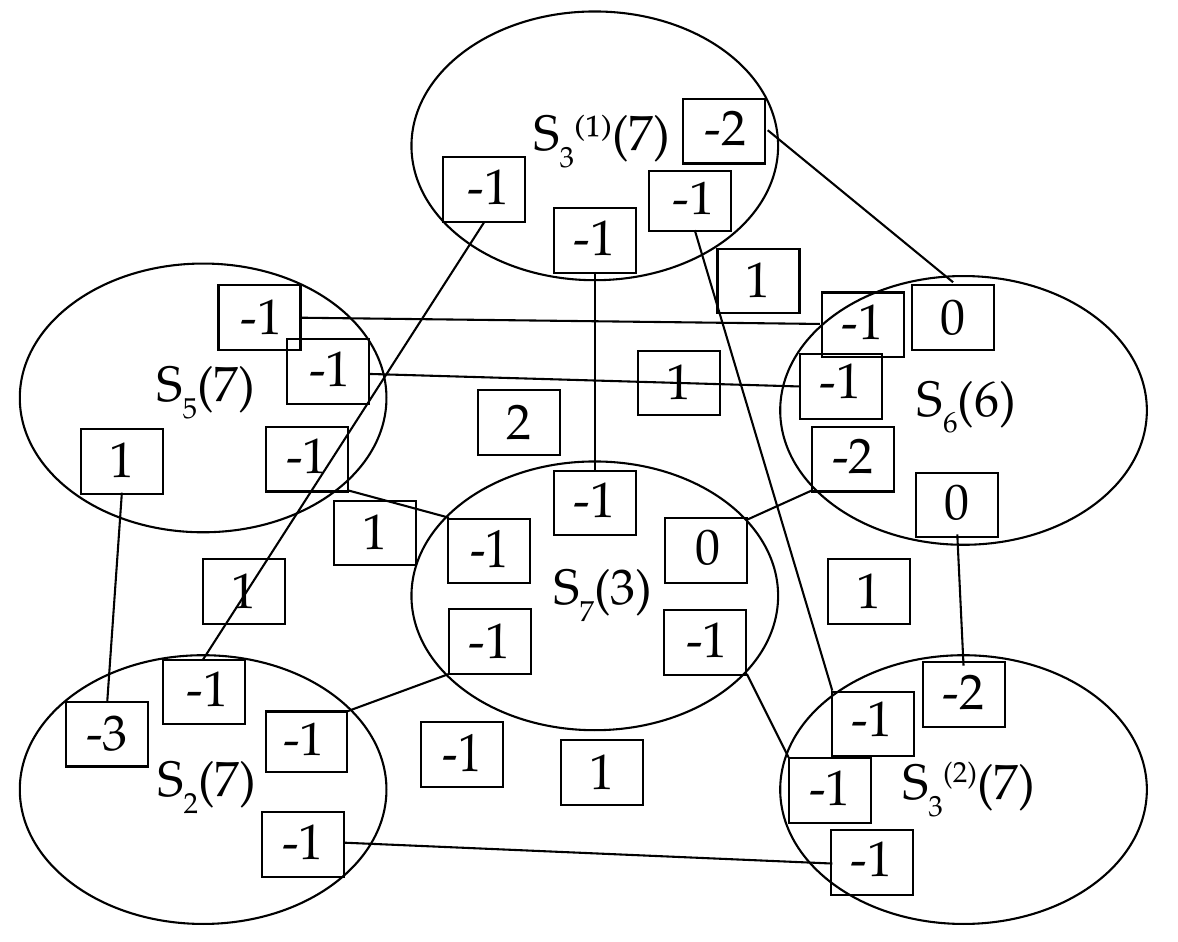}
\ee
Note that in the diagram we have

$S_7\cdot S_3^{(1)}\cdot S_3^{(2)}=S_7\cdot S_3^{(1)}\cdot S_6=S_7\cdot S_3^{(1)}\cdot S_2=S_7\cdot S_3^{(2)}\cdot S_6=S_7\cdot S_2\cdot S_5=1$,

$S_3^{(1)}\cdot S_2\cdot S_3^{(2)}=-1$,\quad$S_7\cdot S_6\cdot S_5=2$.

For the $G_{F,nA}$, we first write down the additional flavor curves on $S_7$. It can be taken as a type $\mbf{A}_3$ gdP$_6$, with the following $(-2)$-curves:
\be
\text{Flavor\ curves}:\ e_2-e_4\ ,\ h-e_1-e_2-e_5\ ,\ e_1-e_3\,.
\ee
Besides that, for $S_6$, it can be taken as a type $2\mbf{A}_1$ gdP$_3$ with an additional $(-2)$-curve on it. In total, they give rise to $G_{F,nA}=SU(4)\times SU(2)$. 

\subsubsection{Class (120)}
\be
\begin{cases}
& z_1 z_2+z_4^2+z_5^5=0 \\
& z_1 z_4+z_3^2+z_2^n=0\,.
\end{cases}
\ee
$3\leqslant n\leqslant 13$

We have a smooth resolution only for $n=4$.
\be
\begin{array}{||c||c||c|c|c||c|c|c||c|c|c||}
\hline
 &r & d_H & G_F & \h r & \h d_H & f & b_3 & a&c\\
\hline
 n=4&6 & 32 & \color{blue}{SU(6)} & 27& 11 & 5 & 0 & \frac{1577}{24}&\frac{397}{6}\\
\hline
\end{array}
\ee
The resolution sequence is
\be
\ba
&(z_1,z_2,z_3,z_4,z_5;\delta_1)\cr
&(z_1,z_2,z_3,z_4,\delta_1;\delta_3)\cr
&(z_1,z_3,z_4,\delta_1;\delta_2)\cr
&(z_1,z_3,z_4,\delta_3;\delta_5)\cr
&(z_1^{(2)},z_3^{(1)},\delta_2^{(1)},\delta_5^{(1)};\delta_6)\cr
&(z_1,\delta_2;\delta_4)
\ea
\ee
The resolved equation is
\be
\begin{cases}
&z_1 z_2+z_5^5 \delta_1^3 \delta_2^2 \delta_3 \delta_4+z_4^2 \delta_2 \delta_5=0\\
&z_3^2+z_1 z_4 \delta_4+z_2^4\delta_1^2\delta_3^4\delta_5^2=0\,.
\end{cases}
\ee

We can deduce the following intersection diagram:
\be
\includegraphics[height=5cm]{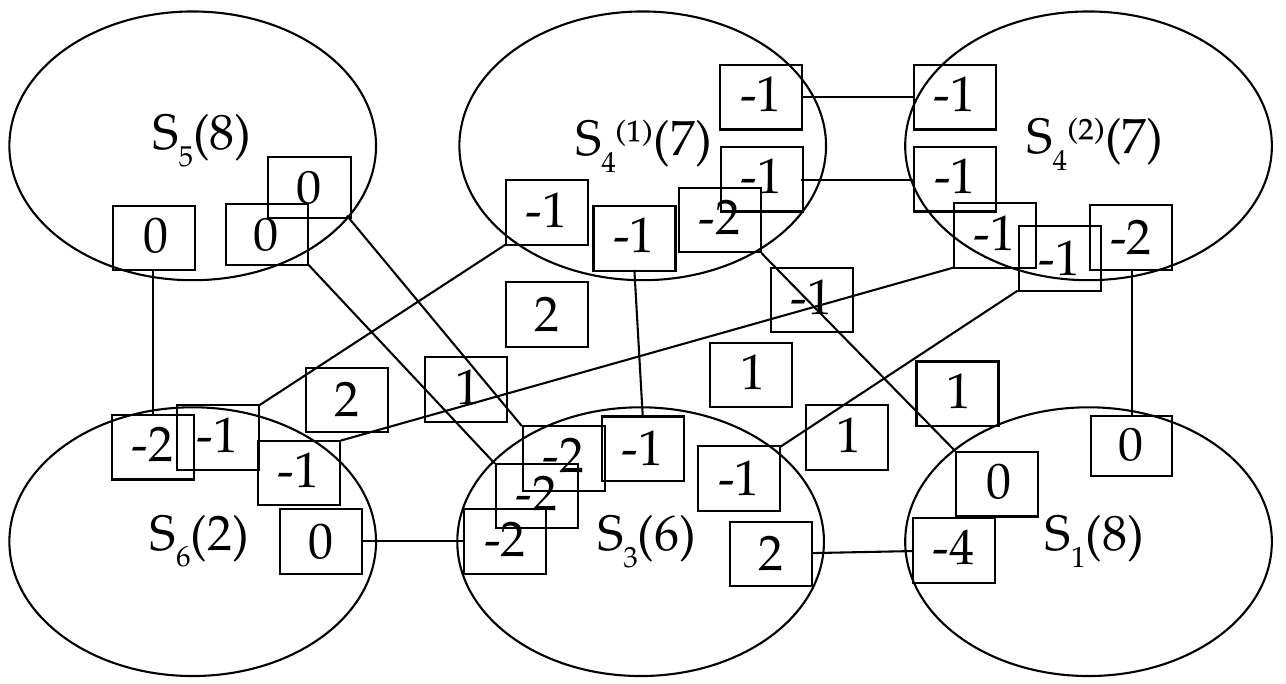}
\ee

Note that in the diagram we have

$S_1\cdot S_3\cdot S_4^{(i)}=S_3\cdot S_6\cdot S_4^{(i)}=1$,
$S_3\cdot S_5 \cdot S_6=S_6 \cdot S_4^{(1)} \cdot S_4^{(2)}=2$,$S_3\cdot S_4^{(1)} \cdot S_4^{(2)}=-1$\quad $(i=1,\dots,2)$\,.

For $G_{F,nA}$, we see that only $S_6$ can contain additional flavor curves. It can be chosen as a type $\mbf{E}_6$ gdP$_7$, with the $(-2)$-curves
\be
\ba
&S_6\cdot S_5:\ e_1-e_2\cr
&\text{Flavor\ curves}:\ e_6-e_7\ ,\ h-e_1-e_2-e_6\ ,\ e_2-e_3\ ,\ e_3-e_4\ ,\ e_4-e_5\,.
\ea
\ee
One can see $G_{F,nA}=SU(6)$.

\subsubsection{Class (256)}
\be
\begin{cases}
& z_1 z_2+z_3 z_4=0 \\
& z_1 z_4+z_3^2+z_2^n z_4+z_5^5=0\,.
\end{cases}
\ee
$2\leqslant n\leqslant 6$

We have a smooth resolution and $r\leq10$ only for $n=2$. 
\be
\begin{array}{||c||c|c|c||c|c|c||c|c|c||}
\hline
 &r & d_H & G_F & \h r & \h d_H & f & b_3 & a&c\\
\hline
 n=2&6 & 27 & \color{blue}{SU(3)\times SU(2)^3} & 22 & 9 & 5 & 0&\frac{1127}{24} & \frac{569}{12}\\
\hline
\end{array}
\ee


The resolution sequence is
\be
\ba
&(z_1,z_2,z_3,z_4,z_5;\delta_1)\cr
&(z_1,z_2,z_3,z_4,\delta_1;\delta_4)\cr
&(z_1^{(3)},z_3^{(2)},z_4^{(1)},\delta_1^{(1)},\delta_4^{(1)};\delta_7)\cr
&(z_1,z_3,z_4,\delta_1;\delta_3)\cr
&(z_1,z_3,\delta_1;\delta_2)\cr
&(z_1,z_3,\delta_4;\delta_6)\cr
&(z_1,\delta_3;\delta_5)
\ea
\ee
The resolved equation is
\be
\begin{cases}
&z_1 z_2+z_3 z_4 \delta_3=0\\
&z_1 z_4 \delta_5+z_5^5 \delta_1^3 \delta_2^2 \delta_3 \delta_4 \delta_5+z_3^2 \delta_2 \delta_6+z_2^2 z_4 \delta_1 \delta_4^2 \delta_6=0\,.
\end{cases}
\ee 
We can deduce the following intersection diagram:
\be
\includegraphics[height=7cm]{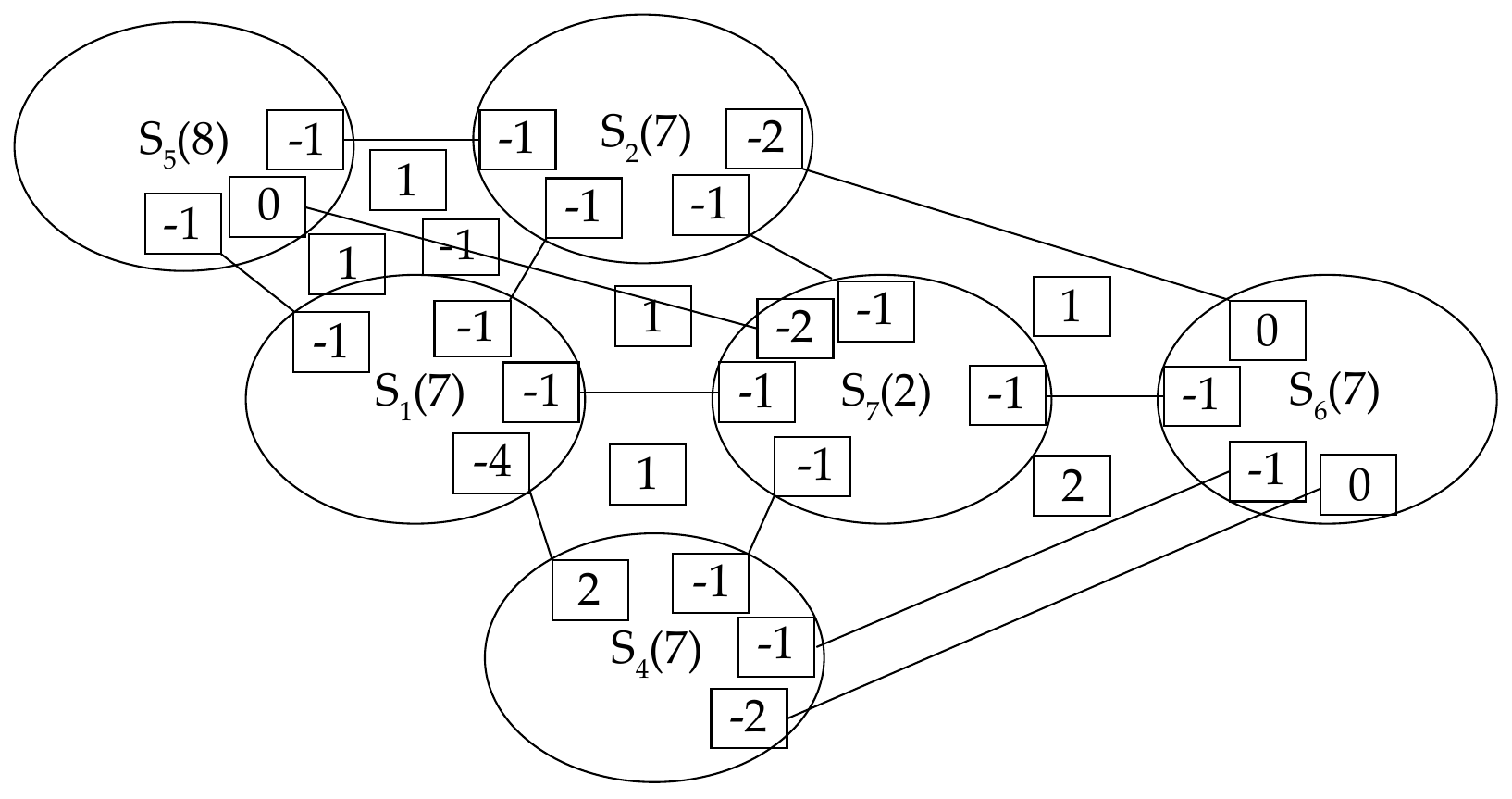}
\ee

For the $G_{F,nA}$, additional flavor curves can live on $S_7$ and $S_6$. For $S_7$, it can be chosen as a gdP$_7$ of type $\mbf{A}_2+3\mbf{A}_1$. The $(-2)$-curves on $S_7$ are
\be
\ba
&S_7\cdot S_5:\ h-e_2-e_6-e_7\cr
&\text{Flavor\ curves}:\ e_2-e_6\ ,\ e_6-e_7\ ,\ e_1-e_5\ ,\ e_3-e_4\,,
\ea
\ee
which contribute to $SU(3)\times SU(2)^2$ flavor symmetry factors. For $S_6$, it can be chosen as a gdP$_2$ with an additional $(-2)$-curve. In total, we have $G_{F,nA}=SU(3)\times SU(2)^3$.
\subsubsection{Class (266)}
\be
\begin{cases}
& z_1 z_2+z_3 z_4=0 \\
& z_1 z_5+z_3^2+z_4^4+z_5^3+z_2^{12}=0\,.
\end{cases}
\ee
\be
\begin{array}{||c|c|c||c|c|c||c|c|c||}
\hline
 r & d_H & G_F & \h r & \h d_H & f & b_3 & a&c\\
\hline
 6 & 51 & E_7\times U(1) & 43 & 14 & 8 & 0 & \frac{611}{6}&\frac{1229}{12}\\
\hline
\end{array}
\ee
We have the 5d magnetic quiver:
\be
 \begin{tikzpicture}[x=.5cm,y=.5cm]
\draw[ligne, black](3,0)--(-3,0);
\draw[ligne, black](0,0)--(0,1);
\draw[ligne, black](1,0)--(1,1);
\node[] at (-5,0) {$\MQfive= $};
\node[bd] at (3,0) [label=below:{{\scriptsize$1$}}] {};
\node[bd] at (2,0) [label=below:{{\scriptsize$5$}}] {};
\node[bd] at (1,0) [label=below:{{\scriptsize$9$}}] {};
\node[bd] at (0,0) [label=below:{{\scriptsize$12$}}] {};
\node[bd] at (-1,0) [label=below:{{\scriptsize$9$}}] {};
\node[bd] at (-2,0) [label=below:{{\scriptsize$6$}}] {};
\node[bd] at (-3,0) [label=below:{{\scriptsize$3$}}] {};
\node[bd] at (0,1) [label=right:{{\scriptsize$6$}}] {};
\node[bd] at (1,1) [label=right:{{\scriptsize$1$}}] {};
\end{tikzpicture} 
\ee
The resolution sequence is
\be
\ba
&(z_1,z_2,z_3,z_4,z_5;\delta_1)\cr
&(z_1^{(3)},z_3^{(2)},z_4^{(1)},z_5^{(1)},\delta_1^{(1)};\delta_4)\cr
&(z_1^{(4)},z_3^{(3)},z_4^{(1)},z_5^{(2)},\delta_4^{(1)};\delta_6)\cr
&(z_1,z_3,z_5,\delta_1;\delta_3)\cr
&(z_1,z_3,z_5,\delta_4;\delta_5)\cr
&(z_1,z_3,\delta_1;\delta_2)
\ea
\ee
The resolved equation is
\be
\begin{cases}
&z_1 z_2+z_3 z_4=0\\
&z_1 z_5+z_3^2\delta_2+z_4^4\delta_1^2\delta_2\delta_4^2+z_5^3\delta_1\delta_3^2\delta_5+z_2^{12}\delta_1^{10}\delta_2^9\delta_3^8\delta_4^6\delta_5^4 =0\,.
\end{cases}
\ee
We read off $G_F=E_7\times U(1)$ from the Hasse diagram associated to $\MQfive$.

We can deduce the following intersection diagram:
\be
\includegraphics[height=6cm]{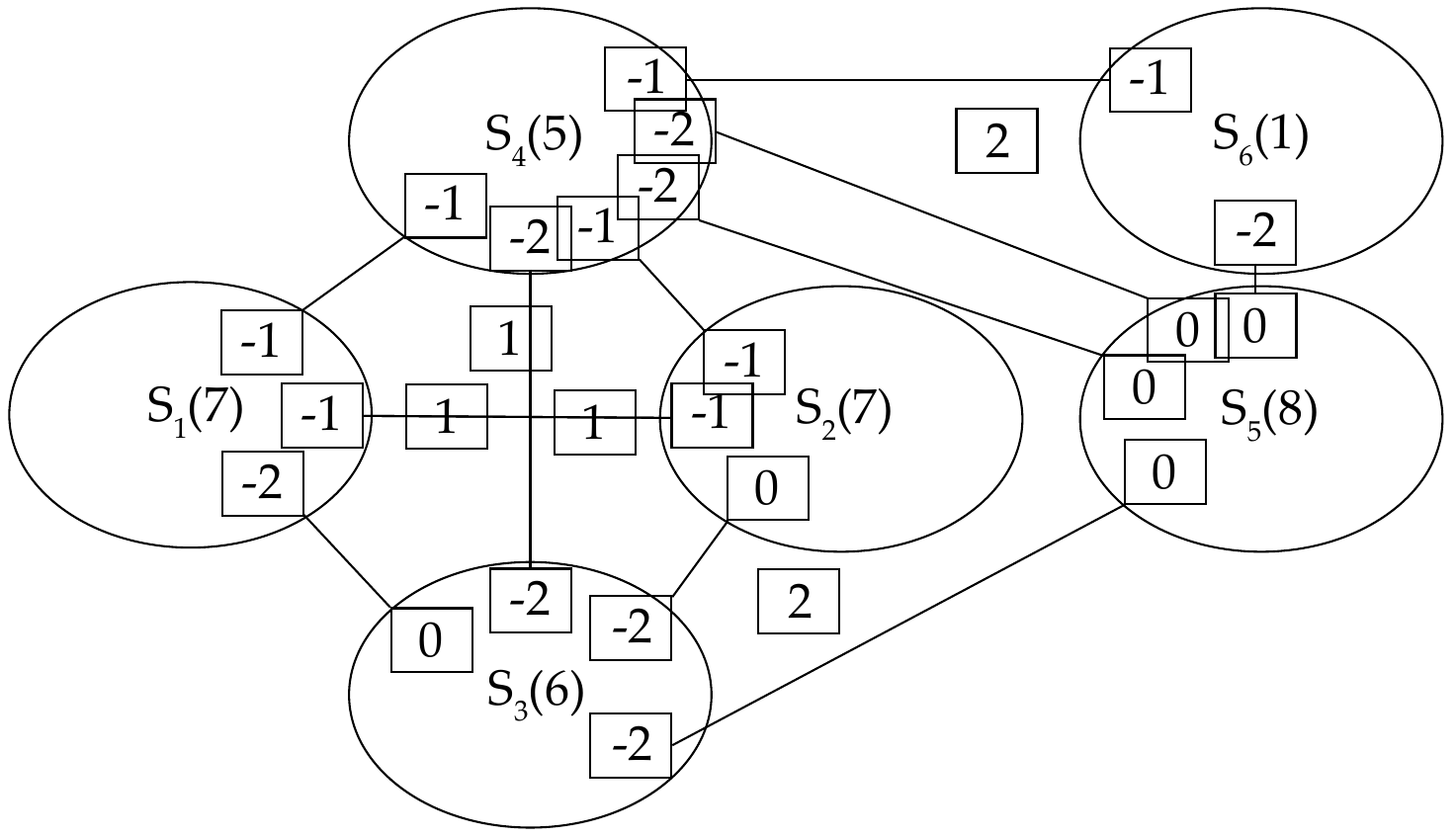}
\ee
\subsubsection{Class (268)}
\be
\begin{cases}
& z_1 z_2+z_3 z_4=0 \\
& z_1 z_5+z_3^2+z_4^3+z_5^4+z_2^{12}=0\,.
\end{cases}
\ee
\be
\begin{array}{||c|c|c||c|c|c||c|c|c||}
\hline
 r & d_H & G_F & \h r & \h d_H & f & b_3 & a&c\\
\hline
 6 & 52 & E_7\times U(1) & 44 & 14 & 8 & 0 & \frac{1261}{12}&\frac{317}{3}\\
\hline
\end{array}
\ee
We have the 5d magnetic quiver:
\be
 \begin{tikzpicture}[x=.5cm,y=.5cm]
\draw[ligne, black](3,0)--(-3,0);
\draw[ligne, black](0,0)--(0,1);
\draw[ligne, black](-1,0)--(-1,1);
\node[] at (-5,0) {$\MQfive= $};
\node[bd] at (3,0) [label=below:{{\scriptsize$1$}}] {};
\node[bd] at (2,0) [label=below:{{\scriptsize$4$}}] {};
\node[bd] at (1,0) [label=below:{{\scriptsize$7$}}] {};
\node[bd] at (0,0) [label=below:{{\scriptsize$10$}}] {};
\node[bd] at (-1,0) [label=below:{{\scriptsize$12$}}] {};
\node[bd] at (-2,0) [label=below:{{\scriptsize$8$}}] {};
\node[bd] at (-3,0) [label=below:{{\scriptsize$4$}}] {};
\node[bd] at (0,1) [label=right:{{\scriptsize$1$}}] {};
\node[bd] at (-1,1) [label=right:{{\scriptsize$6$}}] {};
\end{tikzpicture} 
\ee
We can get the hasse diagram with six layers:
\be
\begin{tikzpicture}
\node (1) [hasse] at (0,3) {};
\node (2) [hasse] at (0,2) {};
\node (3) [hasse] at (-1,1) {};
\node (4) [hasse] at (1,1) {};
\node (5) [hasse] at (-1,0) {};
\node (6) [hasse] at (1,0) {};
\node (7) [hasse] at (0,-1) {};
\node (8) [hasse] at (0,-2) {};
\node (9) [hasse] at (0,-3) {};
\draw (1) edge [] node[label=right:$\mathfrak{d}_5$] {} (2);
\draw (2) edge [] node[label=right:$\mathfrak{d}_7$] {} (3);
\draw (2) edge [] node[label=right:$\mathfrak{a}_6$] {} (4);
\draw (3) edge [] node[label= right:$\mathfrak{a}_1$] {} (5);
\draw (4) edge [] node[label=right:$\mathfrak{a}_6$] {} (5);
\draw (4) edge [] node[label=right:$\mathfrak{e}_6$] {} (6);
\draw (5) edge [] node[label=right:$\mathfrak{d}_5$] {} (7);
\draw (6) edge [] node[label=right:$\mathfrak{a}_2$] {} (7);
\draw (7) edge [] node[label=right:$\mathfrak{d}_6$] {} (8);
\draw (8) edge [] node[label=right:$\mathfrak{e}_7$] {} (9);
\end{tikzpicture}
\ee
The resolution sequence is
\be
\ba
&(z_1,z_2,z_3,z_4,z_5;\delta_1)\cr
&(z_1^{(3)},z_3^{(2)},z_4^{(1)},z_5^{(1)},\delta_1^{(1)};\delta_4)\cr
&(z_1^{(5)},z_3^{(3)},z_4^{(2)},z_5^{(1)},\delta_4^{(1)};\delta_6)\cr
&(z_1^{(2)},z_3^{(1)},z_4^{(1)},\delta_1^{(1)};\delta_3)\cr
&(z_1^{(2)},z_3^{(1)},z_4^{(1)},\delta_4^{(1)};\delta_5)\cr
&(z_1,z_3,\delta_1;\delta_2)
\ea
\ee
The resolved equation is
\be
\begin{cases}
&z_1 z_2+z_3 z_4=0\\
&z_1 z_5+z_3^2\delta_2+z_5^4\delta_1^2\delta_2\delta_4^2+z_4^3\delta_1\delta_3^2\delta_5+z_2^{12}\delta_1^{10}\delta_2^9\delta_3^8\delta_4^6\delta_5^4 =0\,.
\end{cases}
\ee
We can deduce the following intersection diagram:
\be
\includegraphics[height=6cm]{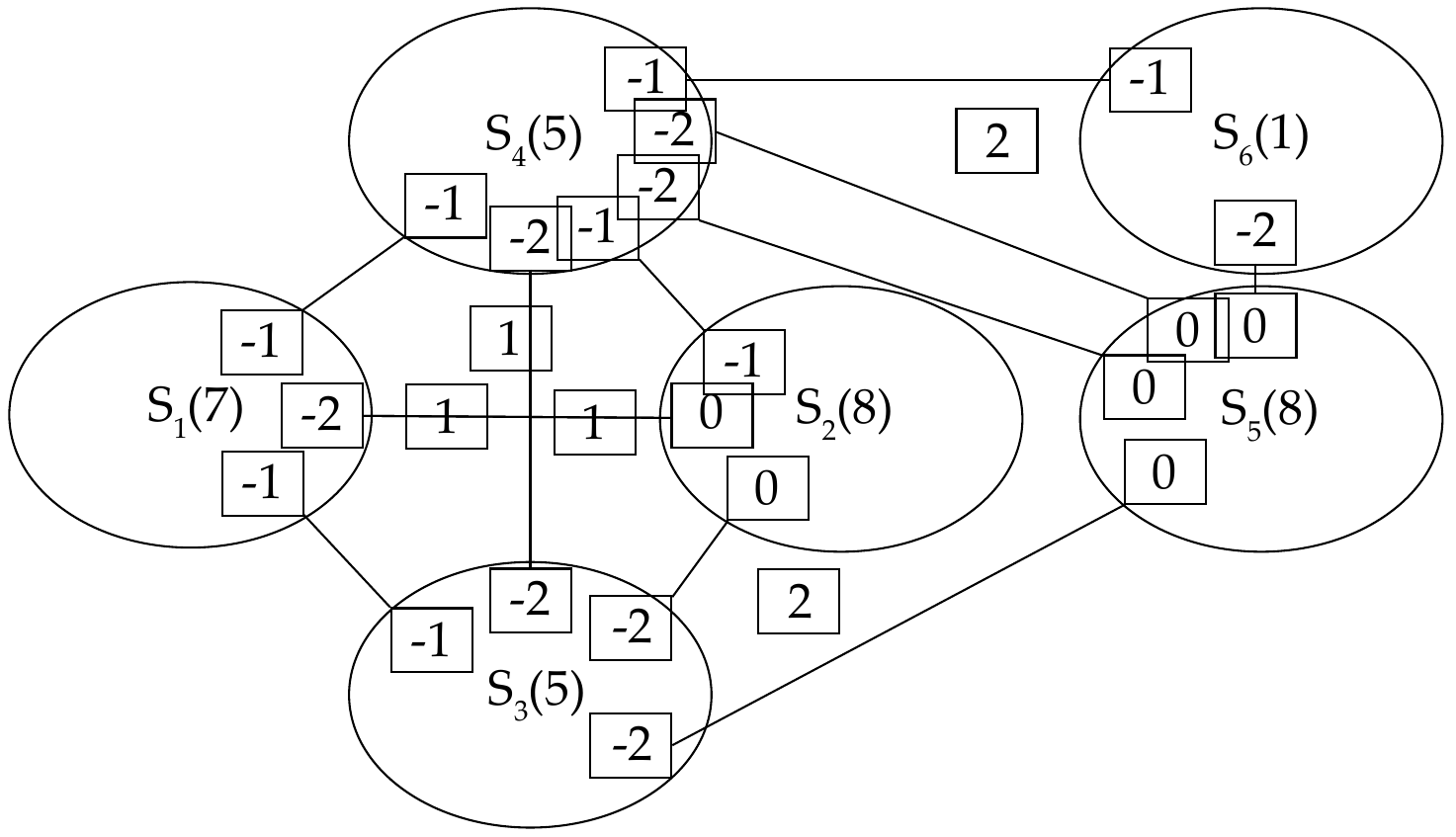}
\ee
\subsubsection{Class (289)}
\be
\begin{cases}
& z_1 z_2+z_3 z_4=0 \\
& z_1 z_5+z_2 z_3^2+z_4^3+z_5^3+z_2^9=0\,.
\end{cases}
\ee
\be
\begin{array}{||c|c|c||c|c|c||c|c|c||}
\hline
 r & d_H & G_F & \h r & \h d_H & f & b_3 & a&c\\
\hline
 6 & 36 & E_6\times U(1)^2 & 28 & 14 & 8 & 0 & \frac{637}{12}&\frac{161}{3}\\
\hline
\end{array}
\ee
We have the 5d magnetic quiver:
\be
 \begin{tikzpicture}[x=.5cm,y=.5cm]
\draw[ligne, black](2,0)--(-3,0);
\draw[ligne, black](0,0)--(0,2);
\draw[ligne, black](-1,0)--(-1,1);
\node[] at (-5,0) {$\MQfive= $};
\node[bd] at (2,0) [label=below:{{\scriptsize$3$}}] {};
\node[bd] at (1,0) [label=below:{{\scriptsize$6$}}] {};
\node[bd] at (0,0) [label=below:{{\scriptsize$9$}}] {};
\node[bd] at (-1,0) [label=below:{{\scriptsize$7$}}] {};
\node[bd] at (-2,0) [label=below:{{\scriptsize$4$}}] {};
\node[bd] at (-3,0) [label=below:{{\scriptsize$1$}}] {};
\node[bd] at (0,1) [label=right:{{\scriptsize$5$}}] {};
\node[bd] at (0,2) [label=right:{{\scriptsize$1$}}] {};
\node[bd] at (-1,1) [label=right:{{\scriptsize$1$}}] {};
\end{tikzpicture} 
\ee
From the Hasse diagram (which we would not explicitly draw here), we read off the flavor symmetry $G_F=E_6\times U(1)^2$.

The resolution sequence is
\be
\ba
&(z_1,z_2,z_3,z_4,z_5;\delta_1)\cr
&(z_1^{(2)},z_3^{(1)},z_4^{(1)},z_5^{(1)},\delta_1^{(1)};\delta_6)\cr
&(z_1^{(3)},z_3^{(2)},z_4^{(1)},z_5^{(1)},\delta_6^{(1)};\delta_8)\cr
&(z_1,z_3,\delta_1;\delta_5)\cr
&(z_1,z_3,\delta_6;\delta_7)\cr
&(z_1,z_4,\delta_1;\delta_4)\cr
&(z_1,\delta_1;\delta_3)\cr
&(z_5,\delta_1;\delta_2)
\ea
\ee
The resolved equation is
\be
\begin{cases}
&z_1 z_2 \delta_3+z_3 z_4=0\\
&z_1 z_5+z_5^3 \delta_1 \delta_2^3 \delta_6+ z_4^3 \delta_1 \delta_4^3 \delta_6+z_2 z_3^2 \delta_1 \delta_5^2 \delta_7+z_2^9 \delta_1^7 \delta_2^6 \delta_3^6 \delta_4^6 \delta_5^6 \delta_6^4 \delta_7^3=0\,.
\end{cases}
\ee

We can deduce the following intersection diagram:
\be
\includegraphics[height=7cm]{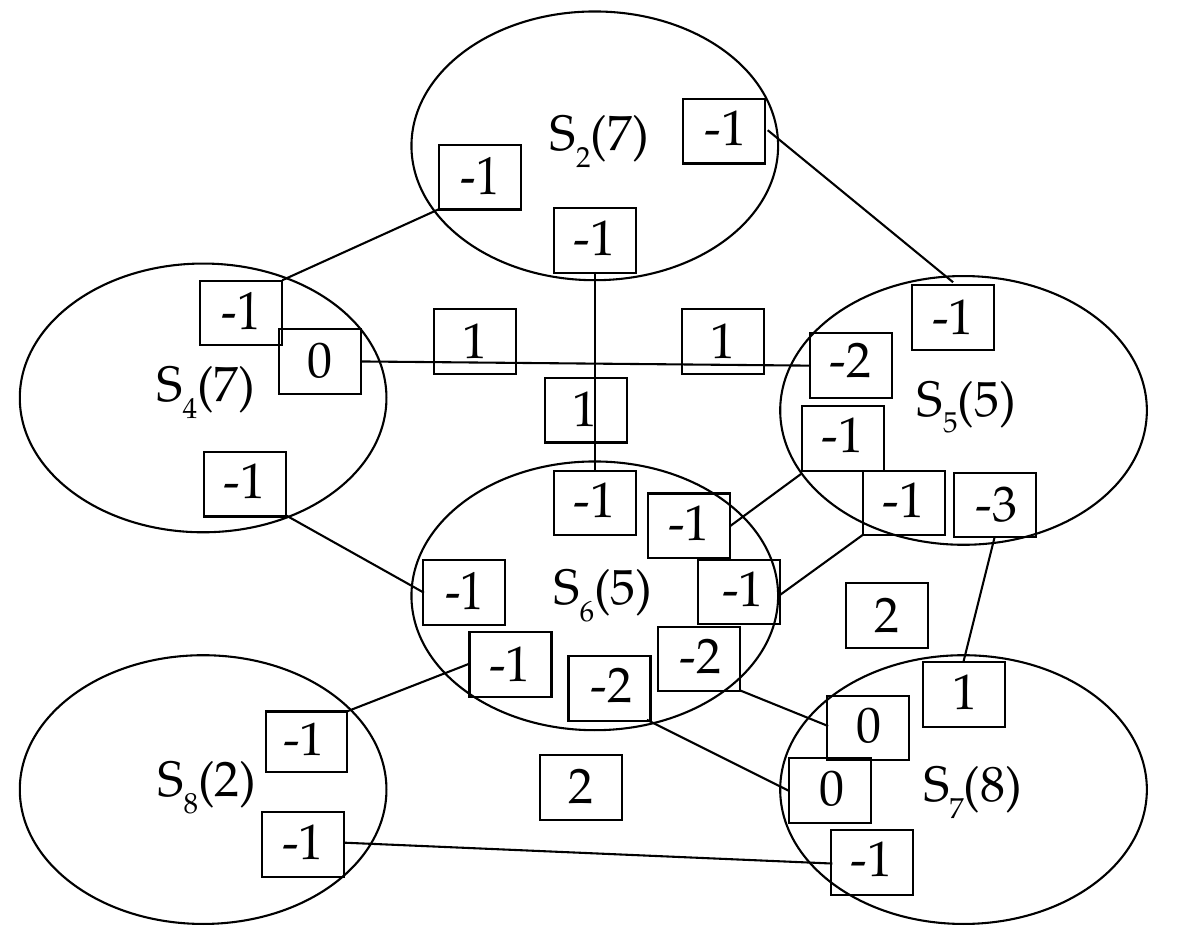}
\ee
Note that in the diagram we have

$S_6\cdot S_5\cdot S_4=S_6\cdot S_5\cdot S_2=S_6\cdot S_4\cdot S_2=S_6\cdot S_4\cdot S_2=1$,

$S_6\cdot S_8 \cdot S_7=S_6 \cdot S_5 \cdot S_7=2$.

\subsection{Rank-7 examples}
\subsubsection{Class (12)}
\be
\begin{cases}
&z_1 z_3+z_4 z_5=0\\
&z_1^2+z_2^2+z_3^{10}+z_3 z_4^3+z_3 z_5^3=0\,.
\end{cases}
\ee
\be
\begin{array}{||c|c|c||c|c|c||c|c|c||}
\hline
 r & d_H & G_F & \h r & \h d_H & f & b_3 & a&c\\
\hline
 7 & 41 & E_6\times U(1)^3 & 32 & 16 & 9 & 0 & \frac{197}{3}&\frac{199}{3}\\
\hline
\end{array}
\ee
We have the 5d magnetic quiver:
\be
 \begin{tikzpicture}[x=.5cm,y=.5cm]
\draw[ligne, black](3,0)--(-3,0);
\draw[ligne, black](0,0)--(0,1);
\draw[ligne, black](1,2)--(0,1);
\draw[ligne, black](-1,2)--(0,1);
\node[] at (-5,0) {$\MQfive= $};
\node[bd] at (3,0) [label=below:{{\scriptsize$1$}}] {};
\node[bd] at (2,0) [label=below:{{\scriptsize$4$}}] {};
\node[bd] at (1,0) [label=below:{{\scriptsize$7$}}] {};
\node[bd] at (0,0) [label=below:{{\scriptsize$10$}}] {};
\node[bd] at (-1,0) [label=below:{{\scriptsize$7$}}] {};
\node[bd] at (-2,0) [label=below:{{\scriptsize$4$}}] {};
\node[bd] at (-3,0) [label=below:{{\scriptsize$1$}}] {};
\node[bd] at (0,1) [label=right:{{\scriptsize$6$}}] {};
\node[bd] at (1,2) [label=right:{{\scriptsize$1$}}] {};
\node[bd] at (-1,2) [label=right:{{\scriptsize$1$}}] {};
\end{tikzpicture} 
\ee
The resolution sequence is
\be
\ba
&(z_1,z_2,z_3,z_4,z_5;\delta_1)\cr
&(z_1^{(2)},z_2^{(2)},z_4^{(1)},z_5^{(1)},\delta_1^{(1)};\delta_5)\cr
&(z_1^{(2)},z_2^{(2)},z_4^{(1)},z_5^{(1)},\delta_5^{(1)};\delta_6)\cr
&(z_1,z_2,z_4,\delta_1;\delta_4)\cr
&(z_1,z_2,z_5,\delta_1;\delta_3)\cr
&(z_1,z_2,\delta_1;\delta_2)
\ea
\ee
The resolved equation is
\be
\begin{cases}
&z_1 z_3 \delta_2+z_4 z_5=0\\
&z_1^2+z_2^2+z_3 z_5^3 \delta_1^2 \delta_3^3 \delta_5+z_3 z_4^3 \delta_1^2 \delta_4^3 \delta_5+z_3^{10} \delta_1^8 \delta_2^6 \delta_3^6 \delta_4^6 \delta_5^4=0\,.
\end{cases}
\ee 
From the Hasse diagram (which we would not explicitly draw here), we read off the flavor symmetry $G_F=E_6\times U(1)^3$

We can deduce the following intersection diagram:
\be
\includegraphics[height=7cm]{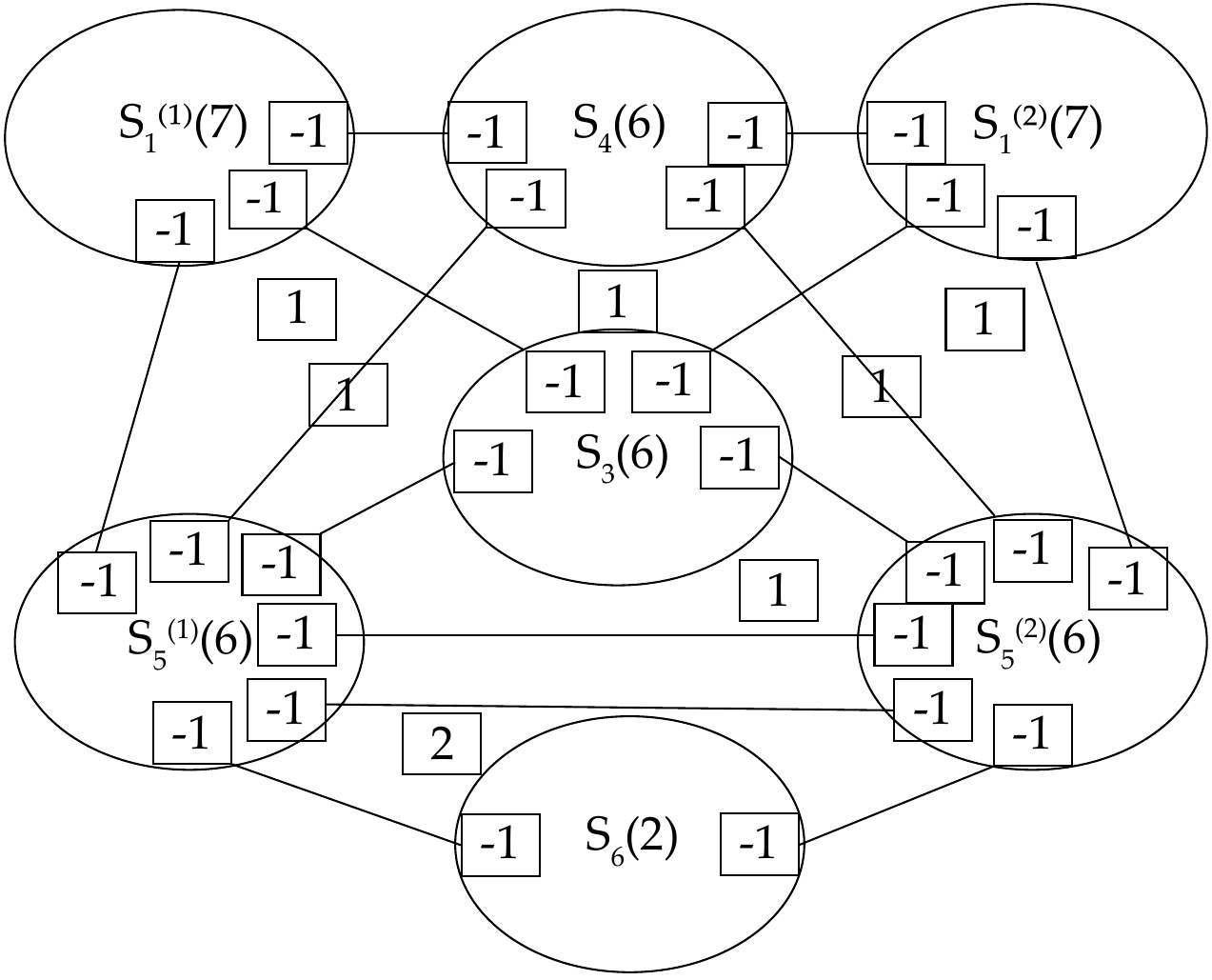}
\ee

Note that in the diagram we have

$S_1^{(1)}\cdot S_5^{(1)}\cdot S_4=S_1^{(1)}\cdot S_5^{(1)}\cdot S_3=S_1^{(2)}\cdot S_5^{(2)}\cdot S_4=S_1^{(2)}\cdot S_5^{(2)}\cdot S_3=S_5^{(1)}\cdot S_5^{(2)}\cdot S_3=S_5^{(1)}\cdot S_5^{(2)}\cdot S_4=1$,

$S_5^{(1)}\cdot S_5^{(2)}\cdot S_6=2$.
\subsection{Rank-9 examples}
\subsubsection{Class (37)}
\be
\begin{cases}
& z_1 z_3+z_2 z_4=0 \\
& z_1^2+z_2^2+z_4^3+z_3^3+z_5^n=0\,.
\end{cases}
\ee
$4\leqslant n \leqslant5$, here we take $n=4$.
\be
\begin{array}{||c|c|c||c|c|c||c|c|c||}
\hline
 r & d_H & G_F & \h r & \h d_H & f & b_3 & a&c\\
\hline
 9 & 25 & \color{blue}{SU(2)^3\times U(1)} & 21 & 13 & 4 & 0 & \frac{1231}{24}&\frac{311}{6}\\
\hline
\end{array}
\ee


The resolution sequence is
\be
\ba
&(z_1,z_2,z_3,z_4,z_5;\delta_1)\cr
&(z_1,z_2,z_3,z_4,\delta_1;\delta_3)\cr
&(z_1,z_2,\delta_1,\delta_3;\delta_5)\cr
&(z_1,z_2,\delta_1,\delta_5;\delta_6)\cr
&(z_1,z_2,\delta_1;\delta_2)\cr
&(\delta_1,\delta_2;\delta_4)
\ea
\ee
The resolved equation is
\be
\begin{cases}
&z_1 z_3+z_2 z_4=0\\
&z_1^2 \delta_2+z_2^2 \delta_2+z_5^4 \delta_1^2 \delta_2 \delta_4^2+z_3^3 \delta_1 \delta_3^2 \delta_5+z_4^{3} \delta_1 \delta_3^2 \delta_5=0\,.
\end{cases}
\ee 
We can deduce the following intersection diagram:
\be
\includegraphics[height=7cm]{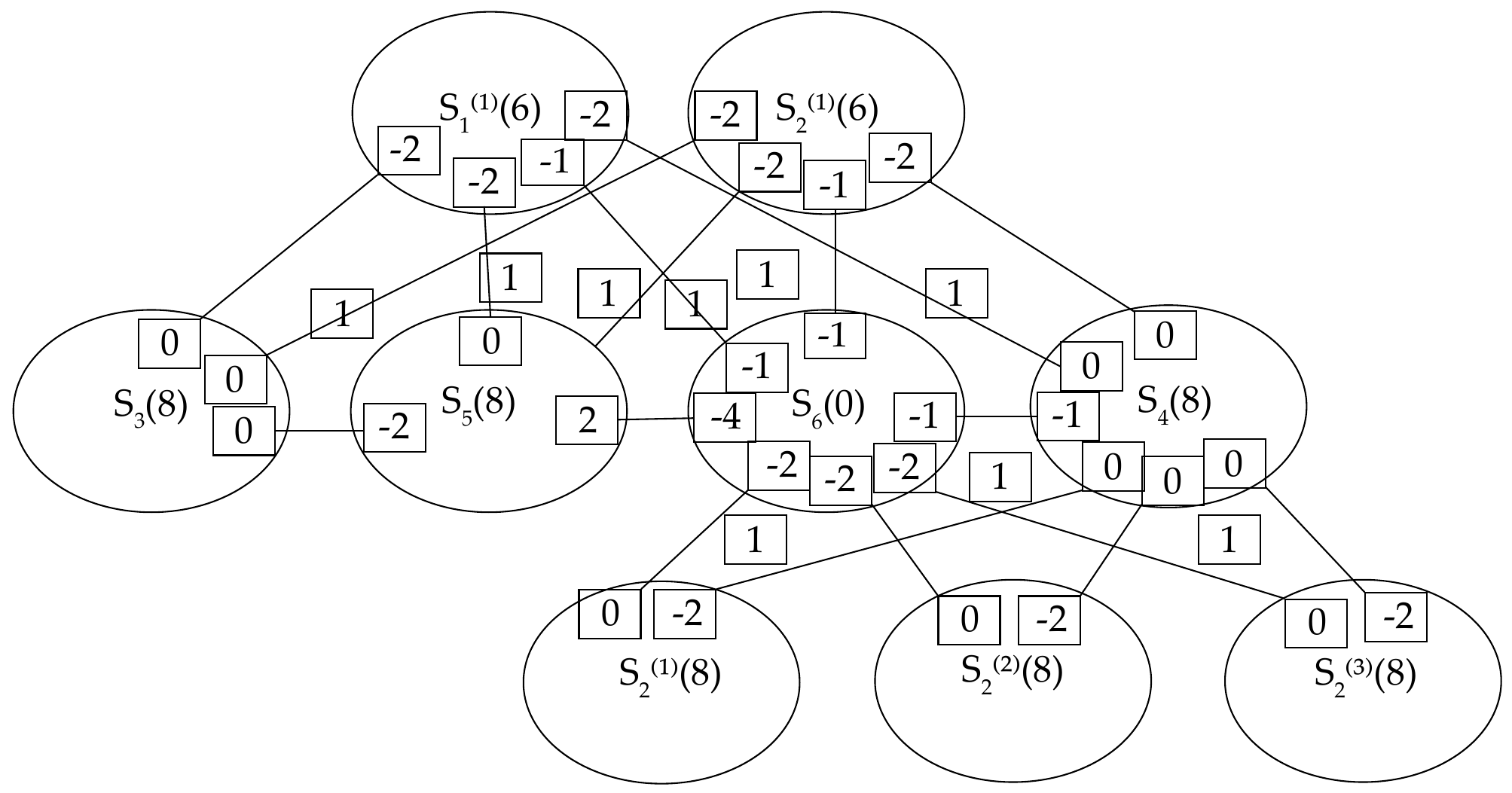}
\ee

For $G_{F,nA}$, we find that only $S_6$ can have additional flavor curves. We write down $(-2)$ and $(-4)$-curves on $S_6$:
\be
\ba
&S_6\cdot S_5:\ 2h-\sum_{i=1}^8 e_i\cr
&S_6\cdot S_2^{(1)}:\ e_1-e_2\cr
&S_6\cdot S_2^{(2)}:\ e_3-e_4\cr
&S_6\cdot S_2^{(3)}:\ e_5-e_6\cr
&\text{Flavor\ curves}:\ h-e_1-e_2-e_9\ ,\ h-e_3-e_4-e_9\ ,\ h-e_5-e_6-e_9\,.
\ea
\ee
Hence $G_{F,nA}=SU(2)^3$.

Note that this case is equivalent to the Class (186) with $n=3$:
\be
\begin{cases}
& z_1 z_2+z_3 z_4=0 \\
& z_1 z_3+z_2^n+z_4^3+z_5^4=0\,.
\end{cases}
\ee

\subsubsection{Class (51)}
\be
\begin{cases}
& z_2 z_3+z_4 z_5=0 \\
& z_1^2+z_2^2 z_5+z_3^4 z_5+z_2 z_4^2+z_5^4+z_3^2 z_4^2=0\,.
\end{cases}
\ee

\be
\begin{array}{||c|c|c||c|c|c||c|c|c||}
\hline
 r & d_H & G_F & \h r & \h d_H & f & b_3 & a&c\\
\hline
 9 & 33 & \color{blue}{SU(4)\times U(1)^2} & 28 & 14 & 5 & 0 & \frac{1027}{12}&\frac{517}{6}\\
\hline
\end{array}
\ee

The resolution sequence is
\be
\ba
&(z_1,z_2,z_3,z_4,z_5;\delta_1)\cr
&(z_1^{(2)},z_2^{(1)},z_4^{(1)},z_5^{(1)},\delta_1^{(1)};\delta_6)\cr
&(z_1,z_2,z_4,\delta_1;\delta_5)\cr
&(z_1^{(2)},z_2^{(1)},\delta_1^{(1)},\delta_5^{(1)},\delta_6^{(1)};\delta_{11})\cr
&(z_1,z_2,z_5,\delta_1;\delta_4)\cr
&(z_1,z_2,\delta_1,\delta_6;\delta_{10})\cr
&(z_1,z_2,\delta_1;\delta_3)\cr
&(z_1,\delta_1,\delta_5;\delta_9)\cr
&(z_1,\delta_1;\delta_2)\cr
&(z_2,\delta_6;\delta_8)\cr
&(\delta_1,\delta_2;\delta_7)\cr
\ea
\ee
The resolved equation is
\be
\begin{cases}
&z_2 z_3 \delta_3+z_4 z_5 \delta_6=0\\
&z_1^2 \delta_2+z_2 z_4^2 \delta_1 \delta_5^2 \delta_8 \delta_9+z_3^2 z_4^2 \delta_1^2 \delta_2 \delta_5^2 \delta_7^2 \delta_9^2+z_2^2 z_5 \delta_1 \delta_3 \delta_4^2 \delta_5 \delta_8^2 \delta_{10}\\
&+z_3^4 z_5 \delta_1^3 \delta_2^2 \delta_3 \delta_4^2 \delta_5 \delta_7^4 \delta_9^2 \delta_{10}+z_5^4 \delta_1^2 \delta_2 \delta_4^4 \delta_6^2 \delta_7^2 \delta_8^2 \delta_{10}^2=0\,.
\end{cases}
\ee 
We can deduce the following intersection diagram:
\be
\includegraphics[height=7cm]{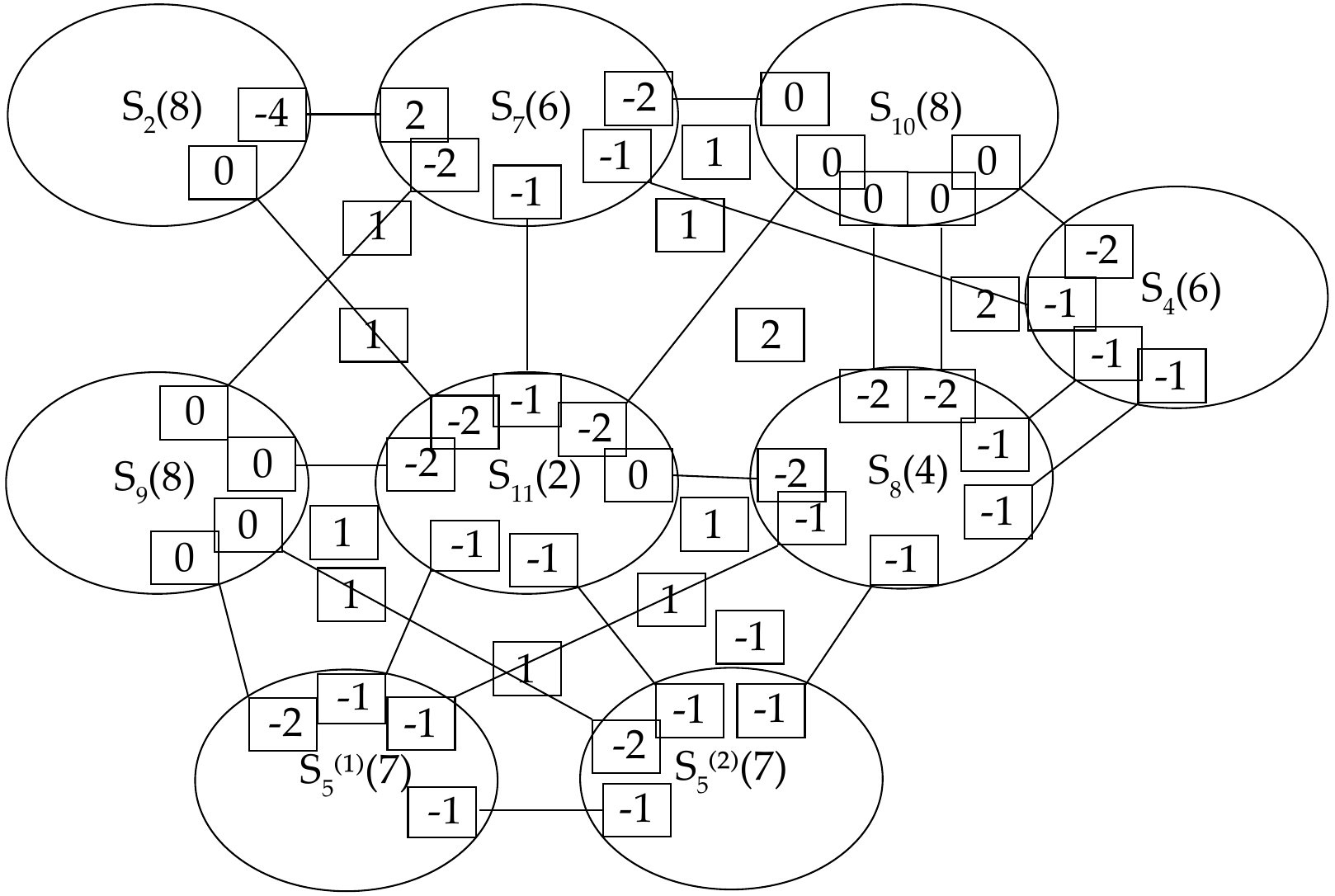}
\ee

Note that in the diagram we have

$S_5^{(1)}\cdot S_5^{(2)}\cdot S_{11}=S_5^{(1)}\cdot S_{11}\cdot S_9=S_5^{(1)}\cdot S_{11}\cdot S_8=S_5^{(2)}\cdot S_{11}\cdot S_9=S_5^{(2)}\cdot S_{11}\cdot S_8=S_{11}\cdot S_{10}\cdot S_8=S_{11}\cdot S_{10}\cdot S_7=S_{11}\cdot S_{9}\cdot S_7=S_{11}\cdot S_{2}\cdot S_7=S_{4}\cdot S_{10}\cdot S_7=1$,

$S_{11}\cdot S_{10}\cdot S_{8}=S_{4}\cdot S_{10}\cdot S_{8}=2$,\quad$S_{5}^{(1)}\cdot S_{5}^{(2)} \cdot S_{8}=-1$\,.

For $G_{F,nA}$, we look at $S_{11}$, which can be taken as a type $\mbf{A}_3+3\mbf{A}_1$ gdP$_7$, with $(-2)$-curves
\be
\ba
&S_{11}\cdot S_9:\ h-e_5-e_6-e_7\cr
&S_{11}\cdot S_2:\ e_6-e_7\cr
&S_{11}\cdot S_{10}:\ 2h-e_1-e_2-e_3-e_4-e_6-e_7\cr
&\text{Flavor curves}:\ e_1-e_2\ ,\ h-e_1-e_3-e_5\ ,\ e_3-e_4\,.
\ea
\ee
Hence $G_{F,nA}=SU(4)$.

\subsubsection{Class (153)}
\be
\begin{cases}
& z_1 z_2+z_4^3+z_5^4=0 \\
& z_1 z_4+z_3^2+z_2^n=0\,
\end{cases}
\ee

$3\leqslant n \leqslant6$

We have a smooth resolution only for $n=3$. 
\be
\begin{array}{||c||c|c|c||c|c|c||c|c|c||}
\hline
 &r & d_H & G_F & \h r & \h d_H & f & b_3 & a&c\\
\hline
 n=3&9 & 27 & \color{blue}{SU(2)^4} & 23 & 13 & 4 & 0 & \frac{1423}{24}&\frac{359}{6}\\
\hline
\end{array}
\ee



The resolution sequence is
\be
\ba
&(z_1,z_2,z_3,z_4,z_5;\delta_1)\cr
&(z_1,z_2,z_3,z_4,\delta_1;\delta_5)\cr
&(z_1^{(2)},z_3^{(1)},\delta_1^{(1)},\delta_5^{(1)};\delta_7)\cr
&(z_1^{(2)},z_3^{(1)},\delta_1^{(1)},\delta_7^{(1)};\delta_8)\cr
&(z_1,z_3,\delta_1;\delta_4)\cr
&(z_1,\delta_1,\delta_4;\delta_6)\cr
&(z_1,\delta_1;\delta_3)\cr
&(z_2,\delta_1;\delta_2)
\ea
\ee
The resolved equation is
\be
\begin{cases}
&z_1 z_2+z_5^4 \delta_1^2 \delta_2 \delta_3 \delta_4 \delta_6^2+z_4^3 \delta_1 \delta_5^2 \delta_7=0\\
&z_1 z_4 \delta_3+z_3^2 \delta_4+z_2^3 \delta_1 \delta_2^4 \delta_3 \delta_5^2 \delta_7=0\,.
\end{cases}
\ee 
We can deduce the following intersection diagram:
\be
\includegraphics[height=8cm]{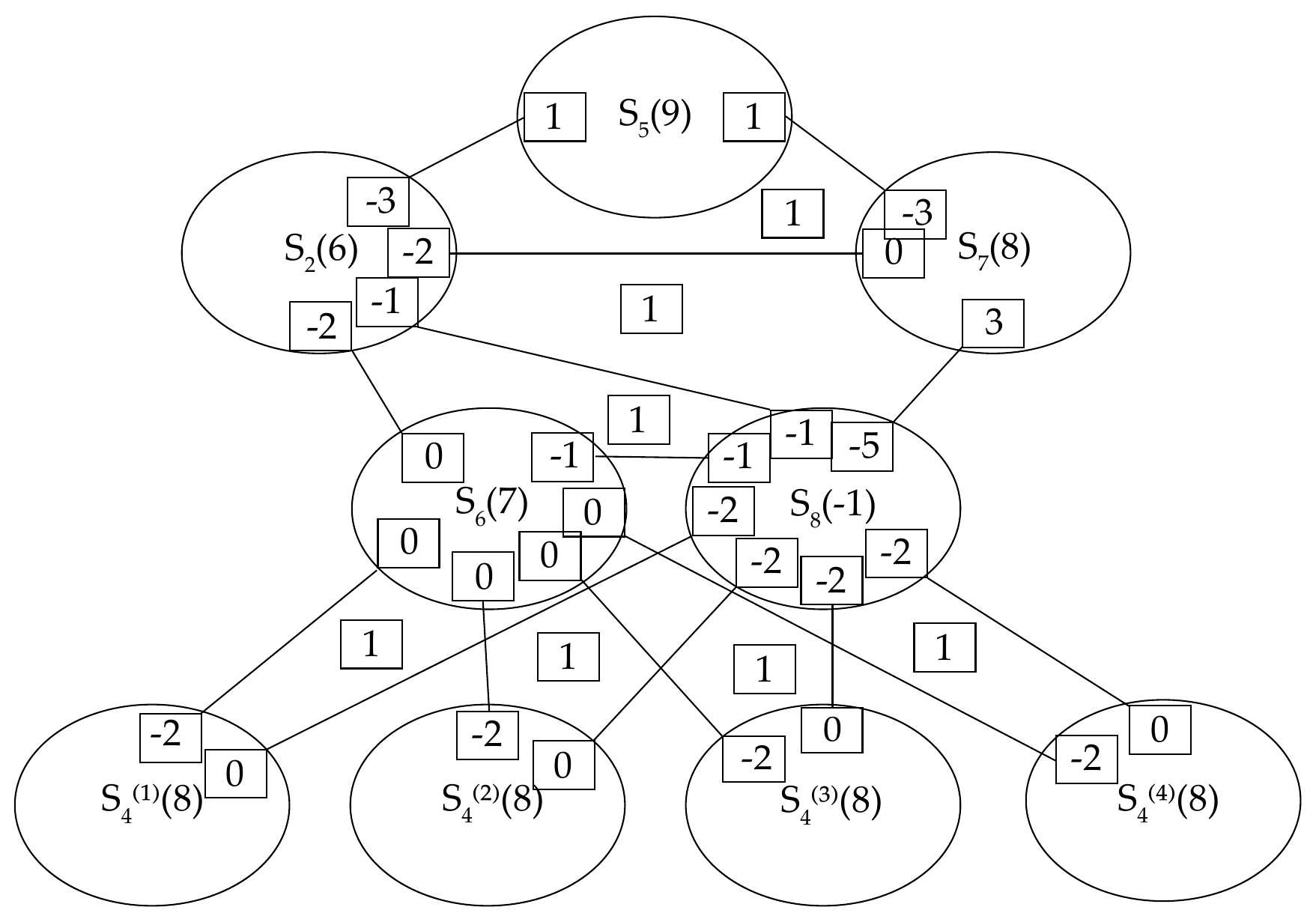}
\ee

For $G_{F,nA}$, the additional flavor curves come from $S_8$. We list the $(-2)$ and $(-5)$-curves on $S_8$:
\be
\ba
&S_8\cdot S_7:\ 2h-\sum_{i=1}^9 e_i\cr
&S_8\cdot S_4^{(1)}:\ e_1-e_2\cr
&S_8\cdot S_4^{(2)}:\ e_3-e_4\cr
&S_8\cdot S_4^{(3)}:\ e_5-e_6\cr
&S_8\cdot S_4^{(5)}:\ e_7-e_8\cr
&\text{Flavor\ curves}:\ h-e_1-e_2-e_{10}\ ,\ h-e_3-e_4-e_{10}\ ,\ h-e_5-e_6-e_{10}\ ,\ h-e_7-e_8-e_{10}\,,
\ea
\ee
hence $G_{F,nA}=SU(2)^4$.

Class (75) has a smooth resolution only for $n=4$. In that case, it is the same theory as Class (153)(with $n=3$).

\subsubsection{Class (287)}
\be
\begin{cases}
& z_1 z_2+z_3 z_4+z_5^3=0 \\
& z_1 z_5+z_2 z_3^2+z_4^3+z_2^{15}=0\,.
\end{cases}
\ee
\be
\begin{array}{||c|c|c||c|c|c||c|c|c||}
\hline
 r & d_H & G_F & \h r & \h d_H & f & b_3 & a&c\\
\hline
 9 & 63 & E_7\times U(1) & 55 & 17 & 8 & 0 & \frac{3803}{24}&\frac{955}{6}\\
\hline
\end{array}
\ee
We have the 5d magnetic quiver:
\be
 \begin{tikzpicture}[x=.5cm,y=.5cm]
\draw[ligne, black](2,0)--(-3,0);
\draw[ligne, black](0,0)--(0,2);
\draw[ligne, black](-1,0)--(-1,1);
\node[] at (-5,0) {$\MQfive= $};
\node[bd] at (2,0) [label=below:{{\scriptsize$5$}}] {};
\node[bd] at (1,0) [label=below:{{\scriptsize$10$}}] {};
\node[bd] at (0,0) [label=below:{{\scriptsize$15$}}] {};
\node[bd] at (-1,0) [label=below:{{\scriptsize$12$}}] {};
\node[bd] at (-2,0) [label=below:{{\scriptsize$8$}}] {};
\node[bd] at (-3,0) [label=below:{{\scriptsize$4$}}] {};
\node[bd] at (0,1) [label=right:{{\scriptsize$8$}}] {};
\node[bd] at (0,2) [label=right:{{\scriptsize$1$}}] {};
\node[bd] at (-1,1) [label=right:{{\scriptsize$1$}}] {};
\end{tikzpicture} 
\ee
We can read $G_F=E_7\times U(1)$ from the bottom layer of the Hasse diagram.

The resolution sequence is
\be
\ba
&(z_1,z_2,z_3,z_4,z_5;\delta_1)\cr
&(z_1^{(2)},z_3^{(1)},z_4^{(1)},z_5^{(1)},\delta_1^{(1)};\delta_6)\cr
&(z_1^{(3)},z_3^{(2)},z_4^{(1)},z_5^{(1)},\delta_6^{(1)};\delta_9)\cr
&(z_1^{(5)},z_3^{(3)},z_4^{(2)},z_5^{(1)},\delta_9^{(1)};\delta_{11})\cr
&(z_1^{(2)},z_3^{(1)},z_4^{(1)},\delta_6^{(1)};\delta_{8})\cr
&(z_1^{(2)},z_3^{(1)},z_4^{(1)},\delta_9^{(1)};\delta_{10})\cr
&(z_1,z_3,\delta_1;\delta_5)\cr
&(z_1,z_3,\delta_6;\delta_7)\cr
&(z_1,z_4,\delta_1;\delta_4)\cr
&(z_1,\delta_1;\delta_3)\cr
&(z_5,\delta_1;\delta_2)
\ea
\ee
The resolved equation is
\be
\begin{cases}
&z_3 z_4+z_1 z_2\delta_3+z_5^3\delta_1 \delta_2^4 \delta_3 \delta_6^2\delta_7\delta_9^2=0\\
&z_1 z_5+z_2 z_3^2\delta_1 \delta_5^2\delta_7+z_4^3\delta_1\delta_4^3\delta_6\delta_8^2\delta_{10}+z_2^{15}\delta_1^{13}\delta_3^{12}\delta_4^{12}\delta_5^{12}\delta_6^{10}\delta_7^9\delta_8^8\delta_9^6\delta_{10}^4=0\,.
\end{cases}
\ee
We can deduce the following intersection diagram:
\be
\includegraphics[height=8cm]{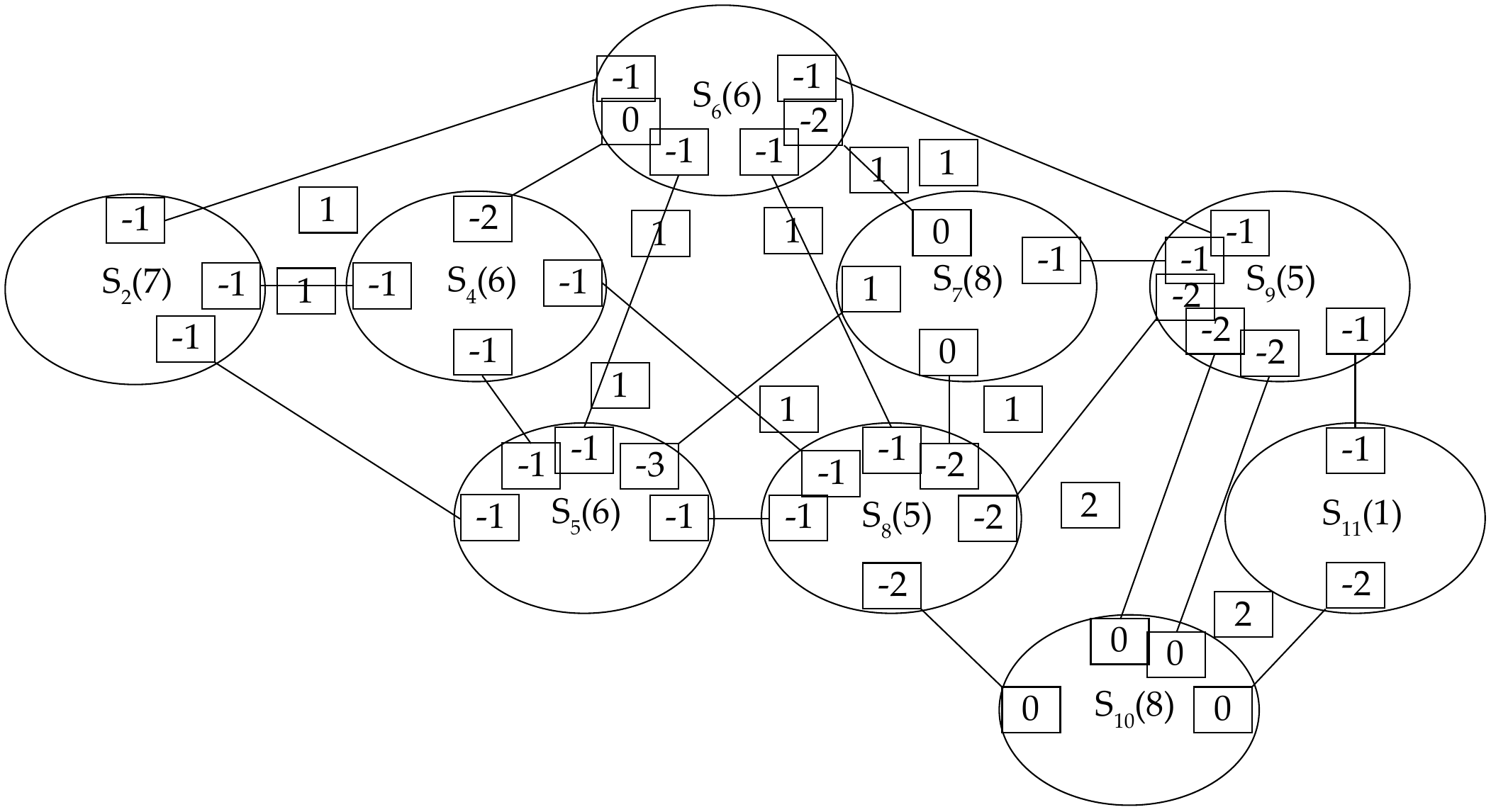}
\ee
\subsection{Rank-10 examples}
\subsubsection{Class (10)}
\be
\begin{cases}
&z_1 z_3+z_4 z_5=0\\
&z_1^2+z_2^2+z_3^{16}+z_3 z_4^3+z_5^4=0\,.
\end{cases}
\ee 
\be
\begin{array}{||c|c|c||c|c|c||c|c|c|c||}
\hline
 r & d_H & G_F & \h r & \h d_H & f & b_3 & a&c\\
\hline
 10 & 68 & E_7\times U(1)^2 & 59 & 19 & 9 & 0 & \frac{4285}{24}&\frac{538}{3}\\
\hline
\end{array}
\ee
We have the 5d magnetic quiver:
\be
 \begin{tikzpicture}[x=.5cm,y=.5cm]
\draw[ligne, black](3,0)--(-3,0);
\draw[ligne, black](0,0)--(0,1);
\draw[ligne, black](1,2)--(0,1);
\draw[ligne, black](-1,2)--(0,1);
\node[] at (-5,0) {$\MQfive= $};
\node[bd] at (3,0) [label=below:{{\scriptsize$1$}}] {};
\node[bd] at (2,0) [label=below:{{\scriptsize$6$}}] {};
\node[bd] at (1,0) [label=below:{{\scriptsize$11$}}] {};
\node[bd] at (0,0) [label=below:{{\scriptsize$16$}}] {};
\node[bd] at (-1,0) [label=below:{{\scriptsize$12$}}] {};
\node[bd] at (-2,0) [label=below:{{\scriptsize$8$}}] {};
\node[bd] at (-3,0) [label=below:{{\scriptsize$4$}}] {};
\node[bd] at (0,1) [label=right:{{\scriptsize$9$}}] {};
\node[bd] at (1,2) [label=right:{{\scriptsize$1$}}] {};
\node[bd] at (-1,2) [label=right:{{\scriptsize$1$}}] {};
\end{tikzpicture} 
\ee
From the Hasse diagram, we read off $G_{F}=E_7\times U(1)^2$.

The resolution sequence is
\be
\ba
&(z_1,z_2,z_3,z_4,z_5;\delta_1)\cr
&(z_1^{(2)},z_2^{(2)},z_4^{(1)},z_5^{(1)},\delta_1^{(1)};\delta_5)\cr
&(z_1^{(2)},z_2^{(2)},z_4^{(1)},z_5^{(1)},\delta_5^{(1)};\delta_7)\cr
&(z_1^{(3)},z_2^{(3)},z_4^{(2)},z_5^{(1)},\delta_7^{(1)};\delta_9)\cr
&(z_1,z_2,z_4,\delta_1;\delta_4)\cr
&(z_1,z_2,z_4,\delta_5;\delta_6)\cr
&(z_1,z_2,z_4,\delta_7;\delta_8)\cr
&(z_1,z_2,z_5,\delta_1;\delta_3)\cr
&(z_1,z_2,\delta_1;\delta_2)\cr
\ea
\ee
The resolved equation is
\be
\begin{cases}
&z_4 z_5+z_1 z_3 \delta_2=0\\
&z_1^2+z_2^2+z_5^4\delta_1^2\delta_3^4\delta_5^2\delta_7^2+z_3 z_4^3\delta_1^2
\delta_4^3\delta_5\delta_6^2\delta_8+z_3^{16}\delta_1^{14}\delta_2^{12}\delta_3^{12}\delta_4^{12}\delta_5^{10}\delta_6^{8}\delta_7^{6}\delta_8^{4}=0\,.
\end{cases}
\ee
We can deduce the following intersection diagram:
\be
\includegraphics[height=6cm]{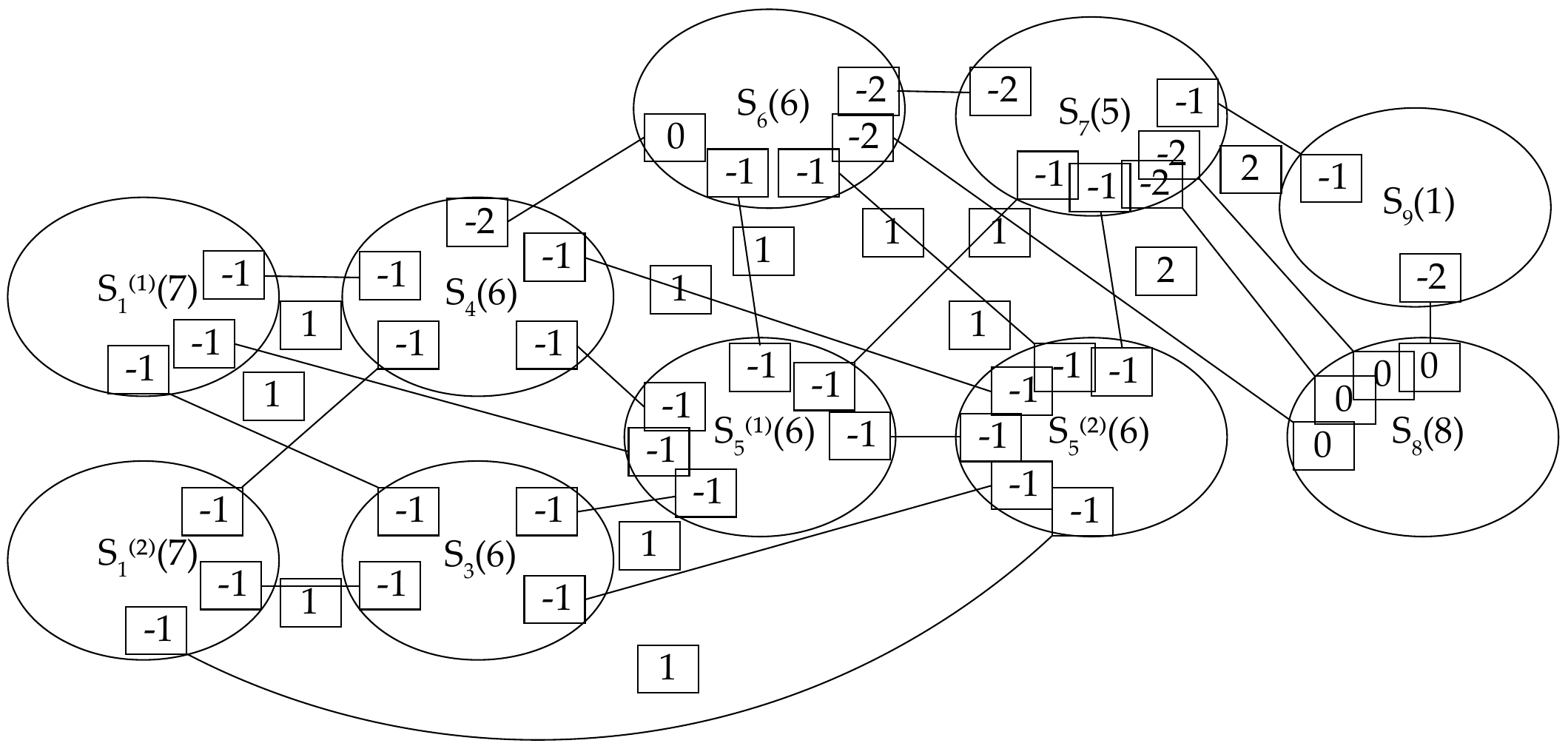}
\ee

Note that in the diagram we have

$S_1^{(1)}\cdot S_5^{(1)}\cdot S_{4}=S_1^{(1)}\cdot S_5^{(1)}\cdot S_{3}=S_1^{(2)}\cdot S_5^{(2)}\cdot S_{4}=S_1^{(2)}\cdot S_5^{(2)}\cdot S_{3}=S_5^{(1)}\cdot S_5^{(2)}\cdot S_{7}=S_5^{(1)}\cdot S_5^{(2)}\cdot S_{3}=S_5^{(1)}\cdot S_7\cdot S_6=S_5^{(1)}\cdot S_4\cdot S_6=S_5^{(2)}\cdot S_7\cdot S_6=S_5^{(2)}\cdot S_4\cdot S_6=1$,

$S_{7}\cdot S_{9}\cdot S_{8}=S_{7}\cdot S_{6}\cdot S_{8}=2$.
\subsubsection{Class (76)}
\be
\begin{cases}
& z_1 z_2+z_3^2+z_4^3+z_5^3=0 \\
& z_1 z_4+z_2^3 z_4+z_2 z_3 z_5=0\,
\end{cases}
\ee

\be
\begin{array}{||c|c|c||c|c|c||c|c|c||}
\hline
 r & d_H & G_F & \h r & \h d_H & f & b_3 & a&c\\
\hline
 10 & 29 & \color{blue}{SU(2)^3\times U(1)^2} & 24 & 15 & 5 & 0 & \frac{521}{8}&\frac{263}{4}\\
\hline
\end{array}
\ee



The resolution sequence is
\be
\ba
&(z_1,z_2,z_3,z_4,z_5;\delta_1)\cr
&(z_1,z_3,z_4,\delta_1;\delta_5)\cr
&(z_1,z_3,\delta_1;\delta_4)\cr
&(z_1^{(2)},z_5^{(1)},\delta_4^{(1)},\delta_5^{(1)};\delta_{11})\cr
&(z_1,z_5,\delta_5;\delta_7)\cr
&(z_1,\delta_1,\delta_4;\delta_8)\cr
&(z_1,\delta_1,\delta_5;\delta_9)\cr
&(z_1,\delta_4,\delta_{11};\delta_{12})\cr
&(z_1,\delta_1;\delta_3)\cr
&(z_1,\delta_4;\delta_6)\cr
&(z_1,\delta_8;\delta_{10})\cr
&(z_4,\delta_1;\delta_2)\cr
\ea
\ee
The resolved equation is
\be
\begin{cases}
&z_3^2 \delta_4 \delta_5 \delta_6+z_1 z_2 \delta_3 \delta_6 \delta_{10}+z_5^3 \delta_1 \delta_2 \delta_3 \delta_7^2 \delta_{11}+z_4^3 \delta_1 \delta_2^4 \delta_3 \delta_5^3 \delta_7^2 \delta_9^3 \delta_{11}=0\\
&z_1 z_4+z_2 z_3 z_5\delta_1 \delta_4 \delta_8+z_2^3 z_4\delta_1^2 \delta_2^2 \delta_3 \delta_4 \delta_5 \delta_8^2 \delta_9^2 \delta_{10}=0\,.
\end{cases}
\ee 
We can deduce the following intersection diagram:
\be
\includegraphics[height=8cm]{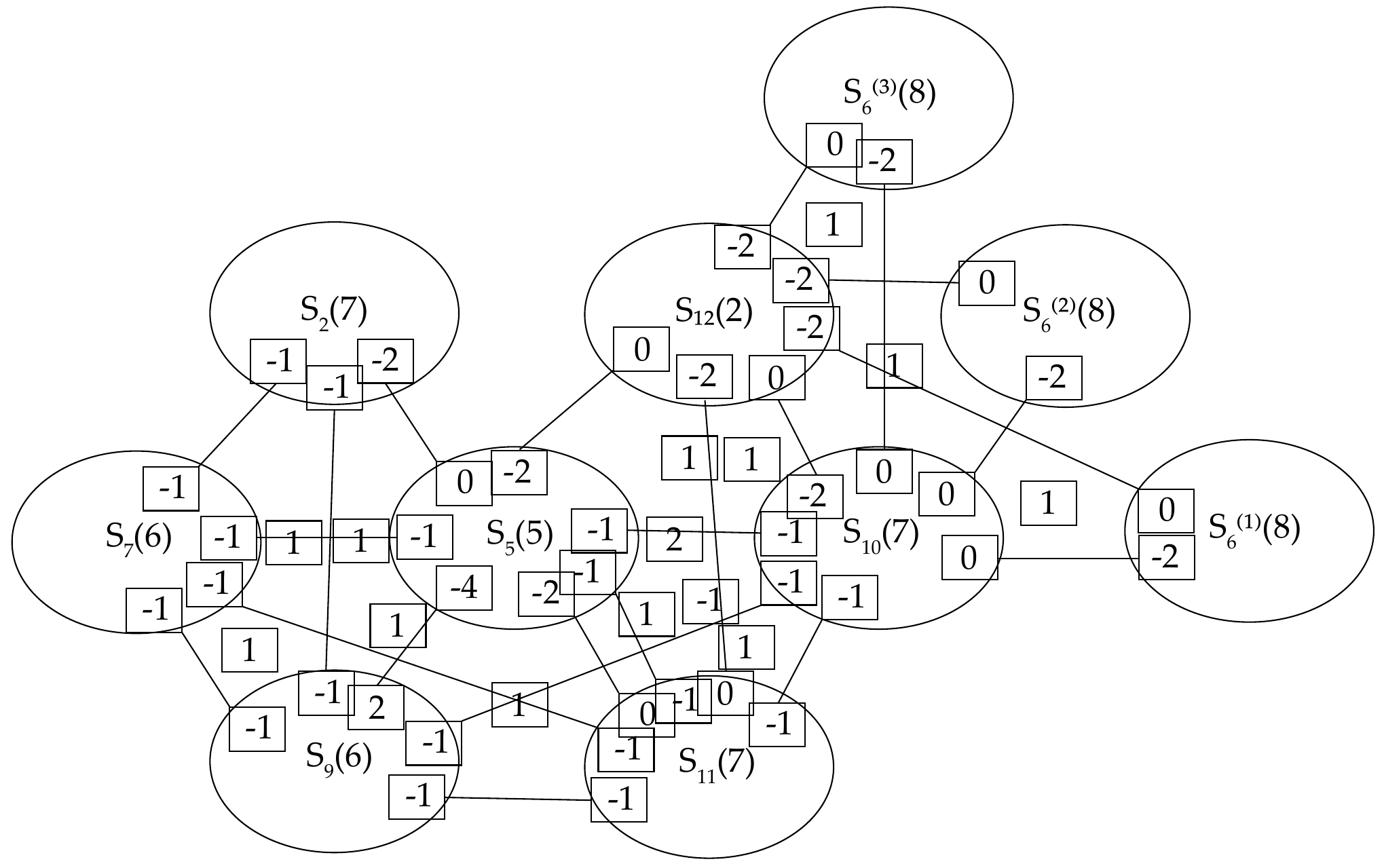}
\ee

Note that in the diagram we have
$S_5\cdot S_7\cdot S_{11}=S_5 \cdot S_{11}\cdot S_9=S_5 \cdot S_9 \cdot S_{10}=S_5\cdot S_9\cdot S_2=S_5\cdot S_{12}\cdot S_{10}=S_{11}\cdot S_7\cdot S_9=S_{11}\cdot S_9\cdot S_{10}=S_{11}\cdot S_{12}\cdot S_{10}=S_{7}\cdot S_{9}\cdot S_2=S_{12}\cdot S_6^{(i)}\cdot S_{10}=1$,

$S_5 \cdot S_{11}\cdot S_{12}=2$,\quad$S_5 \cdot S_{11}\cdot S_{10}=-1$.

To get $G_{F,nA}$, we observe that $S_{12}$ can be a type $7\mbf{A}_1$ gdP$_7$, with the following $(-2)$-curves:
\be
\ba
&S_{12}\cdot S_{11}:\ 2h-e_1-e_2-e_3-e_4-e_5-e_6\cr
&S_{12}\cdot S_6^{(1)}:\ e_1-e_6\cr
&S_{12}\cdot S_6^{(2)}:\ h-e_4-e_5-e_7\cr
&S_{12}\cdot S_6^{(3)}:\ h-e_2-e_3-e_7\cr
&\text{Flavor\ curves}:\ e_2-e_3\ ,\ e_4-e_5\ ,\ h-e_1-e_6-e_7\,.
\ea
\ee
Hence $G_{F,nA}=SU(2)^3$.
\subsubsection{Class (280)}
\be
\begin{cases}
& z_1 z_2+z_3 z_4=0 \\
& z_1 z_5+z_3^2+z_2 z_4^3+z_5^4+z_2^{16}=0\,.
\end{cases}
\ee

\be
\begin{array}{||c|c|c||c|c|c||c|c|c||}
\hline
 r & d_H & G_F & \h r & \h d_H & f & b_3 & a&c\\
\hline
 10 & 70 & E_7 \times U(1)^2 & 61 & 19 & 9 & 0 & \frac{4489}{24}&\frac{1127}{6}\\
\hline
\end{array}
\ee
We have the 5d magnetic quiver:
\be
 \begin{tikzpicture}[x=.5cm,y=.5cm]
\draw[ligne, black](4,0)--(-3,0);
\draw[ligne, black](0,0)--(0,1);
\draw[ligne, black](1,0)--(1,1);
\node[] at (-5,0) {$\MQfive= $};
\node[bd] at (4,0) [label=below:{{\scriptsize$1$}}] {};
\node[bd] at (3,0) [label=below:{{\scriptsize$6$}}] {};
\node[bd] at (2,0) [label=below:{{\scriptsize$11$}}] {};
\node[bd] at (1,0) [label=below:{{\scriptsize$16$}}] {};
\node[bd] at (0,0) [label=below:{{\scriptsize$13$}}] {};
\node[bd] at (-1,0) [label=below:{{\scriptsize$9$}}] {};
\node[bd] at (-2,0) [label=below:{{\scriptsize$5$}}] {};
\node[bd] at (-3,0) [label=below:{{\scriptsize$1$}}] {};
\node[bd] at (0,1) [label=right:{{\scriptsize$1$}}] {};
\node[bd] at (1,1) [label=right:{{\scriptsize$8$}}] {};
\end{tikzpicture} 
\ee
We can read $G_F=E_7\times U(1)^2$ from the bottom layer of the Hasse diagram.

The resolution sequence is
\be
\ba
&(z_1,z_2,z_3,z_4,z_5;\delta_1)\cr
&(z_1^{(3)},z_3^{(2)},z_4^{(1)},z_5^{(1)},\delta_1^{(1)};\delta_7)\cr
&(z_1^{(3)},z_3^{(2)},z_4^{(1)},z_5^{(1)},\delta_7^{(1)};\delta_{10})\cr
&(z_1^{(5)},z_3^{(3)},z_4^{(2)},z_5^{(1)},\delta_{10}^{(1)};\delta_{12})\cr
&(z_1,z_3,z_5,\delta_1;\delta_3)\cr
&(z_1^{(2)},z_3^{(1)},z_4^{(1)},\delta_1^{(1)};\delta_{5})\cr
&(z_1^{(2)},z_3^{(1)},z_4^{(1)},\delta_7^{(1)};\delta_{9})\cr
&(z_1^{(2)},z_3^{(1)},z_4^{(1)},\delta_{10}^{(1)};\delta_{11})\cr
&(z_1,z_3,\delta_1;\delta_2)\cr
&(z_1,z_3,\delta_2;\delta_6)\cr
&(z_1,z_3,\delta_7;\delta_8)\cr
&(z_1,\delta_2;\delta_4)
\ea
\ee
The resolved equation is
\be
\begin{cases}
&z_3 z_4+z_1 z_2\delta_4=0\\
&z_1 z_5+z_3^2\delta_2 \delta_6^2\delta_8+z_5^4\delta_1^2\delta_2\delta_3^4\delta_7^2\delta_8\delta_{10}^2+z_2 z_4^3 \delta_1^2\delta_2\delta_5^{3}\delta_7\delta_9^{2}\delta_{11}+z_2^{16}\delta_1^{14}\delta_2^{13}\delta_3^{12}\delta_4^{12}\delta_5^{12}\delta_6^{12}\delta_7^{10}\delta_8^9\delta_9^8\delta_{10}^6\delta_{11}^4=0\,.
\end{cases}
\ee
We can deduce the following intersection diagram:
\be
\includegraphics[height=8cm]{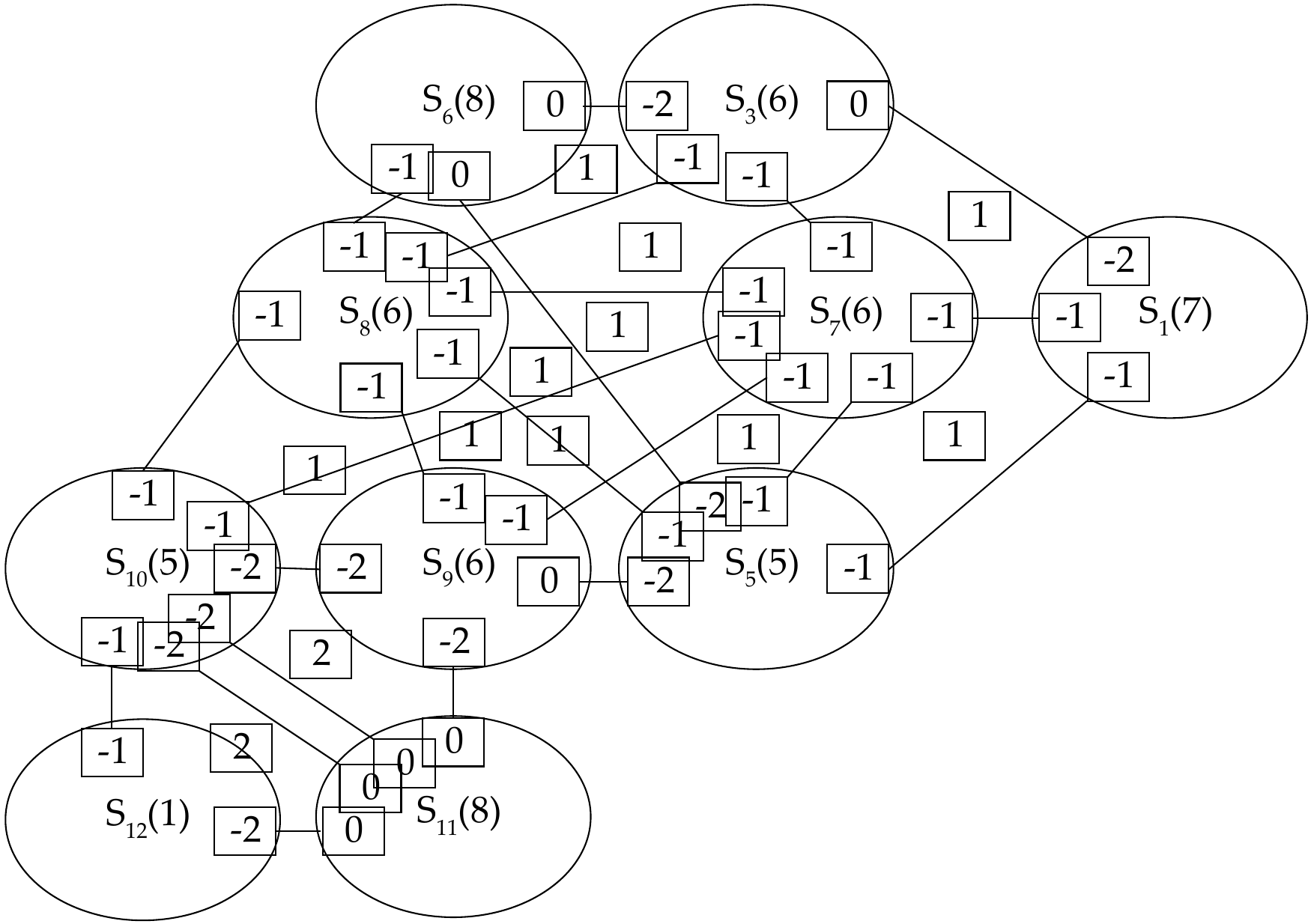}
\ee

Note that in the diagram we have
$S_1\cdot S_7\cdot S_{3}=S_1 \cdot S_{7}\cdot S_5=S_7 \cdot S_9 \cdot S_{10}=S_7\cdot S_{10}\cdot S_8=S_7\cdot S_{3}\cdot S_{8}=S_{7}\cdot S_5\cdot S_9=S_{10}\cdot S_9\cdot S_{8}=S_{3}\cdot S_{6}\cdot S_{8}=S_{5}\cdot S_{9}\cdot S_8=S_{5}\cdot S_6\cdot S_{8}=1$,

$S_{10} \cdot S_{12}\cdot S_{11}=S_{10} \cdot S_{9}\cdot S_{11}=2$.


\section{Discussions}
\label{sec:discussions}

In this paper we systematically studied the 5d SCFTs $\FT$ from M-theory on ICIS $X$, and checked the relation to the 4d $\mc{N}=2$ SCFTs $\FTfour$ from IIB on $X$. We established the crepant resolution algorithm for $X$, and studied the Coulomb branch properties of $\FT$ from the triple intersection numbers of compact divisors on the resolved CY3 $\widetilde{X}$. We also studied the UV flavor symmetry enhancement $G_F$ from the additional flavor curves. For the Higgs branch, we constructed the magnetic quiver $\MQfive$ using the 5d-4d relation conjecture for a number of cases. We also checked the relation between the smoothness of $\widetilde{X}$ and the integrality of the 4d CB spectrum for $\FTfour$. We found the counter-example Class (209) with a smooth resolution and fractional 4d CB spectrum.

There are a number of future questions to investigate.
\begin{itemize}
\item For many classes of $X$, there is not a Lagrangian description for $\FTfour$, even when the 5d SCFT $\FT$ is well-defined from the smooth crepant resolution. In these cases, $\FT$ should have a well-defined Higgs branch. It would be interesting to investigate the properties of the magnetic quiver and discover the relation to $\FTfour$.
\item Generalize the story to the cases of isolated non-complete intersection singularities, which is the most general class of isolated canonical threefold singularities. A full geometric classification of 5d SCFTs from M-theory on CY3 can be achieved if such general cases can be classified.
\item Study the relation between M-theory on $X$ (in particular for the cases with smooth crepant resolution) and the IIB brane web constructions. It would be interesting to see which IHS and ICIS cases can be mapped to a generalized toric polytope (GTP) construction~\cite{Benini:2009gi,vanBeest:2020kou,VanBeest:2020kxw,vanBeest:2021xyt,Bourget:2023wlb,Franco:2023flw}.
\end{itemize}


\section*{Acknowledgements}

We thank Cyril Closset, Amihay Hanany, Chunhao Li, Deshuo Liu, Sakura Schafer-Nameki, Marcus Sperling, Dan Xie, Wenbin Yan for discussions. 
The work is supported by National Natural Science Foundation of China under Grant No. 12175004, by Peking University under startup Grant No. 7100603667 and by Young Elite Scientists Sponsorship Program by CAST (2022QNRC001, 2023QNRC001). 

\appendix

\section{Tables: Smoothable Models}
\label{app:tables}

In this section we list the cases of $X$ with a smooth resolved $\widetilde{X}$ and smooth divisors. In table~\ref{smoothableModels1} we list the infinite sequences with $b_3>0$, while in table~\ref{smoothableModels2}, \ref{smoothableModels3}, \ref{smoothableModels4}, \ref{smoothableModels5}, \ref{smoothableModels6}, \ref{smoothableModels7}, \ref{smoothableModels8} we list the sporadic cases with $r\leqslant 10$ and $b_3\geqslant 0$. The classification of ICIS singularities and meaning of $n_i$ follow from \cite{Chen:2016bzh}. The physical quantities are
\begin{itemize}
\item The rank $r$ of the 5d SCFT $\FT$;
\item The quarternionic Higgs branch dimension $d_H$ of the 5d SCFT $\FT$;
\item The Coulomb branch rank $\widehat{r}$ of the 4d $\mc{N}=2$ SCFT $\FTfour$;
\item The quarternionic Higgs branch dimension $\widehat{d}_H$ of the $\mc{N}=2$ SCFT $\FTfour$;
\item The central charges $a$ and $c$ of the $\mc{N}=2$ SCFT $\FTfour$;
\item The flavor rank $f$ of the 4d $\mc{N}=2$ SCFT $\FTfour$, which also matches the flavor rank of the 5d SCFT when $b_3=0$;
\item The $b_3$ of the resolved CY3 $\widetilde{X}$;
\item The Coulomb branch spectrum of the 4d $\mc{N}=2$ SCFT $\FTfour$.
\end{itemize}

\begin{sidewaystable}[htp]
\centering
\be
\begin{array}{|c|c|c|c|c|c|c|c|c|c|c|}
\hline
 \text{Class} & \begin{cases}& f_1 \\& f_2\,\end{cases}& \{n_i\} & r &d_H & \widehat{r} & \widehat{d}_H  &a&c& f &b_3\cr \hline \hline
1 &  
\begin{cases}
&z_1^2+ z_2^2+z_3^2+z_4^ 2+ z_5^n=0 \\
&z_1^2+2 z_2^2+3z_3^2+4z_4^ 2+ 5z_5^n=0\,.
\end{cases} 
&n=2k,k\geqslant1 
& k & 8k-1 &8k-6  & k+5 &3k^2-\frac{5}{4} &3k^2-1 &5&2k-2  

\\ 

\hline
119&\begin{cases}
&z_1z_2+z_4^2+z_5^4=0 \\
&z_1z_4+z_3^2+z_2^n=0\,.
\end{cases}&
n=6k-1,k\geqslant 1 &6k-2 &30k-2  &30k-3  &6k-1  & \frac{1440k^2-222k+25}{24}&\frac{360k^2-54k+5}{6}  &1 &8   
 \\
\cline{3-11}
~& ~&
n=6k+2,k\geqslant 1 &6k+2 &30k+15  &30k+10  &6k+7  & 60k^2+\frac{203}{4}k+\frac{43}{4}& 60k^2+51k+11 &5&2  
 \\
\hline
126&\begin{cases}
&z_1z_2+z_4^2+z_5^4=0 \\
&z_1z_4+z_3^2+z_2^nz_4+z_2^nz_5^2=0\,.
\end{cases}&
n=4k-1,k\geqslant 1 &6k-2 &30k-2  &30k-3  &6k-1  & \frac{1440k^2-222k+25}{24}&\frac{360k^2-54k+5}{6}  &1 &8   
 \\
\cline{3-11}
~& ~&
n=4k+1,k\geqslant 1 &6k+2 &30k+15  &30k+10  &6k+7  & 60k^2+\frac{203}{4}k+\frac{43}{4}& 60k^2+51k+11 &5&2  
 \\
\hline
152&\begin{cases}
&z_1z_2+z_4^3+z_5^3=0 \\
&z_1z_4+z_3^2+z_2^n=0\,.
\end{cases}&
n=4k-1,k\geqslant 1 &6k-3 & 18k &18k-2  &  6k-1&\frac{288k^2-39 k+10}{12} &\frac{72k^2-9k+2}{3}  &2&6  
 \\
\hline
154&\begin{cases}
&z_1z_2+z_4^3+z_5^3=0 \\
&z_1z_4+z_3^2+z_2^nz_5=0\,.
\end{cases}&
n=3k-1,k\geqslant 1 &6k-3 & 18k &18k-2  &  6k-1&\frac{288k^2-39k+10}{12} &\frac{72k^2-9k+2}{3}  &2&6 
 \\
\hline
183&\begin{cases}
&z_1z_2+z_3 z_4=0 \\
&z_1z_3+z_2^n+z_4^3+z_5^3=0\,.
\end{cases}&
n=3k,k\geqslant 1 &6k-2 & 12k+6 &12k+1  &  6k+3&\frac{144k^2+69k+4}{12} &\frac{144k^2+72k+5}{12}  &5&2
 \\
\hline
202&\begin{cases}
&z_1z_2+z_3 z_4=0 \\
&z_1z_3+z_2^n z_5+z_4^3+z_5^3=0\,.
\end{cases}&
n=2k,k\geqslant 1 &6k-2 & 12k+6 &12k+1  &  6k+3&\frac{144k^2+69k+4}{12} &\frac{144k^2+72k+5}{12}  &5&2
 \\
\hline
229&\begin{cases}
&z_1z_2+z_3z_4=0 \\
&z_1z_4+z_3^2+z_2^n+z_5^4=0\,.
\end{cases}&
n=2k,k\geqslant 2&3k-1 &9k+6  &9k+1  &3k+4  &6k^2+\frac{35}{8}k+\frac{3}{8} &6k^2+\frac{9}{2}k+\frac{1}{2}  &5&2
 \\
\hline
245&\begin{cases}
&z_1z_2+z_3z_4=0 \\
&z_1z_4+z_3^2+z_2^nz_5+z_5^4=0\,.
\end{cases}&
n=3k,k\geqslant 1&6k-1 &18k+6  &18k+1  &6k+4  &\frac{192k^2+70k+3}{8} &\frac{48k^2+18k+1}{2}  &5&2
 \\
\hline
255&\begin{cases}
&z_1z_2+z_3z_4=0 \\
&z_1z_4+z_3^2+z_2^nz_4+z_5^4=0\,.
\end{cases}&
n=3k+1,k\geqslant 1&6k+2 &18k+15  &18k+10  &6k+7  &\frac{96k^2+131k+43}{4} &24k^2+33k+11  &5&2
 \\
\hline
   \end{array}
\ee
\caption{Fully smooth infinite sequence models with $b_3>0$.\label{smoothableModels1}}
\end{sidewaystable}

\begin{sidewaystable}[htp]
\centering
$$
\begin{array}{|c|c|c|c|c|c|c|c|c|c|c|c|}
\hline
\text{Class} & \{n_i\} & \begin{cases}& f_1 \\& f_2\,\end{cases} & r &d_H & \widehat{r} & \widehat{d}_H  &a&c& f &b_3&  \text{CB Spectrum}\cr \hline \hline
1&n=2 &\begin{cases}&z_1^2 + z_2^2 + z_3^2 + z_4^2 + z_5^2 \\& z_1^2 + 2 z_2^2 + 3 z_3^2 + 4 z_4^2 + 5 z_5^2\,\end{cases}
& 1 & 7& 2 &  6& \frac{7}{4} &2 &5 & 0   &
   \left\{ 2^2\right\} \\
\hline
1&n=4 &\begin{cases}&z_1^2 + z_2^2 + z_3^2 + z_4^2 + z_5^4 \\& z_1^2 + 2 z_2^2 + 3 z_3^2 + 4 z_4^2 + 5 z_5^4\,\end{cases}
& 2 & 21& 10 &  13& \frac{43}{4} &11 &5 & 14   &
   \left\{ 2^6, 3^2, 4^2\right\} \\
\hline
7&n=4 &\begin{cases}&z_1 z_3 + z_4 z_5 \\& z_1^2 + z_2^2 + z_3^4 + z_4^2 + 2 z_5^4\,\end{cases}
& 2 &16  & 9 &9  & \frac{77}{8} &10 &7 &  0  &
   \left\{2^5, 3^3, 4\right\} \\
\hline
41 &  & \begin{cases}& z_2^2+z_3^2+z_4^2+z_5^2 \\& z_1^2+z_2^3+z_3^3+z_4^3\,\end{cases}
& 2 & 16 & 15 & 3 & \frac{437}{24} &\frac{109}{6} &1 & 8 & 
   \left\{ 2^9, 4^5, 6\right\} \\
\hline
59&n=4 &\begin{cases}&z_1 z_2 + z_3^2 + z_4^2 + z_5^4\\&3 z_2^4 + 2 z_3^2 + z_4^2 + z_1 z_5\,\end{cases}
& 2 &18  & 11 & 9& \frac{97}{8} &\frac{25}{2} &7&0  &
   \left\{ 2^6, 3^3, 4^2\right\} \\
\hline
220 & n=3& \begin{cases}& z_1 z_2 + z_3 z_4 + z_5^3\\& z_2^3 + z_2 z_3 + z_1 z_4 + z_4^3 + z_1 z_5\,\end{cases}
& 2 & 13 & 6 & 9 & \frac{47}{8}  &\frac{25}{4} & 7&  0  &
   \left\{ 2^4, 3^2\right\} \\
\hline
   \end{array}
   $$
\caption{Fully smooth models with 5d CB rank $r=1,2$.\label{smoothableModels2}}
\end{sidewaystable}


\begin{sidewaystable}[htp]
\centering
$$
\begin{array}{|c|c|c|c|c|c|c|c|c|c|c|c|}
\hline
\text{Class} & \{n_i\} & \begin{cases}& f_1 \\& f_2\,\end{cases} & r &d_H & \widehat{r} & \widehat{d}_H  &a&c&f& b_3 &  \text{CB Spectrum}\cr \hline \hline
1&n=6 &\begin{cases}&z_1^2+z_2^2+z_3^2+z_4^2+z_5^6\\&z_1^2+2z_2^2+3z_3^2+4z_4^2+5z_5^6\,\end{cases}
& 3 & 23& 18 & 8& \frac{103}{4} &26 &5& 4   &
   \left\{2^6, 3^6, 4^2, 5^2, 6^2\right\} \\
\hline
7&n= 6&\begin{cases}&z_1 z_3+z_4 z_5\\&z_1^2+z_2^2+z_3^6+z_4^2+2z_5^6\,\end{cases}
& 3 &29 & 20&12 & \frac{55}{2} &28 &9&  0  &
   \left\{ 2^7, 3^6, 4^4, 5^2, 6\right\} \\
\hline
15& &\begin{cases}&z_1 z_3 + z_4 z_5\\&z_1^2 + z_2^2 + z_3^6 + z_4^3 + z_5^3\,\end{cases}
& 3 & 23& 16 &10 & \frac{263}{12} &\frac{67}{3} &7&  0  &
   \left\{ 2^6, 3^4, 4^4, 5, 6\right\} \\
\hline
59&n= 6&\begin{cases}&z_1 z_2+z_3^2+z_4^2+z_5^6\\&3z_2^6+2z_3^2+z_4^2+z_1 z_5\,\end{cases}
& 3 &33 & 24&12 & \frac{69}{2} &35 &9& 0 &  
   \left\{ 2^8, 3^7, 4^4, 5^3, 6^2\right\} \\
\hline
75&n=3&\begin{cases}&z_1 z_2+z_3^2+z_4^3\\&z_2^3+z_1 z_4+z_5^3\,\end{cases}
& 3& 18& 16 &5 & \frac{259}{12} &\frac{65}{3} &2&  6&  
   \left\{ 2^8, 3, 4^5, 6^2\right\} \\
\hline
118&n=8&\begin{cases}&z_1 z_2+z_4^2+z_5^3\\&z_2^8+z_3^2+z_1 z_4\,\end{cases}
& 3&33 & 26 & 10& \frac{527}{12} &\frac{133}{3} &7&  0&
   \left\{2^7,3^6,4^5,5^3,6^3,7,8\right\} \\
\hline
220&n=4&\begin{cases}&z_1 z_2+z_3 z_4+z_5^4\\&z_2^4+z_2 z_3+z_1 z_4+z_4^4+z_1 z_5\,\end{cases}
& 3 & 21& 12& 12& \frac{27}{2} &14 &9&0  &  
   \left\{ 2^6, 3^4, 4^2\right\} \\
\hline
228&n=6&\begin{cases}&z_1 z_2+z_3 z_4\\&z_2^6+z_3^2+z_1 z_4+z_5^3\,\end{cases}
& 3 &24 & 17& 10& \frac{71}{3} &\frac{289}{12} &7&0  &  
   \left\{ 2^6, 3^5, 4^3, 5^2, 6\right\} \\
\hline

\end{array}
$$
\caption{Fully smooth models with 5d CB rank $r=3$.\label{smoothableModels3}}
\end{sidewaystable}


\begin{sidewaystable}[htp]
\centering
$$
\begin{array}{|c|c|c|c|c|c|c|c|c|c|c|c|}
\hline
\text{Class} & \{n_i\} & \begin{cases}& f_1 \\& f_2\,\end{cases} & r &d_H & \widehat{r} & \widehat{d}_H  &a&c&f& b_3  & \text{CB Spectrum}\cr \hline \hline
1&n=8&\begin{cases}&z_1^2+z_2^2+z_3^2+z_4^2+z_5^8\\&z_1^2+2z_2^2+3z_3^2+4z_4^2+5z_5^8\,\end{cases}
& 4 &31 & 26& 9& \frac{187}{4} &47 &5& 6 &  
   \left\{2^6, 3^6, 4^6, 5^2, 6^2, 7^2, 8^2\right\} \\
\hline
24& &\begin{cases}&z_1 z_3 + z_4^2 + z_2 z_5\\&z_1^2 + z_2^2 + z_3^4 + z_5^4\,\end{cases}
& 4 &22& 19&7 & \frac{743}{24}&\frac{187}{6} &3& 4 &  
   \left\{ 2^7, 3^2, 4^5, 5, 6^3, 8\right\} \\
\hline
35&n=3 &\begin{cases}&z_1 z_3+z_2 z_5\\&z_1^2+z_2^2+z_3^3+z_4^3+z_5^3\,\end{cases}
& 4 &18& 13&9 & \frac{217}{12}&\frac{221}{12} &5&2  &  
   \left\{2^5,3^3,4^3,5,6\right\} \\
\hline
59&n=8 &\begin{cases}&z_1 z_2+z_3^2+z_4^2+z_5^8\\&3z_2^8+2z_3^2+z_4^2+z_1 z_5\,\end{cases}
& 4 &52& 41& 15& \frac{583}{8}&\frac{147}{2} &11&0  &  
   \left\{2^{10},3^9,4^8,5^5,6^4,7^3,8^2\right\} \\
\hline
116&n_2=n_5=4 &\begin{cases}&z_1 z_2+z_4 z_5\\&2z_3^2+z_1 z_4+3z_2^4z_4+4z_1 z_5^4+5z_2^4z_5^4\,\end{cases}
& 4 &46 & 35&15 & \frac{475}{8}&60 &11& 0 &  
   \left\{2^9,3^8,4^7,5^5,6^3,7^2,8\right\} \\
\hline
119&n=5 &\begin{cases}&z_1 z_2+z_4^2+z_5^4\\&z_2^5+z_3^2+z_1 z_4\,\end{cases}
& 4 &28& 27& 5& \frac{1243}{24}&\frac{311}{6} &1& 8 &  
   \left\{2^9,3,4^8,6^5,8^3,10\right\} \\
\hline
220&n=5 &\begin{cases}&z_1 z_2+z_3 z_4+z_5^5\\&z_2^5+z_2 z_3+z_1 z_4+z_4^5+z_1 z_5\,\end{cases}
& 4 &31& 20& 15& \frac{205}{8}&\frac{105}{4} &11& 0 &  
  \left\{2^8,3^6,4^4,5^2\right\} \\
\hline
\end{array}
$$
\caption{Fully smooth models with 5d CB rank $r=4$.\label{smoothableModels4}}
\end{sidewaystable}


\begin{sidewaystable}[htp]
\centering
$$
\begin{array}{|c|c|c|c|c|c|c|c|c|c|c|c|}
\hline
\text{Class} & \{n_i\} & \begin{cases}& f_1 \\& f_2\,\end{cases} & r &d_H & \widehat{r} & \widehat{d}_H  &a&c&f& b_3  & \text{CB Spectrum}\cr \hline \hline
1&n=10&\begin{cases}&z_1^2+z_2^2+z_3^2+z_4^2+z_5^{10}\\&z_1^2+2z_2^2+3z_3^2+4z_4^2+5z_5^{10}\,\end{cases}
& 5 & 39& 34& 10& \frac{295}{4}&74 &5& 8 &  
  \left\{2^6,3^6,4^6,5^6,6^2,7^2,8^2,9^2,10^2\right\} \\
\hline
2& &\begin{cases}&z_1^2 + z_2^2 + z_3^2 + z_4^3 + z_5^3\\& z_1^2 + 2 z_2^2 + 3 z_3^2 + 4 z_4^3 + 
 5 z_5^3\,\end{cases}
& 5 &19 & 13&11 & \frac{473}{24}&\frac{121}{6} &6& 0 &  
  \left\{2^4,3^3,4^4,6^2\right\} \\
\hline
58& &\begin{cases}&z_1 z_2 + z_3^2 + z_4^2 + z_5^3\\&z_1 z_3 + z_2^2 z_4 + z_4 z_5^2\,\end{cases}
& 5 &21 & 15&11 & \frac{581}{24}&\frac{74}{3} &6& 0 &  
  \left\{2^4,3^4,4^3,5^2,6,7\right\} \\
\hline
59&n=10 &\begin{cases}&z_1 z_2+z_3^2+z_4^2+z_5^{10}\\&3z_2^{10}+2z_3^2+z_4^2+z_1 z_5\,\end{cases}
& 5 &75 & 62& 18& \frac{525}{4}&132 &13& 0 &  
  \left\{2^{12},3^{11},4^{10},5^9,6^6,7^5,8^4,9^3,10^2\right\} \\
\hline
60&n=3 &\begin{cases}&z_1 z_2+z_3^2+z_4^2+z_5^3\\&3z_2^7+2z_2 z_3^2+z_2 z_4^2+z_1 z_5\,\end{cases}
& 5 &29 & 21& 13& \frac{799}{24}&\frac{203}{6} &8& 0 &  
  \left\{2^6,3^5,4^5,5^2,6^2,7\right\} \\
\hline
220&n=6 &\begin{cases}&z_1 z_2+z_3 z_4+z_5^6\\&z_2^6+z_2 z_3+z_1 z_4+z_4^6+z_1 z_5\,\end{cases}
& 5 &43 & 30&18 & \frac{173}{4}&44 &13& 0 &  
  \left\{2^{10},3^8,4^6,5^4,6^2\right\} \\
\hline
229&n=4 &\begin{cases}&z_1 z_2+z_3 z_4\\&z_2^4+z_3^2+z_1 z_4+z_5^4\,\end{cases}
& 5 &24& 19&10 & \frac{265}{8}&\frac{67}{2} &5& 2 &  
  \left\{2^5,3^4,4^4,5^2,6^2,7,8\right\} \\
\hline
   \end{array}
$$
\caption{Fully smooth models with 5d CB rank $r=5$.\label{smoothableModels5}}
\end{sidewaystable}


\begin{sidewaystable}[htp]
\centering
$$
\begin{array}{|c|c|c|c|c|c|c|c|c|c|c|c|}
\hline
\text{Class} & \{n_i\} & \begin{cases}& f_1 \\& f_2\,\end{cases} & r &d_H & \widehat{r} & \widehat{d}_H  &a&c&f& b_3  & \text{CB Spectrum}\cr \hline \hline
1&n=12 &\begin{cases}&z_1^2+z_2^2+z_3^2+z_4^2+z_5^{12}\\&z_1^2+2z_2^2+3z_3^2+4z_4^2+5z_5^{12}\,\end{cases}
& 6 &47 & 42&11 & \frac{427}{4}&107 &5&10  &  
  \left\{2^6,3^6,4^6,5^6,6^6,7^2,8^2,9^2,10^2,11^2,12^2\right\} \\
\hline
16& &\begin{cases}&z_1 z_3 + z_4 z_5\\&z_1^2 + z_2^2 + z_3^{12} + z_4^3 + z_5^4\,\end{cases}
& 6 &50 & 42&14 & \frac{1183}{12}&\frac{595}{6} &8& 0 &  
  \left\{2^7,3^7,4^6,5^5,6^5,7^4,8^3,9^2,10,11,12\right\} \\
\hline
23&n=2 &\begin{cases}&z_1 z_3+z_4^2+z_2 z_5^2\\&z_1^2+z_2^2+z_3^3+z_5^6\,\end{cases}
& 6 &28 & 26&8 & \frac{243}{4}&61 &2& 4 &  
  \left\{2^6,3^2,4^6,5,6^4,7,8^3,10^2,12\right\} \\
\hline
31& &\begin{cases}&z_1 z_3 + z_4^2 + z_5^4\\&z_1^2 + z_2^2 + z_3^2 z_4 + z_5^5\,\end{cases}
& 6 &26 & 21&11 & \frac{1049}{24}&\frac{265}{6} &5& 0 &  
  \left\{2^4,3^4,4^4,5^2,6^3,7,8^2,10\right\} \\
\hline
35&n=4 &\begin{cases}&z_1 z_3+z_2 z_5\\&z_1^2+z_2^2+z_3^4+z_4^3+z_5^4\,\end{cases}
& 6 &25 & 24&7 & \frac{433}{8}&\frac{217}{4} &1& 8 &  
  \left\{2^5,3^6,5^3,6^5,8,9^3,12\right\} \\
\hline
57& &\begin{cases}&z_1 z_2 + z_3^2 + z_4^2 + z_5^4\\&z_2^3 + z_1 z_3 + z_2 z_5^3\,\end{cases}
& 6 &24 & 19&11 & \frac{893}{24}&\frac{113}{3} &5& 0 &  
  \left\{2^4,3^4,4^3,5^3,6^2,7,8,9\right\} \\
\hline
59& n=12&\begin{cases}&z_1 z_2+z_3^2+z_4^2+z_5^{12}\\&3z_2^{12}+2z_3^2+z_4^2+z_1 z_5\,\end{cases}
& 6 &102 & 87&21 & \frac{1709}{8}&\frac{429}{2} &15& 0 &  
  \left\{2^{14},3^{13},4^{12},5^{11},6^{10},7^7,8^6,9^5,10^4,11^3,12^2\right\} \\
\hline
77&n=4 &\begin{cases}&z_1 z_2+z_3^2+z_5^4\\&z_2^4+z_4^3+z_1 z_5\,\end{cases}
& 6 &28 & 27& 7& \frac{523}{8}&\frac{131}{2} &1& 8 &  
  \left\{2^5,3^6,4,5^3,6^5,8^2,9^3,12^2\right\} \\
\hline
120&n=4 &\begin{cases}&z_1 z_2+z_4^2+z_5^5\\&z_2^4+z_3^2+z_1 z_4\,\end{cases}
& 6 &32 & 27& 11& \frac{1577}{24}&\frac{397}{6} &5& 0 &  
  \left\{2^4,3^4,4^5,5^3,6^3,7^2,8^2,9,10^2,12\right\} \\
\hline
220& n=7&\begin{cases}&z_1 z_2+z_3 z_4+z_5^7\\&z_2^7+z_2 z_3+z_1 z_4+z_4^7+z_1 z_5\,\end{cases}
& 6 &57 & 42& 21& \frac{539}{8}&\frac{273}{4} &15& 0 &  
  \left\{2^{12},3^{10},4^8,5^6,6^4,7^2\right\} \\
\hline
256&n=2 &\begin{cases}&z_1 z_2+z_3 z_4\\&z_3^2+z_1 z_4+z_2^2z_4+z_5^5\,\end{cases}
& 6 &27 & 22& 11& \frac{1127}{24}&\frac{569}{12} &5& 0 &  
  \left\{2^4,3^4,4^4,5^3,6^2,7^2,8,9,10\right\} \\
\hline
266& &\begin{cases}&z_1 z_2 + z_3 z_4\\& z_2^{12} + z_3^2 + z_4^4 + z_1 z_5 + z_5^3\,\end{cases}
& 6 &51 & 43&14 & \frac{611}{6}&\frac{1229}{12} &8& 0 &  
  \left\{2^7,3^7,4^6,5^6,6^5,7^3,8^3,9^3,10,11,12\right\} \\
\hline
289& &\begin{cases}&z_1 z_2 + z_3 z_4\\& z_2^9 + z_2 z_3^2 + z_4^3 + z_1 z_5 + z_5^3\,\end{cases}
& 6 &36 & 28&14 & \frac{637}{12}&\frac{161}{3} &8& 0 &  
  \left\{2^6,3^6,4^5,5^4,6^3,7^2,8,9\right\} \\
\hline
   \end{array}
   $$
\caption{Fully smooth models with 5d CB rank $r=6$.\label{smoothableModels6}}
\end{sidewaystable}


\begin{sidewaystable}[htp]
\centering
$$
\begin{array}{|c|c|c|c|c|c|c|c|c|c|c|c|}
\hline
\text{Class} & \{n_i\} & \begin{cases}& f_1 \\& f_2\,\end{cases} & r &d_H & \widehat{r} & \widehat{d}_H  &a&c&f& b_3  & \text{CB Spectrum}\cr \hline \hline
12& &\begin{cases}&z_1 z_3 + z_4 z_5\\& z_1^2 + z_2^2 + z_3^{10} + z_3 z_4^3 + z_3 z_5^3\,\end{cases}
& 7 &41 & 32& 16& \frac{197}{3}&\frac{199}{3} &9&0  &  
  \left\{2^6,3^6,4^6,5^4,6^4,7^3,8,9,10\right\} \\
\hline
43& n=4&\begin{cases}&z_2 z_3+z_4 z_5\\&z_1^2+z_2^3+z_3^4+z_4^3+z_5^4\,\end{cases}
& 7 &27 & 23& 11& \frac{429}{8}&54 &4& 4 &  
  \left\{2^4,3^4,4^3,5^3,6^3,7,8^2,9^2,12\right\} \\
\hline
231&n=3 &\begin{cases}&z_1 z_2+z_3 z_4\\&z_2^3+z_3^2+z_1 z_4+z_5^6\,\end{cases}
& 7 & 30& 26& 11& \frac{773}{12}&\frac{389}{6} &4& 2 &  
  \left\{2^4,3^4,4^4,5^3,6^3,7^2,8^2,9,10,11,12\right\} \\
\hline
3& &\begin{cases}&z_1^2 + z_2^2 + z_3^2 + z_4^3 + z_5^4\\& z_1^2 + 2 z_2^2 + 3 z_3^2 + 4 z_4^3 + 
 5 z_5^4\,\end{cases}
& 8 & 24& 23& 9& \frac{1357}{24}&\frac{341}{6} &1& 4 &  
  \left\{2^5,3^4,4,5^2,6^5,8^2,9^2,12^2\right\} \\
\hline
119& n=8&\begin{cases}&z_1 z_2+z_4^2+z_5^4\\&z_2^8+z_3^2+z_1 z_4\,\end{cases}
& 8 &45 & 40& 13& \frac{243}{2}&122 &5& 2 &  
  \left\{2^5,3^4,4^6,5^3,6^5,7^3,8^3,9^2,10^3,11,12^2,13,14,16\right\} \\
\hline
202& n=3&\begin{cases}&z_1 z_2+z_3 z_4\\&z_1 z_3+z_4^3+z_2^3z_5+z_5^3\,\end{cases}
& 8 &25 & 18&15 & \frac{285}{8}&\frac{145}{4} &7& 0 &  
  \left\{2^3,3^4,4^4,5^2,6^2,7^2,9\right\} \\
\hline
229&n=6 &\begin{cases}&z_1 z_2+z_3 z_4\\&z_2^6+z_3^2+z_1 z_4+z_5^4\,\end{cases}
& 8 &33 & 28&13 & \frac{135}{2}&68 &5&2  &  
  \left\{2^5,3^3,4^6,5^2,6^4,7^2,8^2,9,10^2,12\right\} \\
\hline
   \end{array}
   $$
\caption{Fully smooth models with 5d CB rank $r=7,8$.\label{smoothableModels7}}
\end{sidewaystable}


\begin{sidewaystable}[htp]
\centering
$$
\begin{array}{|c|c|c|c|c|c|c|c|c|c|c|c|}
\hline
\text{Class} & \{n_i\} & \begin{cases}& f_1 \\& f_2\,\end{cases} & r &d_H & \widehat{r} & \widehat{d}_H  &a&c&f& b_3  & \text{CB Spectrum}\cr \hline \hline
29& &\begin{cases}&z_1 z_3 + z_4^2 + z_5^3\\& z_1^2 + z_2^2 + z_3^6 + z_4^3\,\end{cases}
& 9 &35 & 33&11 & \frac{865}{8}&\frac{217}{2} &2&4  &  
  \left\{2^4,3^5,4^2,5^3,6^4,7,8^3,9^3,10,11,12^3,14,15,18\right\} \\
\hline
37&n=4 &\begin{cases}&z_1 z_3+z_2 z_4\\&z_1^2+z_2^2+z_3^3+z_4^3+z_5^4\,\end{cases}
& 9 &25 & 21&13 & \frac{1231}{24}&\frac{311}{6} &4& 0 &  
  \left\{2^4,3^2,4^4,5^2,6^3,7,8^2,9,10,12\right\} \\
\hline
56& &\begin{cases}&z_1 z_2 + z_3^2 + z_4^2 + z_5^3\\& z_2^5 + 2 z_1 z_3 + z_2 z_4^2\,\end{cases}
& 9 &29 & 27&11 & \frac{613}{8}&77 &2&4  &  
  \left\{2^4,3^5,4,5^3,6^4,7,8^2,9^3,11,12^2,15\right\} \\
\hline
75& n=4&\begin{cases}&z_1 z_2+z_3^2+z_4^3\\&z_2^3+z_1 z_4+z_5^4\,\end{cases}
& 9 &27 & 23&13 & \frac{1423}{24}&\frac{359}{6} &4&0  &  
  \left\{2^4,3^2,4^4,5^3,6^3,8^3,9^2,12^2\right\} \\
\hline
101& n=4&\begin{cases}&z_1 z_2+z_3^2+z_5^3\\&z_2^4z_3+z_4^3+z_1 z_5\,\end{cases}
& 9 &37 & 37&9 & \frac{3383}{24}&\frac{847}{6} &0& 8 &  
  \left\{2^4,3^5,4,5^4,6^5,8^2,9^4,11^2,12^4,14,15^2,18^2,21\right\} \\
\hline
152& n=7&\begin{cases}&z_1 z_2+z_4^3+z_5^3\\&z_2^7+z_3^2+z_1 z_4\,\end{cases}
& 9 &36 & 34&11 & \frac{271}{3}&\frac{272}{3} &2& 6 &  
  \left\{2^8,4^8,5,6^6,8^5,10^3,12^2,14\right\} \\
\hline
42& &\begin{cases}&z_2^2 + z_3^2 + z_4^2 + z_5^3\\& z_1^2 + z_2^3 + z_3^3 + z_4^3\,\end{cases}
& 10 &27 & 27&10 & \frac{689}{8}&\frac{173}{2} &0& 2 &  
  \left\{2^6,3,4^3,6^6,8^4,10,12^4,14,18\right\} \\
\hline
76& &\begin{cases}&z_1 z_2 + z_3^2 + z_4^3 + z_5^3\\& z_1 z_4 + z_2^3 z_4 + z_2 z_3 z_5\,\end{cases}
& 10 &29 & 24&15 & \frac{521}{8}&\frac{263}{4} &5&0  &  
  \left\{2^3,3^3,4^4,5^3,6^2,7^2,8^2,9^2,10,12,13\right\} \\
\hline
119&n=11 &\begin{cases}&z_1 z_2+z_4^2+z_5^4\\&z_2^{11}+z_3^2+z_1 z_4\,\end{cases}
& 10 &58 & 57&11 & \frac{5341}{24}&\frac{1337}{6} &1& 8 &  
  \left\{2^9,4^9,5,6^8,8^8,10^6,12^5,14^4,16^3,18^2,20,22\right\} \\
\hline
183&n=6 &\begin{cases}&z_1 z_2+z_3 z_4\\&z_2^6+z_1 z_3+z_4^3+z_5^3\,\end{cases}
& 10 &30 & 25&15 & \frac{359}{6}&\frac{725}{12} &5& 2 &  
  \left\{2^5,3^2,4^5,5^3,6^3,7,8^3,9,10,12\right\} \\
\hline
   \end{array}
   $$
\caption{Fully smooth models with 5d CB rank $r=9,10$.\label{smoothableModels8}}
\end{sidewaystable}

\section{Computation of link topology for ICIS}
\label{app:link-top}

To compute the quantity $H_2(\Sigma_5,\mb{Z})=\mb{Z}^f\oplus(\frak{f})^2$ in (\ref{link-top}), we briefly summarize the algorithm developed in \cite{RANDELL1975347}.

Let
\be
f_i(z_1,...,z_{n+m})=\sum_{j=1}^{n+m}\alpha_{ij}z_j^{a_{ij}},i=1,...,m
\ee
be a collection of complex polynomials. Let $V=\bigcap_{i=1}^m V_i$, where $V_i$ is the locus of zeroes of $f_i$. 

With $d_i= \text{lcm} \left(a_{i 1}, \ldots, a_{i, n+m}\right)$ and $q_{i j}=d_i / a_{i j}$
we suppose that
(i) $V$ is the complete intersection of the $V_i$;
(ii) $V$ has an isolated singularity at the origin;
(iii) $q_{i j}$ is independent of $i$. (and so we define $q_j\equiv q_{i j}$ ).
If (i), (ii), (iii) hold, we will say that $V$ is a generalized Brieskorn variety and $K=V \cap S^{2(n+m)-1} \subset C^{n+m}$ is a generalized Brieskorn manifold. 

Now we state the algorithm computing $H_{n-1}(K,\mb{Z})$:
Let $J=\{1,2, \ldots, n+m\}$ and let $J_s$ denote any ordered $s$-element subset of $J$. That is, if $J_s=\left\{j_1, \ldots, j_s\right\}$ we require $j_1<j_2<\cdots<j_s$. We let $I_t=\left\{i_1, \ldots, i_t\right\}$ be any ordered $t$-element subset of $J_s$. $A=\left(a_{i j}\right)$ is an $m \times(n+m)$ matrix. Let $A\left(J_s\right)$ be the $m \times s$ matrix consisting of columns $j_1, \ldots, j_s$ of $A$. Let $K\left(J_s\right)=\left\{z \in K \mid z_j=0\right.$ for $\left.j \notin J_s\right\}$.

For some particular $J_s$, we define $\kappa(J_s)=\kappa\left(K\left(J_s\right)\right)=\kappa\left(A\left(J_s\right)\right)=\operatorname{rank}  H_{s-m-1}\left(K\left(J_s\right)\right)$ and $\kappa(\emptyset)=1$. To compute $\kappa(J_s)$, we consider the matrix $A\left(J_s\right)$. Same as before, define $d_1= \text{lcm} \left(a_{1 1}, \ldots, a_{1, s}\right)$ and $q_{j}=d_1 / a_{1 j}$. Let $t_k= \text{gcd} \left(q_1, \ldots, \hat{q}_k, \ldots, q_{s}\right)$, $q_j^{\prime}=q_j / \prod_{k \neq j} t_k$, $d_{i}^{\prime}=\text{lcm} \left(a_{i 1} / t_1, \ldots, a_{i, s} / t_{s}\right)$. Let $D^{\prime}=d_1^{\prime} \ldots d_m^{\prime}$ and $Q^{\prime}=q_1^{\prime} \ldots q_{s}^{\prime}$. Finally, let $r_p$ be the coefficient of $x^p$ in $\prod_{i=1}^m\left(1-d_i^{\prime} x+\right.$ $\left.\left(d_i^{\prime}\right)^2 x^2-\cdots\right)$ and let $J_k$ run over all $k$-element subsets of $J_s$. Then 
\be
\begin{aligned}
e\left(J_s\right)= & \frac{D^{\prime}}{Q^{\prime}}\left\{\left[\sum_{k=0}^{s-m-1} \sum_{p=0}^{s-m-1-k}\left(r_p\left(\begin{array}{c}
s-k \\
s-m-1-p-k
\end{array}\right) \sum_{J_k \subset J_s} \prod_{j \in J_k}\left(q_j^{\prime}-1\right)\right)\right]\right. \\
& \left.+
\left[\sum_{\substack{J_k \subset J_s \\
k>m}} \sum_{p=0}^{k-1-m} r_p  \left(\text{gcd} \left\{q_j^{\prime} | j \in J_k\right\}-1\right)\left(\begin{array}{c}
k \\
k-m-1-p
\end{array}\right)\right]\right\}. 
\end{aligned}
\ee
and $\kappa(J_s)=(-1)^{s-m-1}(e(J_s)-s+m)$.

Now come back to the original $A$. Let $C_i(\emptyset)= \text{gcd} \left(a_{i 1}, \ldots, a_{i, n+m}\right)$ and inductively define, for $J_s \subset J$,
\be
C_i\left(J_s\right)=\frac{\text{gcd}\left\{a_{i j} \mid j \notin J_s\right\}}{\prod_{I_t \subset J_s, t<s} C_i\left(I_t\right)} .
\ee
Indeed, $C_i(J_s)$ is independent of $i$ unless $J_s=\emptyset$. So we define $C(J_s)=C_i(J_s)$ and $C(\emptyset)=\prod_{i=1}^m C_i(\emptyset) \cdot \prod_{0<t<m} C\left(I_t\right)^{m-t}$.

We now let $b_j=\prod_{k\left(J_s\right)\geqslant j} C\left(J_s\right)$, where $k(J_s)=\kappa(J_s)$, if $n+m-s$ is odd and $k(J_s)=0$ otherwise. Let $r=\max \left\{k\left(J_s\right) \mid J_s \subset J\right\}$. Then we have $Tor(H_{n-1}(K ; Z))=\mathbf{Z} / b_1 \mathbf{Z} \oplus \cdots \oplus \mathbf{Z} / b_r \mathbf{Z}$.

Apply the algorithm described above, and we can get the results in \ref{sec:linktop}. One subtlety here is that this method is not valid for Class (2). The reason is that there is an additional constraint given in \cite{randell1975index}: consider the matrix $A(J)$ and define $q'_i$ as before, we require that $\text{gcd}\left\{q_i'|i \in I\right\}=1$ for all $m+1$ element subsets $I$ of $J$.

\section{Integral spectrum models with singular divisors}
\label{app:sin-div}
When walking through the landscape of models with integral 4d Coulomb branch spectrum and $r\leqslant 10$, we find some cases which can only admit a smooth resolution with singular compact divisors. They are Class (263), Class (265), Class (123) (with $n=3$) and Class (131) (with $n=6$). 

The first two of them satisfies $b_3=0$. Below we discuss them in detail.
\subsection{Class (263)}
\be
\begin{cases}
& z_1 z_2+z_3 z_4+z_5^2=0 \\
& z_1 z_5+z_3^2+z_4^5+z_2^{20}=0\,.
\end{cases}
\ee
\be
\begin{array}{||c|c|c||c|c|c||c|c|c||}
\hline
 r & d_H & G_F & \h r & \h d_H & f & b_3 & a&c\\
\hline
 7 & 91 & E_8 & 83 & 15 & 8 & 0 & \frac{2439}{8}&\frac{611}{2}\\
\hline
\end{array}
\ee
We have the 5d magnetic quiver:
\be
 \begin{tikzpicture}[x=.5cm,y=.5cm]
\draw[ligne, black](3,0)--(-3,0);
\draw[ligne, black](1,0)--(1,1);
\draw[ligne, black](2,0)--(2,1);
\node[] at (-5,0) {$\MQfive= $};
\node[bd] at (3,0) [label=below:{{\scriptsize$7$}}] {};
\node[bd] at (2,0) [label=below:{{\scriptsize$14$}}] {};
\node[bd] at (1,0) [label=below:{{\scriptsize$20$}}] {};
\node[bd] at (0,0) [label=below:{{\scriptsize$16$}}] {};
\node[bd] at (-1,0) [label=below:{{\scriptsize$12$}}] {};
\node[bd] at (-2,0) [label=below:{{\scriptsize$8$}}] {};
\node[bd] at (-3,0) [label=below:{{\scriptsize$4$}}] {};
\node[bd] at (2,1) [label=right:{{\scriptsize$1$}}] {};
\node[bd] at (1,1) [label=left:{{\scriptsize$10$}}] {};
\end{tikzpicture} 
\ee
We can read $G_F=E_8$ from the bottom layer of the Hasse diagram.

The resolution sequence is
\be
\ba
&(z_1,z_2,z_3,z_4,z_5;\delta_1)\cr
&(z_1^{(4)},z_3^{(3)},z_4^{(1)},z_5^{(2)},\delta_1^{(1)};\delta_4)\cr
&(z_1^{(4)},z_3^{(3)},z_4^{(1)},z_5^{(2)},\delta_4^{(1)};\delta_6)\cr
&(z_1^{(4)},z_3^{(3)},z_4^{(1)},z_5^{(2)},\delta_6^{(1)};\delta_7)\cr
&(z_1,z_3,z_5,\delta_1;\delta_2)\cr
&(z_1,z_3,z_5,\delta_4;\delta_5)\cr
&(z_1,z_3,\delta_2;\delta_3)\,.
\ea
\ee
The resolved equation is
\be
\begin{cases}
&z_1 z_2+z_3 z_4+z_5^2\delta_2\delta_5=0\\
&z_1 z_5+z_3^2\delta_3+z_4^5\delta_1^3\delta_2\delta_4^2\delta_6+z_2^{20}\delta_1^{18}\delta_2^{16}\delta_3^{15}\delta_4^{12}\delta_5^{10}\delta_6^6 =0\,.
\end{cases}
\ee
We can deduce the following intersection diagram:
\be
\includegraphics[height=6cm]{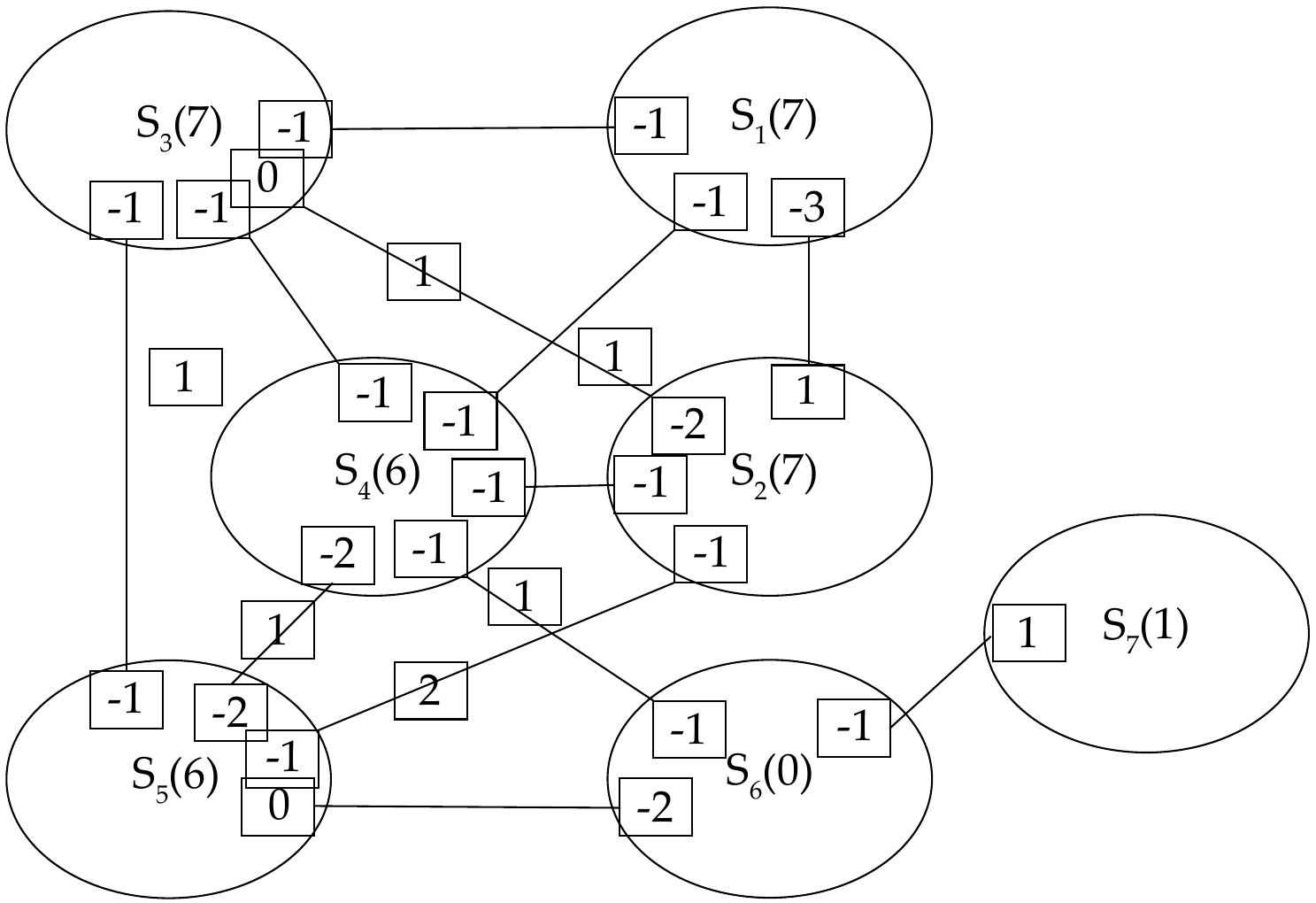}
\ee
Note that in the diagram we have

$S_1\cdot S_4\cdot S_2=S_1\cdot S_4\cdot S_3=S_4 \cdot S_2 \cdot S_5=S_4\cdot S_5\cdot S_3=S_2\cdot S_5\cdot S_3=1$,

$S_4 \cdot S_{6}\cdot S_{5}=2$.

The situation is this case is similar to the example in section 3.2.2 of \cite{Closset:2020scj}. The divisor $S_6:\delta_6=0$ is a $\mb{P}^1$ ruling over a nodal curve $z_3^2-z_3 z_5-z_5^3\delta_5=0$ locally. 

\subsection{Class (265)}
\be
\begin{cases}
& z_1 z_2+z_3 z_4+z_5^4=0 \\
& z_1 z_5+z_3^2+z_4^3+z_2^{24}=0\,.
\end{cases}
\ee
\be
\begin{array}{||c|c|c||c|c|c||c|c|c||}
\hline
 r & d_H & G_F & \h r & \h d_H & f & b_3 & a&c\\
\hline
 10 & 110 & E_8 & 102 & 18 & 8 & 0 & \frac{1785}{4}&447\\
\hline
\end{array}
\ee
We have the 5d magnetic quiver:
\be
 \begin{tikzpicture}[x=.5cm,y=.5cm]
\draw[ligne, black](3,0)--(-3,0);
\draw[ligne, black](-1,0)--(-1,1);
\draw[ligne, black](0,0)--(0,1);
\node[] at (-5,0) {$\MQfive= $};
\node[bd] at (3,0) [label=below:{{\scriptsize$5$}}] {};
\node[bd] at (2,0) [label=below:{{\scriptsize$10$}}] {};
\node[bd] at (1,0) [label=below:{{\scriptsize$15$}}] {};
\node[bd] at (0,0) [label=below:{{\scriptsize$20$}}] {};
\node[bd] at (-1,0) [label=below:{{\scriptsize$24$}}] {};
\node[bd] at (-2,0) [label=below:{{\scriptsize$16$}}] {};
\node[bd] at (-3,0) [label=below:{{\scriptsize$8$}}] {};
\node[bd] at (0,1) [label=right:{{\scriptsize$1$}}] {};
\node[bd] at (-1,1) [label=left:{{\scriptsize$12$}}] {};
\end{tikzpicture} 
\ee
We can read $G_F=E_8$ from the bottom layer of the Hasse diagram.

The resolution sequence is
\be
\ba
&(z_1,z_2,z_3,z_4,z_5;\delta_1)\cr
&(z_1^{(3)},z_3^{(2)},z_4^{(1)},z_5^{(1)},\delta_1^{(1)};\delta_4)\cr
&(z_1^{(5)},z_3^{(3)},z_4^{(2)},z_5^{(1)},\delta_4^{(1)};\delta_7)\cr
&(z_1^{(5)},z_3^{(3)},z_4^{(2)},z_5^{(1)},\delta_7^{(1)};\delta_9)\cr
&(z_1^{(5)},z_3^{(3)},z_4^{(2)},z_5^{(1)},\delta_9^{(1)};\delta_{10})\cr
&(z_1^{(2)},z_3^{(1)},z_4^{(1)},\delta_1^{(1)};\delta_3)\cr
&(z_1^{(2)},z_3^{(1)},z_4^{(1)},\delta_4^{(1)};\delta_5)\cr
&(z_1^{(2)},z_3^{(1)},z_4^{(1)},\delta_7^{(1)};\delta_8)\cr
&(z_1,z_3,\delta_1;\delta_2)\cr
&(z_1,z_3,\delta_5;\delta_6)\,.
\ea
\ee
The resolved equation is
\be
\begin{cases}
&z_1 z_2+z_3 z_4+z_5^4\delta_1^2\delta_2\delta_4^3\delta_5\delta_7^2\delta_9=0\\
&z_1 z_5+z_3^2\delta_2\delta_6+z_4^3\delta_1\delta_3^2\delta_5\delta_8+z_2^{24}\delta_1^{22}\delta_2^{21}\delta_3^{20}\delta_4^{18}\delta_5^{16}\delta_6^{15}\delta_7^{12}\delta_8^{10}\delta_9^6 =0\,.
\end{cases}
\ee
We can deduce the following intersection diagram:
\be
\includegraphics[height=9cm]{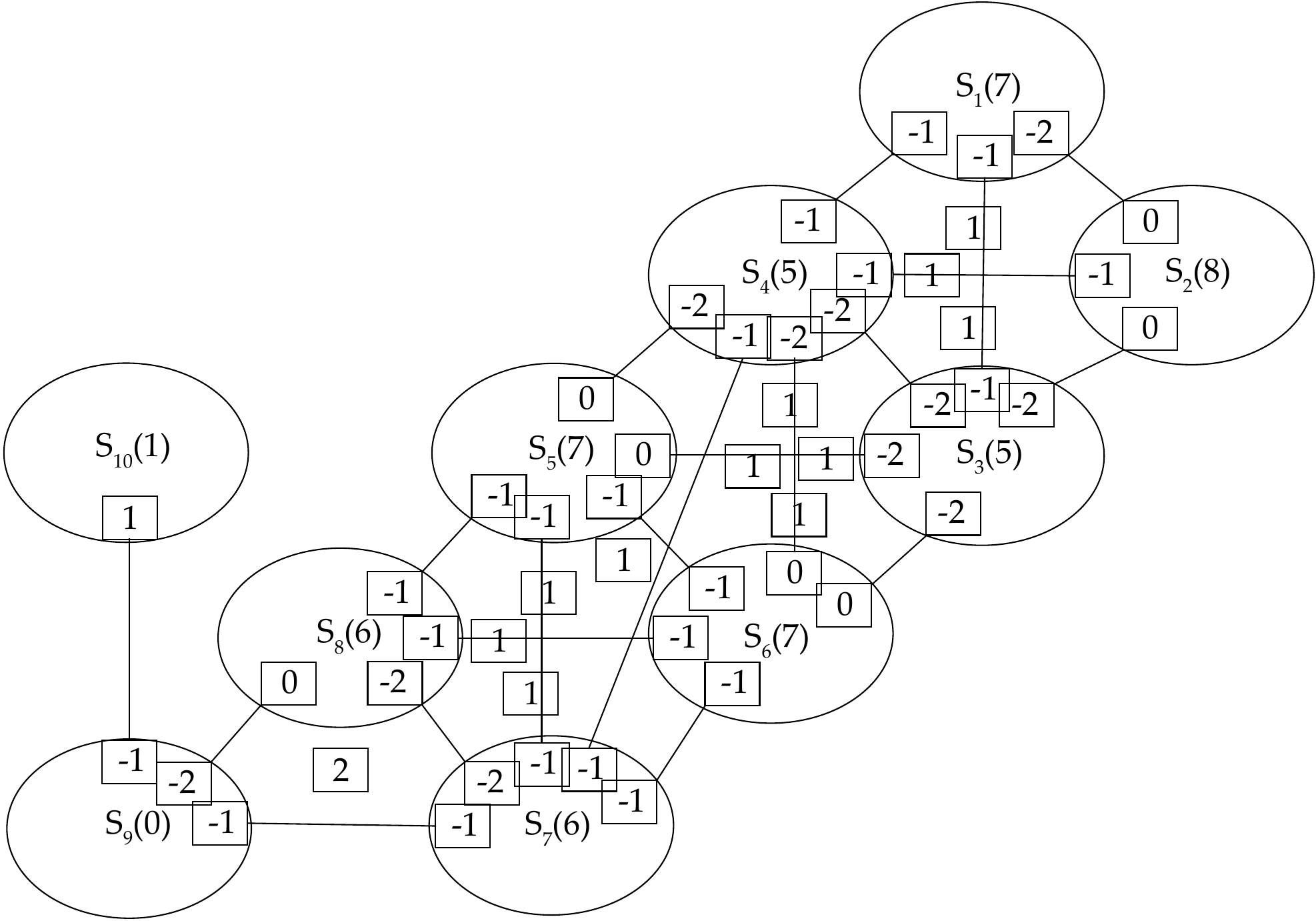}
\ee
Note that in the diagram we have

$S_1\cdot S_4\cdot S_3=S_1\cdot S_4\cdot S_2=S_4 \cdot S_7 \cdot S_5=S_4\cdot S_7\cdot S_6=S_4\cdot S_3\cdot S_5=S_4\cdot S_3\cdot S_2=S_4\cdot S_3\cdot S_6=S_7\cdot S_5\cdot S_8=S_{7}\cdot S_{8}\cdot S_6=S_{3}\cdot S_5\cdot S_6=S_{5}\cdot S_8\cdot S_6=1$,

$S_7 \cdot S_{9}\cdot S_{8}=2$.

The situation is this case is similar to the example in section 3.2.2 of \cite{Closset:2020scj}. The divisor $S_9:\delta_9=0$ is a $\mb{P}^1$ ruling over a nodal curve $z_3^2-z_3 z_4+z_4^3\delta_8=0$ locally. 

\section{An example with $a>c$: Class (41)}
\label{sec:41}

In this section we present an example whose $\FTfour$ satisfies $a>c$:
\be
\begin{cases}
&z_2^2+z_3^2+z_4^2+z_5^2=0 \\
&z_1^2+z_2^3+z_3^3+z_4^3=0\,.
\end{cases}
\ee
\be
\begin{array}{||c|c||c|c|c||c|c|c|c|c||}
\hline
 r & d_H & \h r & \h d_H & f & b_3 & n_v&n_h & a & c\\
\hline
 2 & 16 & 15& 3 & 1 & 8 & 73&72 & \frac{437}{24}&\frac{109}{6}\\
\hline
\end{array}
\ee

The resolution sequence is
\be
\ba
&\{z_1,z_2,z_3,z_4,z_5;\delta_1\}\cr
&\{z_1,\delta_1;\delta_2\}\cr
&\{\delta_1,\delta_2;\delta_3\}\,.
\ea
\ee
The two irreducible exceptional divisors are $S_1:\delta_2=0$, $S_2:\delta_3=0$, which are all smooth. The intersection numbers are
\be
S_1^3=-24\ ,\ S_2^3=8\ ,\ S_1^2\cdot S_2=18\ ,\ S_1\cdot S_2^2=-12\,.
\ee
$S_1$ is a $\mb{P}^1$-fibration over a genus-4 curve $\Sigma=S_1\cdot S_2$. $S_2$ is a Hirzebruch surface. $S_1$ and $S_2$ intersects at $\Sigma$. Hence the $b_3=8$ 3-cycles origin from $\mb{P}^1$ fibration over the 8 1-cycles on the genus-4 curve $\Sigma$.

The 5d SCFT has a IR description of $Sp(2)+2\mbf{AS}+2\mbf{F}$ (or equivalently $G_2+4\mbf{F}$/$SU(3)_5+4\mbf{F}$), see for example Appendix B of \cite{Apruzzi:2019opn}. More precisely, since the resolved CY3 $\widetilde{X}_3$ has a non-zero $b_3$, the classical Coulomb branch has quantum correction from EM2-branes wrapping the 3-cycles. One can alternatively consider the conifold transition to a resolved CY3 $\widetilde{X}'$ without 3-cycles, which describes a 5d SCFT with $r=1$, $f=5$ and IR gauge theory description $Sp(2)+2\mbf{AS}+2\mbf{F}$. Nonetheless, $\widetilde{X}'$ is the resolution of a different (unknown) CY3 singularity.

%

\bibliography{FM}
\bibliographystyle{JHEP}
\end{document}